\documentclass[aps,pra,reprint,twocolumn,superscriptaddress,floatfix,nofootinbib,longbibliography]{revtex4-1}
\usepackage{graphicx,amsmath,amsfonts,amssymb,amsthm,xr}
\usepackage{epsfig,amsmath,amssymb,color,dsfont,upgreek,physics}
\usepackage{mathrsfs}
\usepackage{mathtools}
\usepackage{bbold}
\usepackage{comment}
\usepackage{float}

\usepackage[bookmarks=true,colorlinks,linkcolor=OrangeRed,urlcolor=NavyBlue,citecolor=RoyalBlue]{hyperref}
\usepackage[dvipsnames]{xcolor}

\definecolor{mygold}{rgb}{0.93,0.69,0.13}

\definecolor{mypurple}{rgb}{0.49,0.18,0.56}

\definecolor{mygreen}{rgb}{0,0.5,0}

\definecolor{mygreen}{rgb}{0,0.5,0}
\definecolor{myred}{rgb}{0.7,0,0}

\begin{document}
\title{\texorpdfstring{Stabilizing Gauge Theories in Quantum Simulators: A Brief Review \\
\vspace{0.1cm}
\begin{minipage}{0.7\textwidth}
\centering
\textmd{\small{\it Invited Contribution to Proceedings of the Quantum Simulation for Strong Interactions (QuaSi) Workshops 2021~\cite{quasiproceedingsSite} at the InQubator for Quantum Simulation (IQuS)}}
\end{minipage}
\vspace{-0.1cm}
}{Stabilizing Gauge Theories in Quantum Simulators: A Brief Review}
}
\author{Jad C.~Halimeh}
\email{jad.halimeh@physik.lmu.de}
\affiliation{Department of Physics and Arnold Sommerfeld Center for Theoretical Physics (ASC), Ludwig-Maximilians-Universit\"at M\"unchen, Theresienstra\ss e 37, D-80333 M\"unchen, Germany}
\affiliation{Munich Center for Quantum Science and Technology (MCQST), Schellingstra\ss e 4, D-80799 M\"unchen, Germany}
\author{Philipp Hauke}
\email{philipp.hauke@unitn.it}
\affiliation{INO-CNR BEC Center and Department of Physics, University of Trento, Via Sommarive 14, I-38123 Trento, Italy}
\affiliation{INFN-TIFPA, Trento Institute for Fundamental Physics and Applications, Trento, Italy}

\begin{abstract}
Quantum simulation is at the heart of the ongoing ``second'' quantum revolution, with various synthetic quantum matter platforms realizing evermore exotic condensed matter and particle physics phenomena at high levels of precision and control. The implementation of gauge theories on modern quantum simulators is especially appealing due to three main reasons: (i) it offers a new probe of high-energy physics on low-energy tabletop devices, (ii) it allows exploring condensed matter phenomena that are prominent in gauge theories even without a direct connection to high-energy physics, and (iii) it serves as a banner of experimental benchmarking given the plethora of local constraints arising from the gauge symmetry that need to be programmed and controlled. In order to faithfully model gauge-theory phenomena on a quantum simulator, stabilizing the underlying gauge symmetry is essential. In this brief review, we outline recently developed experimentally feasible methods introduced by us that have shown, in numerical and experimental benchmarks, reliable stabilization of quantum-simulator implementations of gauge theories. We explain the mechanism behind these \textit{linear gauge protection} schemes, and illustrate their power in protecting salient features such as gauge invariance, disorder-free localization, quantum many-body scars, and other phenomena of topical interest. We then discuss their application in experiments based on Rydberg atoms, superconducting qubits, and in particular ultracold neutral atoms in optical superlattices. We hope this review will illustrate some facets of the exciting progress in stabilization of gauge symmetry and in gauge-theory quantum simulation in general.  
\end{abstract}

\date{\today}
\maketitle 
\tableofcontents
\section{Introduction}\label{sec:Intro}
The new quantum revolution is currently in full swing with major advancements in various setups of synthetic quantum matter (SQM), in which levels of control and precision at the resolution of a single atom are being reached \cite{Bloch2008,Bakr2009}. This progress opens the realistic prospect of employing quantum simulators for scientific discovery \cite{Hauke2012,Alexeev_review,klco2021standard}. Of particular recent interest for implementations on SQM devices are gauge theories, which enable accessible investigations into various high-energy physics questions as well as phenomena relevant to condensed matter physics \cite{Pasquans_review,Dalmonte_review,Zohar_review,aidelsburger2021cold,Zohar_NewReview,Bauer_review}. Indeed, gauge theories form a fundamental framework of modern physics. They encode nature's laws through local constraints, which enforce a strict relationship between the distribution of charged matter and the surrounding gauge field \cite{Weinberg_book,Gattringer_book,Zee_book}, and which are manifest in the gauge symmetry underlying these models. 
A paradigmatic example is quantum electrodynamics (QED), where the conservation of gauge symmetry is embodied in the well-known Gauss's law, which forces electric field lines to emerge from positive charges and end in negative charges. While Gauss's law is a postulate in model descriptions of nature, it has to be programmed into quantum simulators of gauge theories.

There are currently two main approaches for realizing quantum simulations of gauge theories:
The first one uses Gauss's law to integrate out redundant degrees of freedom, i.e., it expresses the state of the electric fields through charge configurations or vice versa \cite{Hamer1997,Banuls2013,BanulsPos2013,Saito2015,Banuls2016}. This approach led to the first successful quantum simulations of lattice gauge theories in trapped ions \cite{Martinez2016} and Rydberg atoms \cite{Bernien2017,Surace2020}, which were followed briefly thereafter by various other experiments and refined proposals \cite{Klco2018,Kokail2019,Zohar2019,Klco2020,Ciavarella2021,Atas2021,Nguyen2021}. This approach has advantages in making efficient use of resources and  it ensures the conservation of Gauss's law by hardcoding it into the simulator. However, the resultant interactions can sometimes be of complicated local or nonlocal form, and errors in the quantum devices can lead to unphysical interpretations, such as a particle with infinite attached electric-field string appearing out of the vacuum \cite{Martinez2016,Muschik2017}.
The second approach, which historically was at the basis of the first series of proposals for gauge-theory quantum simulations \cite{Zohar2011,Zohar2012,Banerjee2012,Tagliacozzo2012,Zohar2013,Banerjee2013,Tagliacozzo2013,Hauke2013}, keeps both matter and gauge fields as dynamical degrees of freedom in the quantum simulator. This gives some flexibility in realizations, e.g., through degenerate perturbation theory, and has led to successful experiments in various platforms \cite{Schweizer2019,Goerg2019,Mil2020,Yang2020,Zhou2021,Wang2021,Mildenberger2022}. Maybe even more interestingly, as Gauss's law is not imposed \emph{a priori}, this second approach permits one to test fundamental questions about the possible emergence of gauge invariance in nature \cite{Foerster1980,Poppitz2008,Wetterich2017,Witten2018,Barcelo2021,Bass2022}. 
However, Gauss's law then needs to be stabilized by some other means. 

In this article, we review recent theoretical and experimental progress in stabilizing gauge symmetry in quantum simulators, embarking from our personal and thus necessarily limited point of view. 
We focus mostly on low-dimensional systems, where the bulk of current experimental implementations have been realized and where numerical benchmarks are possible thanks to powerful tensor-network methods, as well as on Abelian symmetries. 
In particular, we review the framework of \emph{linear gauge protection}, where simple single-particle terms proportional to Gauss's law are added to a faulty Hamiltonian, which dynamically stabilize gauge symmetry. 
These are akin to, but experimentally significantly simpler for nonequilibrium quantum simulation, than energy penalties that are quadratic in Gauss's law, which force the low-energy manifold to approximately obey Gauss's law \cite{Zohar2011,Zohar2012,Banerjee2012,Zohar2013,Banerjee2013,Hauke2013,Stannigel2014,Kuehn2014,Kuno2015,Yang2016,Kuno2017,Negretti2017,Dutta2017,Barros2019,Halimeh2020a}.
We discuss what has been shown for linear gauge protection rigorously analytically as well as in numerical studies through exact diagonalization (addressing long times) and infinite matrix product states (addressing systems in the thermodynamic limit).   
We further illustrate the power of linear gauge protection for stabilizing salient phenomena such as disorder-free localization (DFL) and quantum many-body scars (QMBS), and we give a glimpse on recent experimental implementations and proposals.

In this proceedings article, we focus mostly on works we have been personally involved in, but it is worth stressing that the program of ensuring gauge invariance in quantum simulators is a community effort that is seeing a strong current momentum. 
Other closely related approaches for protecting gauge symmetry employ, e.g., engineered dissipation \cite{Stannigel2014}, dynamical decoupling \cite{Kasper2021nonabelian}, or pseudo-random gauge transformations \cite{Lamm2020}. 
Symmetry protection in a general setting, i.e., also of global symmetry, is also a prospect of large current interest \cite{Chubb2017,Tran2021}, and has also been employed beneficially in a trapped-ion simulation of a lattice gauge theory \cite{Nguyen2021}. 
Finally, the complexity of models inspired from the interacting degrees of freedom of gauge theories, but without respecting the underlying gauge symmetry, can give rise to intriguing many-body phenomena and is worthwhile of investigations in its own right \cite{GonzalezCuadra2019,GonzalezCuadra2020}. 

This review is structured as follows: In Sec.~\ref{sec:setting}, we introduce the setting of our problem, which involves \textit{faulty} gauge theories with dynamical matter and gauge fields, and we review previous methods utilized to stabilize them on quantum simulators. In Sec.~\ref{sec:LinPro}, we discuss the concept of linear gauge protection based on local generators and also on local ``pseudogenerators''. We showcase how linear gauge protection stabilizes quantum many-body scars and disorder-free localization in Sec.~\ref{sec:phenomena}. In Sec.~\ref{sec:experiment}, we discuss experimental applications and proposals involving linear gauge protection. Finally, in Sec.~\ref{sec:conc}, we conclude and provide an outlook.

\section{Setting}\label{sec:setting}
\begin{figure}[t!]
    \centering
    \includegraphics[width=\columnwidth]{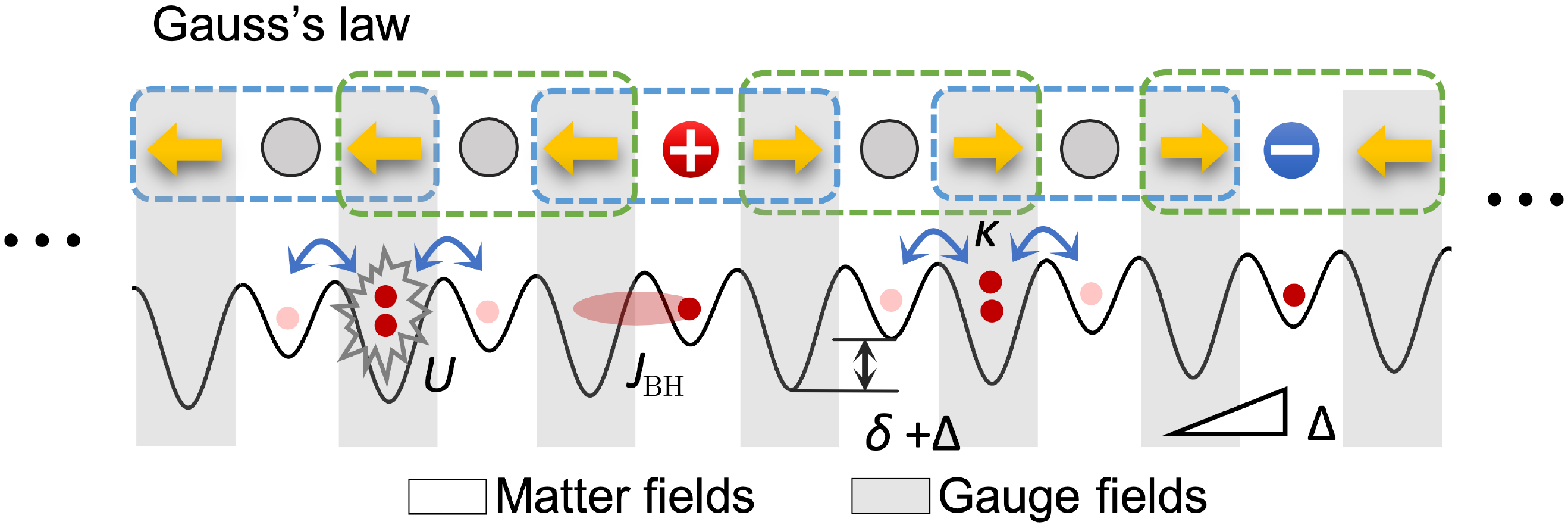}
    \caption{(Color online). An optical-lattice quantum simulator of the spin-$1/2$ $\mathrm{U}(1)$ quantum link model. Shallow (deep) sites represent matter (gauge) sites. A small tilt $\Delta$, a strong on-site interaction strength $U$, and the staggered potential $\delta$ lead to an effective linear gauge protection term in this realization (see text). As a result, at each triple of matter site plus adjacent gauge links, only configurations allowed by the gauge symmetry (Gauss's law) are dynamically accessible. Figure adapted from Ref.~\cite{Zhou2021}.}
    \label{fig:BHM}
\end{figure}

\begin{figure}[t!]
    \centering
    \includegraphics[width=\columnwidth]{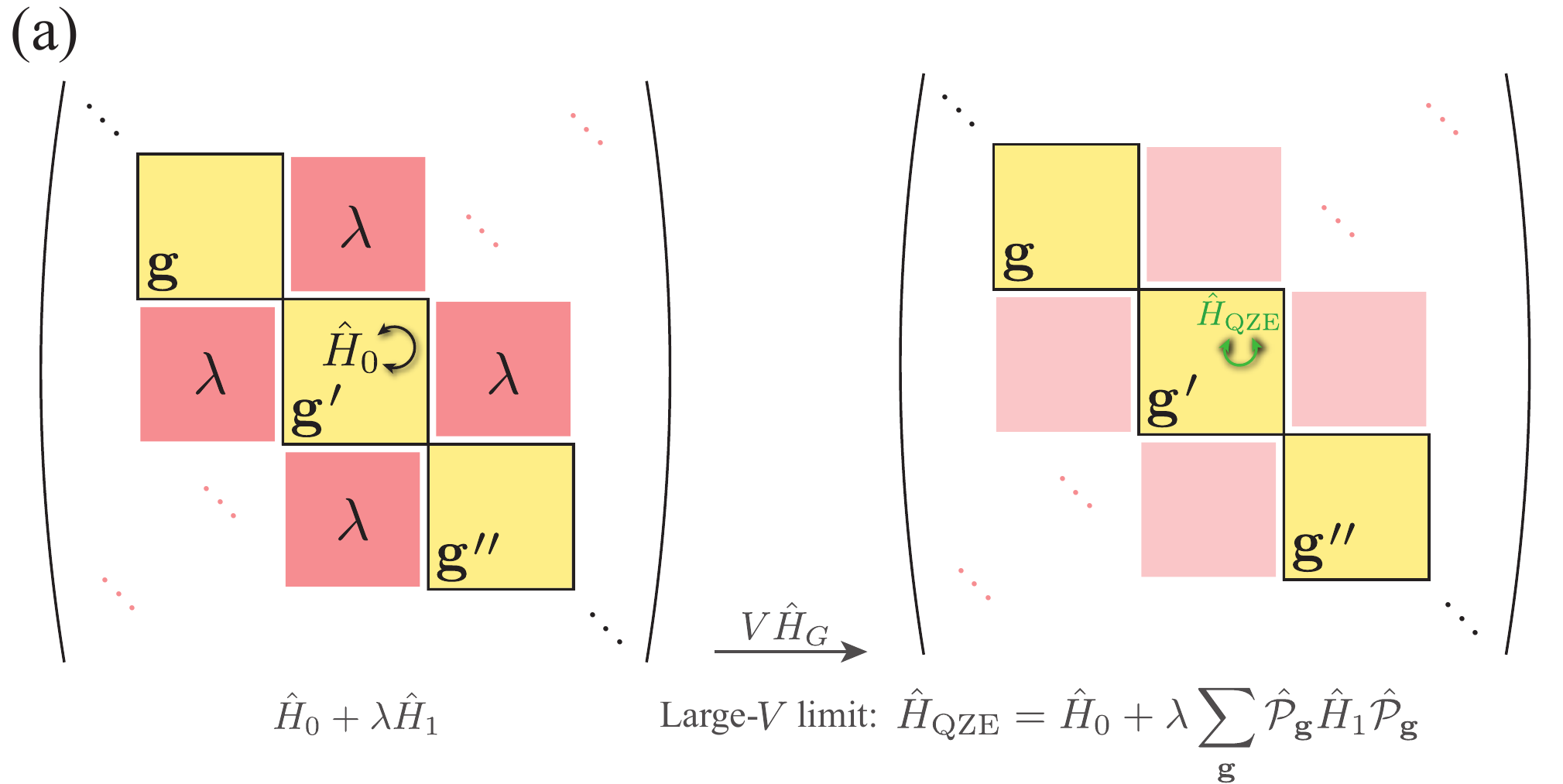}\\
    \includegraphics[width=\columnwidth]{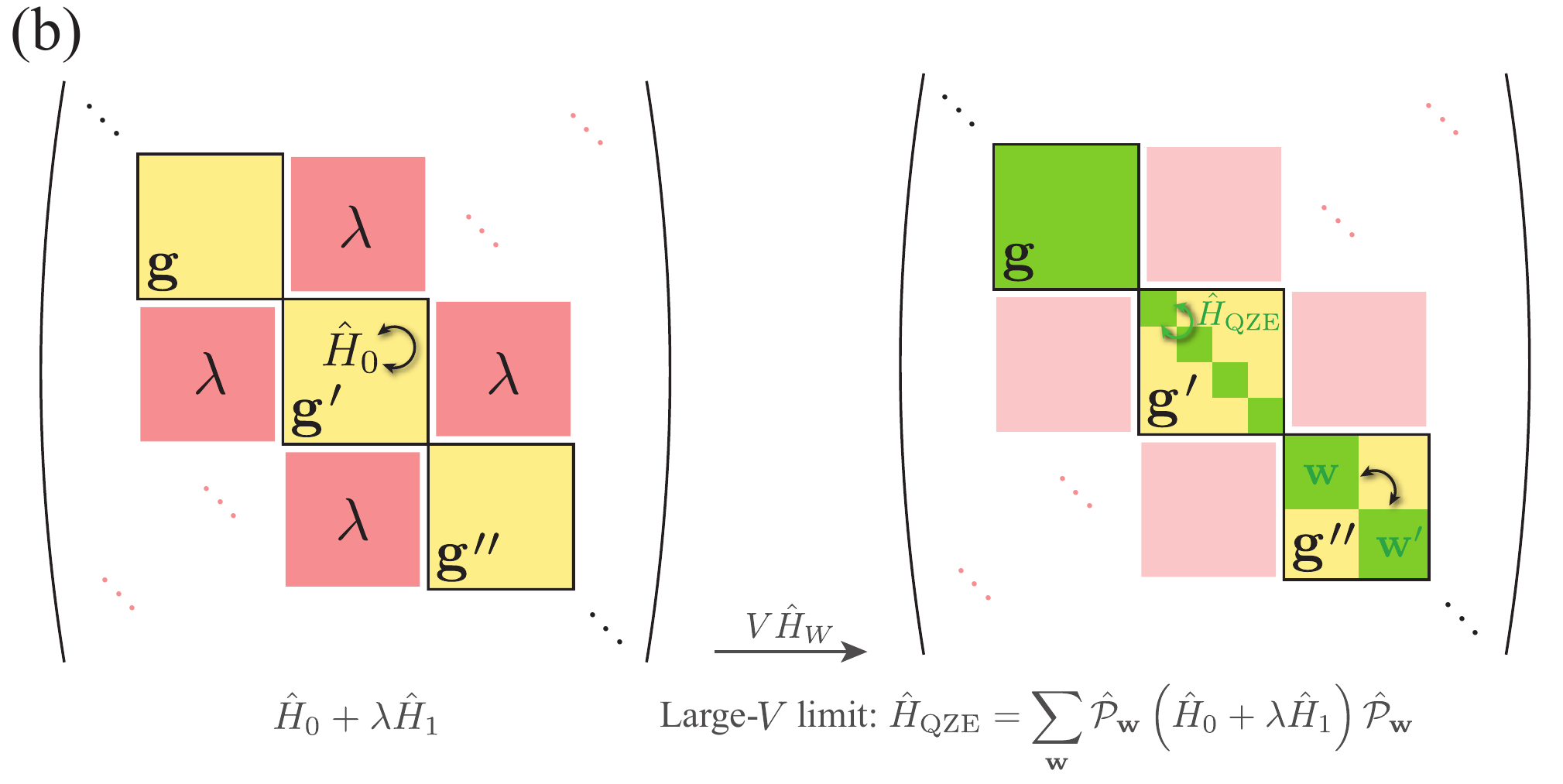}
    \caption{(Color online). Schematic illustration of linear gauge protection in the case of a \textit{faulty} gauge-theory implementation $\hat{H}_0+\lambda\hat{H}_1$ on a synthetic quantum matter platform. Here, $\hat{H}_0$ is the ideal gauge theory and $\lambda\hat{H}_1$ are unavoidable gauge-breaking errors arising from experimental imperfections or subleading orders in the mapping of the ideal theory onto the quantum-simulation platform. Yellow blocks indicate gauge superselection sectors $\mathbf{g}=(g_1,g_2,\ldots,g_L)$ of the gauge symmetry, with projectors $\hat{\mathcal{P}}_\mathbf{g}$, corresponding to conserved eigenvalues $g_j$ of the local symmetry generators $\hat{G}_j$. The error terms $\lambda\hat{H}_1$ generate first-order couplings between adjacent gauge superselection sectors, indicated by red blocks. (a) Upon adding the linear gauge protection $V\hat{H}_G$~\eqref{eq:HG} based on the local generator $\hat{G}_j$, gauge violations are suppressed and a renormalized gauge theory emerges that hosts the same gauge symmetry as $\hat{H}_0$, and which faithfully reproduces the dynamics up to a timescale linear in $V$ at the least, as predicted from the concept of the quantum Zeno dynamics. (b) If instead the linear gauge protection is based on the local pseudogenerator $\hat{W}_j$ as in Eq.~\eqref{eq:HW}, the emergent gauge theory has an enriched symmetry due to $\hat{W}_j$, whose superselection sectors are $\mathbf{w}$ (denoted by green blocks) with projectors $\hat{\mathcal{P}}_\mathbf{w}$. Figure is adapted from Ref.~\cite{Halimeh2021enhancing}.}
    \label{fig:schematic}
\end{figure}

In this work, we focus on Abelian lattice gauge theories (LGTs), described ideally, i.e., in an error-free setting, by Hamiltonian $\hat{H}_0$ [explicit examples are given further below, see Eqs.~\eqref{eq:U1QLM} and \eqref{eq:Z2LGT}]. The system has $L$ lattice sites, where matter fields reside, and $L$ links, where gauge and electric fields exist. Unless otherwise stated, we will assume periodic boundary conditions throughout this work. The gauge symmetry of $\hat{H}_0$ is generated by the operators $\hat{G}_j$ (Gauss's law generators), and gauge invariance is encoded through the commutation relations $\big[\hat{H}_0,\hat{G}_j\big]=0,\,\forall j$. Gauge-invariant states $\ket{\phi}$ are simultaneous eigenstates of all local generators: $\hat{G}_j\ket{\phi}=g_j\ket{\phi},\,\forall j$. The eigenvalues $g_j$ can be seen as background charges, a set of which over the volume of the system defines a gauge superselection sector $\mathbf{g}=(g_1,g_2,\ldots,g_L)$. A given superselection sector imposes a specific relation between matter occupation at site $j$ and the allowed electric-field configurations at its two neighboring links; see Fig.~\ref{fig:BHM}. In the basis of the generators $\hat{G}_j$, the Hamiltonian $\hat{H}_0$ can be block-diagonalized, with each block representing a unique superselection sector; see yellow blocks in Fig.~\ref{fig:schematic}. Due to its gauge symmetry, $\hat{H}_0$ can only drive dynamics within each superselection sector (intrasector dynamics) and cannot couple different superselection sectors (intersector dynamics).

In realistic SQM implementations of LGTs with dynamical matter and gauge fields \cite{Goerg2019,Schweizer2019,Mil2020,Yang2020,Zhou2021,Wang2021,Mildenberger2022} (the second approach mentioned in Sec.~\ref{sec:Intro}), gauge-breaking errors $\lambda\hat{H}_1$ of some strength $\lambda$, even if small, will unavoidably arise. If left uncontrolled, such errors, even when merely perturbative, can drastically modify the system dynamics at sufficiently long times or in its ground state, such that it can no longer be associated with a gauge theory \cite{VanDamme2020,Borla2020}. It is thus essential to devise \textit{experimentally feasible} approaches to stabilize the gauge symmetry in such SQM implementations. In recent years, various proposals have been presented that mitigate such unavoidable experimental errors through energetic constraints \cite{Zohar2011,Zohar2012,Banerjee2012,Zohar2013,Banerjee2013,Hauke2013,Stannigel2014,Kuehn2014,Kuno2015,Yang2016,Kuno2017,Negretti2017,Dutta2017,Barros2019,Halimeh2020a,Lamm2020}. 

Most of the original proposals involved turning the target superselection sector $\mathbf{g}^\text{tar}=(g^\text{tar}_1,g^\text{tar}_2,\ldots,g^\text{tar}_L)$ into a ground-state manifold by introducing a \textit{quadratic protection} term of the general form \cite{Halimeh2020a}
\begin{align}\label{eq:QuadPro}
V\hat{H}_\text{quad}=V\sum_{j=1}^L\big(\hat{G}_j-g^\text{tar}_j\big)^2,
\end{align}
with strength $V$. For a $\mathrm{U}(1)$ lattice gauge theory similar to QED in equilibrium [Eq.~\eqref{eq:U1QLM} below], this has been shown to lead to a \textit{gauge-symmetry violation quantum phase transition} in the system of Hamiltonian $\hat{H}_0+\lambda\hat{H}_1+V\hat{H}_\text{quad}$ at a critical value of $V/\lambda$, above which the system is in a phase characterized by a renormalized gauge theory that retains the critical features of the original model $\hat{H}_0$ \cite{VanDamme2020}, in the spirit of a Higgs transition \cite{Poppitz2008,Kuno2015,Bazavov2015,HeitgerPhDThesis} and similar to what has been discussed for topological states of matter \cite{Hastings2005,Sachdev2018}. 

Such quadratic protection schemes have also been demonstrated through classical simulations and analytic arguments to work reliably in out-of-equilibrium scenarios. A useful figure of merit to quantify the performance of the protection is the gauge violation, defined as the average variance of the Gauss's law generators with respect to the target sector, 
\begin{align}\label{eq:viol}
    \varepsilon(t)=\frac{1}{Lt}\int_0^t ds\sum_{j=1}^L\bra{\psi(s)}\big(\hat{G}_j-g_j^\text{tar}\big)^2\ket{\psi(s)},
\end{align}
where $\ket{\psi(t)}=e^{-i\hat{H}t}\ket{\psi_0}$ and $\hat{H}=\hat{H}_0+\lambda\hat{H}_1+V\hat{H}_\text{quad}$ is the \textit{faulty} theory. At sufficiently large $V$, the gauge violation can be shown to settle into a plateau of value $\propto\lambda^2/V^2$ at a timescale $\propto1/V$, which lasts up to all numerically accessible times for finite systems \cite{Halimeh2020a} and also in the thermodynamic limit \cite{vandamme2021reliability}. This method has even been shown to work well for non-Abelian gauge theories \cite{Halimeh2021gauge}.

However, a major drawback of the quadratic protection scheme is that Eq.~\eqref{eq:QuadPro} is very challenging to realize experimentally. Indeed, $\hat{G}_j^2$ usually includes terms that are at least quadratic (in case $\hat{G}_j$ is composed of at most single-body terms). Furthermore, quadratic protection necessarily energetically isolates a given target superselection sector from all its counterparts, which is not useful in applications where the dynamics is across several sectors simultaneously. In the following, we will review an approach we have developed that solves both these issues \cite{Halimeh2020e}.

\section{Linear gauge protection}\label{sec:LinPro}
Let us first consider a setting where we are interested in the dynamics occurring only within a given target superselection sector $\mathbf{g}^\text{tar}$. In an SQM implementation, this would entail quantum-simulating the quench dynamics of a given LGT, starting in a gauge-invariant initial state $\ket{\psi_0}$ residing in the target gauge sector: $\hat{G}_j\ket{\psi_0}=g_j^\text{tar}\ket{\psi_0},\,\forall j$. In an ideal situation, the time-evolved wave function $\ket{\psi(t)}$ should also reside in the superselection sector $\mathbf{g}^\text{tar}$, which would indeed be the case if the dynamics was propagated by only $\hat{H}_0$. However, as mentioned, in a realistic SQM implementation there will be unavoidable gauge-breaking errors $\lambda\hat{H}_1$ that will cause $\ket{\psi(t)}$ to ``leak'' out of the target sector. An experimentally feasible way to stabilize the gauge symmetry of the model in presence of these errors is to introduce protection terms \textit{linear} in the gauge-symmetry local generator \cite{Halimeh2020e}. 
In case the local generator is itself complicated to implement, a further experimental simplification can be obtained by enforcing the protection through a local \textit{pseudogenerator} (LPG) \cite{Halimeh2021stabilizing}.

\subsection{In terms of the local generator}
If the LGT possesses a local generator that is composed of at most single-body terms, its experimental addition to the Hamiltonian is straightforward and there is no need for an LPG. This is, e.g., the case of the spin-$S$ $\mathrm{U}(1)$ quantum link model (QLM) \cite{Chandrasekharan1997,Wiese_review,Kasper2017}
\begin{align}\nonumber
    \hat{H}_0=\sum_{j=1}^L\bigg[&\frac{J}{2\sqrt{S(S+1)}}\big(\hat{\sigma}^-_j\hat{s}^+_{j,j+1}\hat{\sigma}^-_{j+1}+\text{H.c.}\big)\\\label{eq:U1QLM}
    &+\frac{\mu}{2}\hat{\sigma}^z_j+\frac{\eta^2}{2}\big(\hat{s}^z_{j,j+1}\big)^2\bigg].
\end{align}
Here, the Pauli matrix $\hat{\sigma}^z_j$ at lattice site $j$ represents the matter field (after a Jordan--Wigner transformation from fermions to spins-$1/2$), while the spin-$S$ matrix $\hat{s}^z_{j,j+1}$ denotes the electric field at the link between sites $j$ and $j+1$. 
Further, $J$ sets the ``assisted-tunneling'' strength, corresponding to the generation of particle--anti-particle pairs on adjacent sites with an adjustment of the electric field according to the new charge configuration. Moreover, $\mu$ is the fermionic rest mass, and $\eta$ is the electric field coupling strength and sets the field energy. 
The above $\mathrm{U}(1)$ gauge theory is a spin-$S$ quantum link formulation \cite{Chandrasekharan1997} of the lattice Schwinger model, where the infinite-dimensional gauge fields of the latter are represented by rescaled finite-dimensional spin-$S$ ladder operators $\hat{s}^\pm_{j,j+1}/\sqrt{S(S+1)}$. Equation~\eqref{eq:U1QLM} converges to the lattice Schwinger model in the Kogut--Susskind limit $S\to\infty$. 
Often, even surprisingly small spin representations already allow for reliable extrapolations to the limit of the lattice Schwinger model \cite{Zache2021achieving,Halimeh2021achieving}, a feature that has been observed also in other truncation schemes of the gauge field \cite{Yang2016,Klco2018,Buyens2017,Banuls2018,Banuls2020}. The QLM formulation has the advantage over some of these that it preserves the canonical commutation relations between electric field and parallel transporter (exponential of vector potential) \cite{Chandrasekharan1997,Wiese_review,Kasper2017}. 

The generator of the $\mathrm{U}(1)$ gauge symmetry of Eq.~\eqref{eq:U1QLM} is
\begin{align}\label{eq:Gj}
    \hat{G}_j=(-1)^j\bigg[\hat{s}^z_{j-1,j}+\hat{s}^z_{j,j+1}+\frac{\hat{\sigma}^z_j+\mathds{1}}{2}\bigg],
\end{align}
which can be interpreted as a discretized version of Gauss's law from QED. Importantly, this generator is composed of only single-body terms, which add little experimental overhead in typical SQM setups \cite{Yang2020,Mildenberger2022}. We now introduce the linear gauge protection term
\begin{align}\label{eq:HG}
    V\hat{H}_G=V\sum_{j=1}^Lc_j\hat{G}_j.
\end{align}
The protection sequence $c_j$ is pivotal to the efficacy of the linear gauge protection~\eqref{eq:HG}. Ideally, it must be tailored in such a way that, at sufficiently large $V$, the target sector is energetically completely separated from other superselection sectors. This can be achieved when $c_j$ is composed of rational numbers obeying the \textit{compliance condition}
\begin{align}
    \sum_{j=1}^Lc_j\big(g_j-g_j^\text{tar}\big)=0\iff g_j=g_j^\text{tar},\,\forall j.
\end{align}
Indeed, as proven by the Gauge-Protection Theorem detailed in Ref.~\cite{Halimeh2020e}, such a sequence guarantees a controlled gauge violation~\eqref{eq:viol} up to a timescale exponential in a volume-independent $V$ \cite{abanin2017rigorous}. However, an experimentally nontrivial challenge arises in the case of such a compliant sequence $c_j$, as it would have to grow exponentially with system size for a fixed value of $V$. Thus, although $V$ itself is volume-independent \cite{Halimeh2020e}, $c_j$ is not. While this issue may be secondary for small-size realizations, it poses a problem for large-scale SQM implementations.

Nevertheless, dependent on the error model, one can employ simpler noncompliant periodic sequences, even as simple as $c_j=(-1)^j$. These can be shown analytically through the concept of quantum Zeno dynamics \cite{facchi2002quantum,facchi2004unification,facchi2009quantum,burgarth2019generalized} to stabilize gauge invariance up to timescales at least linear in $V$, during which the dynamics under the faulty LGT $\hat{H}=\hat{H}_0+\lambda\hat{H}_1+V\hat{H}_G$ is reproduced by the effective Zeno Hamiltonian
\begin{align}\label{eq:HQZE0}
    \hat{H}_\text{QZE}^\text{tar}=\hat{H}_0+\lambda\hat{\mathcal{P}}_{\mathbf{g}^\text{tar}}\hat{H}_1\hat{\mathcal{P}}_{\mathbf{g}^\text{tar}},
\end{align}
Here, $\hat{\mathcal{P}}_{\mathbf{g}^\text{tar}}$ is the projector onto the target sector $\mathbf{g}^\text{tar}$, i.e., the dynamics of the effective Zeno Hamiltonian falls into the ideal sector of conserved Gauss's law. 
The linear gauge protection~\eqref{eq:HG} with $c_j=(-1)^j$ has been demonstrated to suppress the gauge violation~\eqref{eq:viol} in the target sector up to all numerically accessible times in both finite systems through exact diagonalization (ED) \cite{Halimeh2020e} and in the thermodynamic limit using infinite matrix product state (iMPS) techniques \cite{vandamme2021reliability}, for experimentally relevant gauge-breaking errors.

It is important to note that Eq.~\eqref{eq:HQZE0} is the effective Hamiltonian when restricting to the target sector $\mathbf{g}^\text{tar}$. However, as we will discuss in Sec.~\ref{sec:DFL}, situations can be of interest where the initial state is a superposition of an extensive number of superselection sectors \cite{Smith2017,Brenes2018}, instead of a single one as above. In such cases, and with the appropriate sequence $c_j$, the Zeno Hamiltonian takes the more general form \cite{Halimeh2021stabilizingDFL}
\begin{align}\label{eq:HQZE}
    \hat{H}_\text{QZE}=\hat{H}_0+\lambda\sum_\mathbf{g}\hat{\mathcal{P}}_\mathbf{g}\hat{H}_1\hat{\mathcal{P}}_\mathbf{g},
\end{align}
where $\hat{\mathcal{P}}_\mathbf{g}$ is the projector onto the superselection sector $\mathbf{g}$. 
Though no longer restricted to a single superselection sector, the effective dynamics---again up to timescales at least linear in $V$---nevertheless does not couple different superselection sectors. 

\subsection{In terms of a local pseudogenerator}\label{sec:LPG}
The generator~\eqref{eq:Gj} of the spin-$S$ $\mathrm{U}(1)$ QLM~\eqref{eq:U1QLM} is ideal for linear gauge protection, given that it is composed of only single-body terms. However, not all gauge theories possess such a simply gauge-symmetry local generator. For example, a paradigmatic model that has recently been at the center of several experiments \cite{Goerg2019,Schweizer2019,Wang2021,Mildenberger2022} is the $\mathbb{Z}_2$ LGT, described by the Hamiltonian \cite{Barbiero2019,Zohar2017,Borla2019,Yang2020fragmentation,kebric2021confinement,Borla2020}
\begin{align}
    \label{eq:Z2LGT}
    \hat{H}_0=\sum_{j=1}^L\Big[J\big(\hat{b}_j^\dagger\hat{\tau}^z_{j,j+1}\hat{b}_{j+1}+\text{H.c.}\big)-h\hat{\tau}^x_{j,j+1}\Big].
\end{align}
The associated $\mathbb{Z}_2$ gauge-symmetry local generator is the three-body term
\begin{align}\label{eq:GjZ2}
    \hat{G}_j=(-1)^{\hat{n}_j}\hat{\tau}^x_{j-1,j}\hat{\tau}^x_{j,j+1},
\end{align}
where $\big[\hat{H}_0,\hat{G}_j\big]=0,\,\forall j$. Here, the hard-core bosonic ladder operators $\hat{b}_j,\hat{b}_j^\dagger$ represent the matter field on site $j$, with $\hat{n}_j=\hat{b}_j^\dagger\hat{b}_j$ the corresponding boson number operator, and the Pauli operator $\hat{\tau}^{z(x)}_{j,j+1}$ represents the gauge (electric) field at the link between the sites $j$ and $j+1$.

Implementing the linear gauge protection scheme~\eqref{eq:HG} with the generator in Eq.~\eqref{eq:GjZ2} in modern SQM devices is challenging as it requires the realization of three-body terms. It is therefore necessary to search for alternative, experimentally more feasible approaches. For this purpose, the local pseudogenerator (LPG) has been proposed, \cite{Halimeh2021stabilizing}
\begin{align}\label{eq:LPG}
    \hat{W}_j=\hat{\tau}^x_{j-1,j}\hat{\tau}^x_{j,j+1}+2g_j^\text{tar}\hat{n}_j,
\end{align}
which obeys the relation
\begin{align}\label{eq:LPGrelation}
    \hat{W}_j\ket{\phi}=g_j^\text{tar}\ket{\phi}\iff \hat{G}_j\ket{\phi}=g_j^\text{tar}\ket{\phi},
\end{align}
where $\{\ket{\phi}\}$ is the set of all gauge-invariant states. An implication of relation~\eqref{eq:LPGrelation} is that $\hat{W}_j$ and $\hat{G}_j$ are identical to each other in the target gauge superselection sector $\mathbf{g}^\text{tar}$. One can therefore employ the linear gauge protection term
\begin{align}\label{eq:HW}
    V\hat{H}_W=V\sum_{j=1}^Lc_j\hat{W}_j,
\end{align}
for which the formalism of Ref.~\cite{Halimeh2020e} can be extended to obtain the Zeno Hamiltonian
\begin{align}\label{eq:QZE2}
\hat{H}_\text{QZE}^\text{tar}=\hat{H}_0+\lambda\hat{\mathcal{P}}_{\mathbf{g}^\text{tar}}\hat{H}_1\hat{\mathcal{P}}_{\mathbf{g}^\text{tar}},
\end{align}
for quenches starting in the target sector \cite{Halimeh2021stabilizing}.

However, and quite intriguingly, for applications when the dynamics involves an extensive number of superselection sectors, an \textit{enriched} Zeno Hamiltonian emerges,
\begin{align}\label{eq:QZE3}
\hat{H}_\text{QZE}=\sum_\mathbf{w}\hat{\mathcal{P}}_\mathbf{w}\big(\hat{H}_0+\lambda\hat{H}_1\big)\hat{\mathcal{P}}_\mathbf{w},
\end{align}
where $\hat{\mathcal{P}}_\mathbf{w}$ is the projector onto the superselection sector $\mathbf{w}$ of the local symmetry associated with the LPG. Unlike Eq.~\eqref{eq:QZE2}, the Hamiltonian in Eq.~\eqref{eq:QZE3} encodes the full structure of the local symmetry associated with the LPG $\hat{W}_j$. This is nontrivial because this symmetry is richer than the $\mathbb{Z}_2$ gauge symmetry generated by  $\hat{G}_j$ of Eq.~\eqref{eq:GjZ2}, and we will discuss later how this can lead to rich physics in the case of DFL. In particular, the local symmetry associated with $\hat{W}_j$ contains the $\mathbb{Z}_2$ gauge symmetry:
\begin{align}
    \big[\hat{H}',\hat{W_j}\big]=0,\,\forall j\Rightarrow\big[\hat{H}',\hat{G_j}\big]=0,\,\forall j,
\end{align}
for a given Hamiltonian $\hat{H}'$, but the converse is generally not true. Noting that $\big[\hat{H}_0,\hat{W}_j\big]\neq0$, this means that $\hat{H}_\text{QZE}$, which commutes with both $\hat{W}_j$ and $\hat{G}_j$, hosts a richer local symmetry than the $\mathbb{Z}_2$ LGT.

\section{Stabilizing exotic far-from-equilibrium phenomena}\label{sec:phenomena}

\begin{figure}[t!]
    \centering
    \includegraphics[width=\columnwidth]{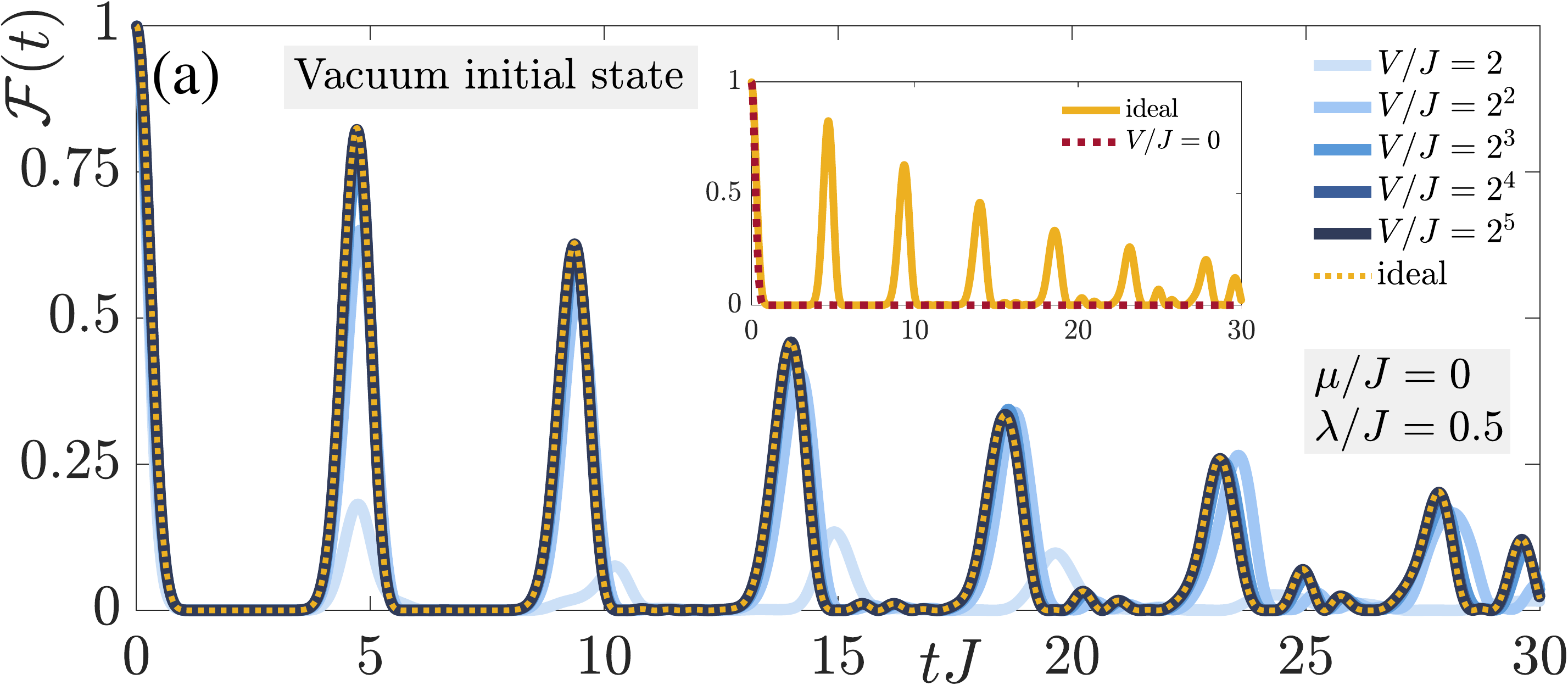}\\
    \vspace{1.1mm}
    \includegraphics[width=\columnwidth]{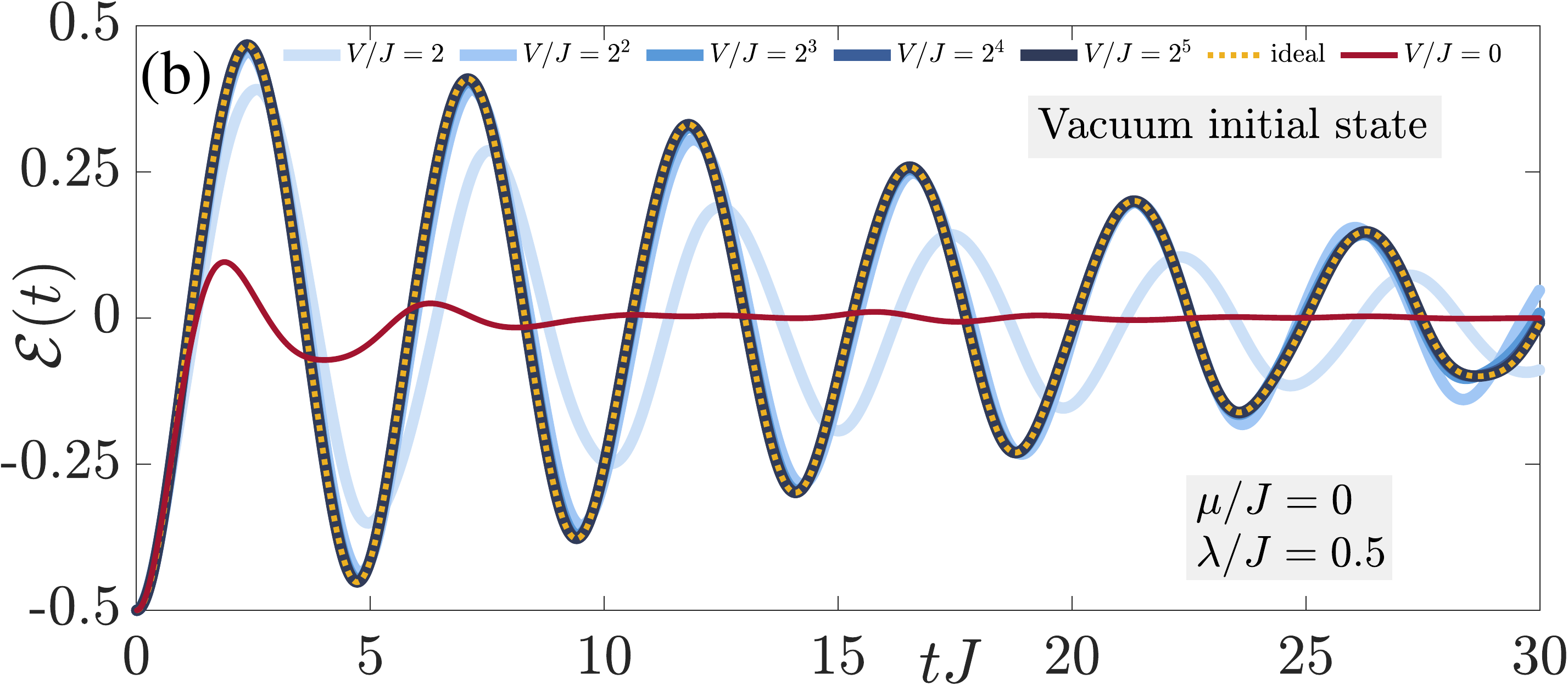}
    \caption{(Color online). Stabilizing quantum many-body scars in the spin-$1/2$ $\mathrm{U}(1)$ quantum link model~\eqref{eq:U1QLM}. The vacuum state is quenched with the faulty theory $\hat{H}_0+\lambda\hat{H}_1+V\hat{H}_G$, with $c_j=(-1)^j$. Once the protection term is sufficiently strong, the ensuing dynamics in (a) fidelity and (b) electric flux show persistent revivals identical to the ideal case up to all investigated evolution times. Results were calculated using Krylov-subspace methods. Figure adapted from Ref.~\cite{Halimeh2022robust}.}
    \label{fig:QMBS}
\end{figure}

The linear gauge protection scheme has been employed to protect dynamics within a target superselection sector for both the $\mathrm{U}(1)$ QLM and $\mathbb{Z}_2$ LGT, where it has demonstrated reliable stability and suppression of gauge violations $\propto\lambda^2/V^2$ up to all numerically accessible times in both finite systems \cite{Halimeh2020e,Halimeh2021stabilizing,Halimeh2022robust} and in the thermodynamic limit \cite{vandamme2021reliability,vandamme2021suppressing}. 
In this way, also local observables remain close to the dynamics of an ideal gauge theory. In this section, we illustrate how linear gauge protection can stabilize also the salient features of many-body scarring and DFL. 

\subsection{Quantum many-body scars}\label{sec:QMBS}
Here, we focus on quantum many-body scarring, which occurs in the spin-$S$ $\mathrm{U}(1)$ QLM \cite{Bernien2017,Surace2020,Desaules2022weak,Desaules2022prominent} and the $\mathbb{Z}_2$ LGT \cite{Iadecola2020,aramthottil2022scar} for quenches in a target superselection sector starting in special initial states. In the case of the $\mathrm{U}(1)$ QLM, the initial state is the bare vacuum of QED (corresponding to the symmetry-broken ground state of Eq.~\eqref{eq:U1QLM} for $S=1/2$ and at $\mu\to\infty$ where no matter particles are present), while in the case of the $\mathbb{Z}_2$ LGT it is an equal superposition of product states at half-filling, where each is composed of $L/4$ pairs of bosons with any two pairs separated by at least one site \cite{Halimeh2022robust}. Scarring is a paradigm of weak ergodicity breaking that arises due to the presence of special nonthermal eigenstates in the quench Hamiltonian that have anomalously low bipartite entanglement entropy and are roughly equally spaced in energy over the entire spectrum \cite{ShiraishiMori,Moudgalya2018,BernevigEnt,lin2018exact,Iadecola2019_2,MotrunichTowers}. These special eigenstates form a \textit{cold subspace} that is weakly connected to the rest of the Hilbert space of the quench Hamiltonian, which leads to a significant delay in thermalization well beyond relevant timescales \cite{Turner2018}.

Scars have been shown to be susceptible to perturbations, and experimentally unrealistic fine-tuning is required in case of no stabilization scheme \cite{Surace2021}. However, it has been shown recently that linear gauge protection stabilizes scars in both the $\mathrm{U}(1)$ QLM and the $\mathbb{Z}_2$ LGT in the presence of gauge-breaking errors \cite{Halimeh2022robust}, as shown in Fig.~\ref{fig:QMBS} for the case of the former for $L=12$ matter sites with open boundary conditions. Using Krylov-subspace methods \cite{Moler2003,EXPOKIT}, we calculate the quench dynamics of the fidelity and electric flux
\begin{subequations}
\begin{align}
    \mathcal{F}(t)&=\big\lvert\braket{\psi_0}{\psi(t)}\big\rvert^2,\\
    \mathcal{E}(t)&=\frac{1}{L-1}\sum_{j=1}^{L-1}(-1)^j\bra{\psi(t)}\hat{s}^z_{j,j+1}\ket{\psi(t)},
\end{align}
\end{subequations}
where $\ket{\psi(t)}=e^{-i\hat{H}t}\ket{\psi_0}$, $\hat{H}=\hat{H}_0+\lambda\hat{H}_1+V\hat{H}_G$ is the faulty gauge theory, and $\ket{\psi_0}$ is a vacuum state. As typical of scarred dynamics, the fidelity exhibits persistent revivals up to all investigated evolution times in the case of the ideal model ($\lambda=V=0$), as seen in the inset of Fig.~\ref{fig:QMBS}(a). Taking errors into account, we find that the revivals vanish even when the error strength is perturbative ($\lambda=0.5J$). The errors used for these results are
\begin{align}\label{eq:H1}
    \lambda\hat{H}_1=\lambda\sum_{j=1}^{L-1}\big(\hat{\sigma}^-_j\hat{\sigma}^-_{j+1}+\hat{\sigma}^+_j\hat{\sigma}^+_{j+1}+\hat{s}^x_{j,j+1}\big),
\end{align}
but our conclusions remain the same for other experimentally relevant gauge-breaking errors.

Upon employing linear gauge protection~\eqref{eq:HG} with $c_j=(-1)^j$, we find that the revivals are almost perfectly restored over all investigated evolution times at experimentally accessible protection strengths $V\gtrsim8J$. This is also reflected in the electric flux, shown in Fig.~\ref{fig:QMBS}(b). Whereas unprotected errors destroy persistent oscillations typical of scarred dynamics of the ideal case, linear gauge protection restores such oscillations qualitatively and quantitatively for moderate values of $V$.

These results not only indicate the efficacy and experimental feasibility of linear gauge protection in stabilizing a very fine-tuned phenomenon such as QMBS, but it also shows the intimate connection between QMBS and gauge invariance, as elucidated in further detail in Ref.~\cite{Halimeh2022robust}.

\begin{figure}[t!]
    \centering
    \includegraphics[width=\columnwidth]{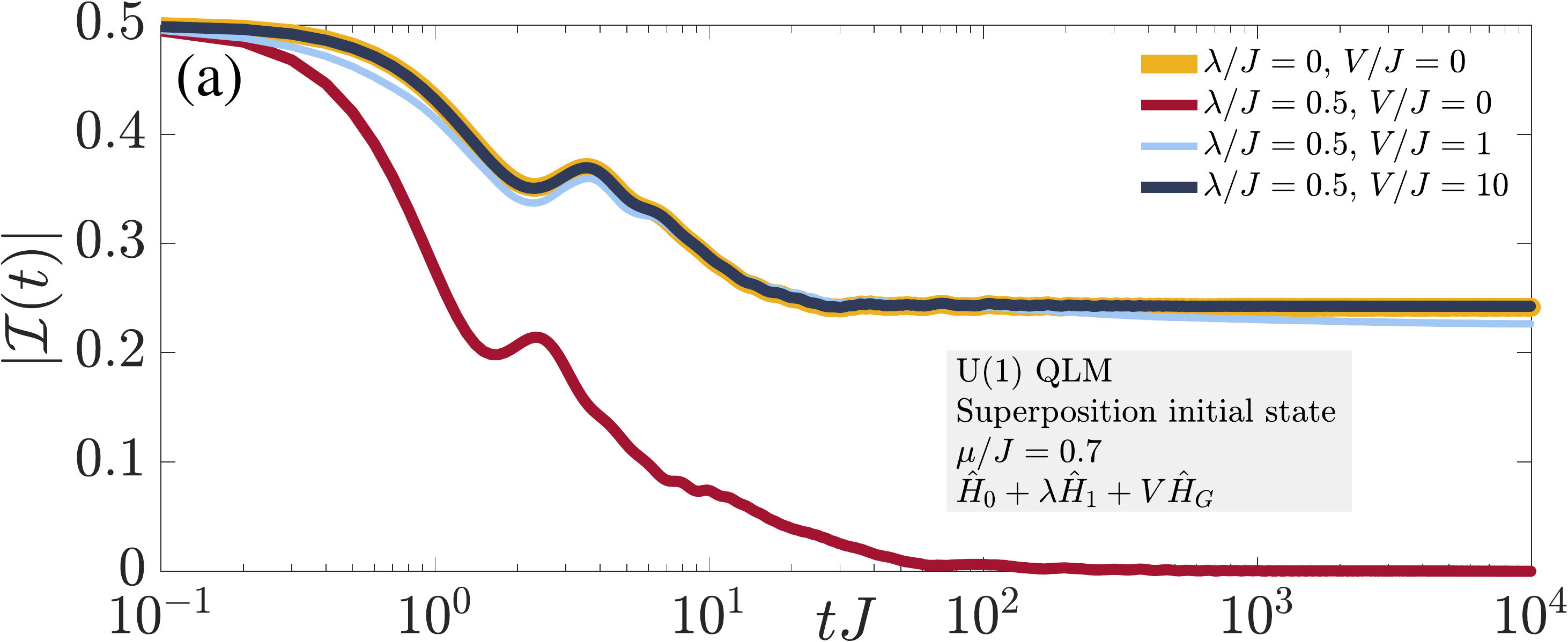}\\
    \vspace{1.1mm}
    \includegraphics[width=\columnwidth]{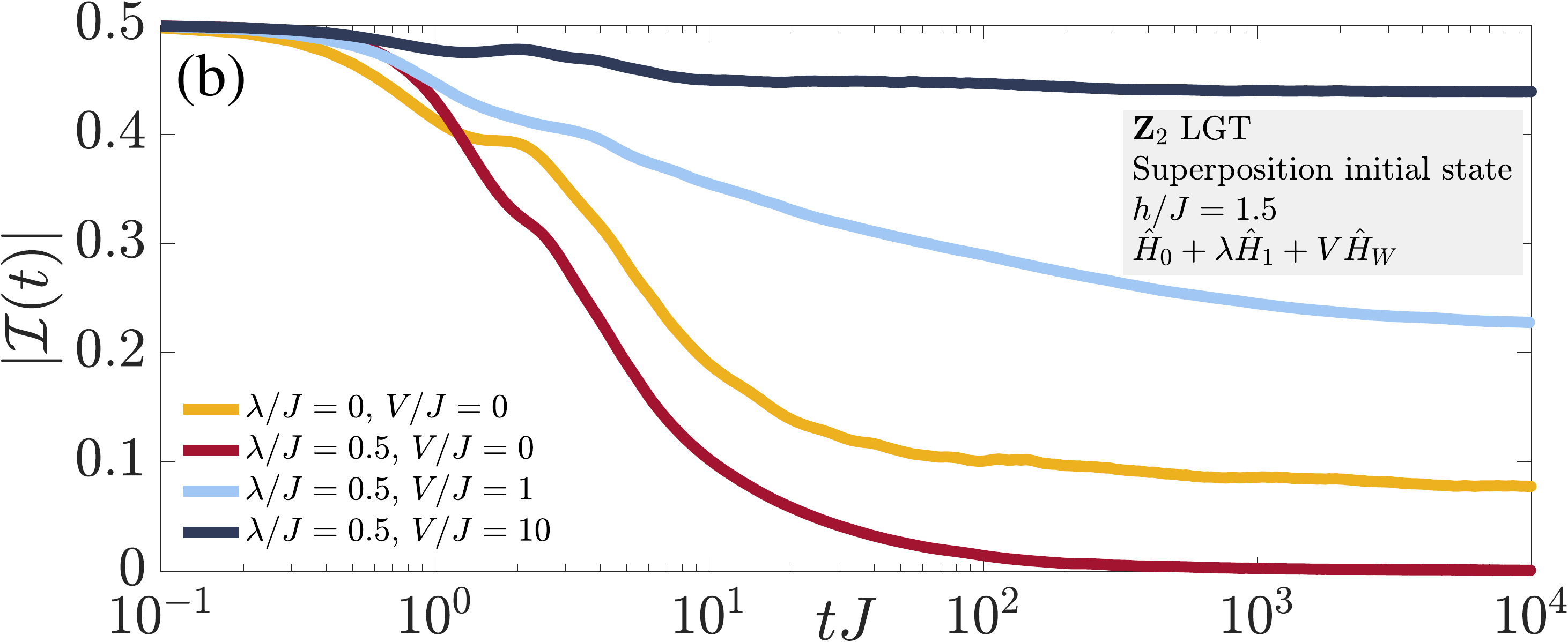}
    \caption{(Color online). Linear gauge protection of disorder-free localization. (a) In the case of the $\mathrm{U}(1)$ QLM, we protect against gauge-breaking errors $\lambda\hat{H}_1$ using Eq.~\eqref{eq:HG}, which involves the local generator $\hat{G}_j$, with $c_j=j(-1)^j$. Since this leads to an emergent gauge theory with the same gauge symmetry as $\hat{H}_0$, the same effective disorder emerges due to the superposition initial state, and a sufficiently strong $V$ stabilizes DFL to the same plateau as in the case of the ideal theory. (b) In the case of the $\mathbb{Z}_2$ LGT, we employ the linear gauge protection~\eqref{eq:HW} based on the local pseudogenerator $\hat{W}_j$ of Eq.~\eqref{eq:LPG}, with $c_j=j$. This leads to an emergent gauge theory with an \textit{enriched} local symmetry containing the $\mathbb{Z}_2$ gauge symmetry of the ideal model (see Fig.~\ref{fig:schematic}). This enriched symmetry due to $\hat{W}_j$ leads to additional superselection sectors in the superposition, inducing a greater effective disorder, and hence enhanced DFL. Results were calculated using Krylov-subspace methods. Figure is adapted from Ref.~\cite{lang2022disorder}.}
    \label{fig:DFL}
\end{figure}

\subsection{Disorder-free localization}\label{sec:DFL}
A paradigm of strong ergodicity breaking, DFL emerges in gauge theories in the wake of quenches starting in initial states that form a superposition of an extensive number of gauge superselection sectors \cite{Smith2017,Brenes2018}. The distinctive feature of DFL is that it can occur in nonintegrable translation-invariant models where no quenched disorder is present. The superposition initial state is also translation-invariant. The localization occurs through the dynamical emergence of an effective disorder over the background charges associated with the superselection sectors $\mathbf{g}$ involved in the superposition. In an ideal situation, this leads to localized dynamics for all evolution times, where for example the imbalance
\begin{align}\label{eq:imbalance}
    \mathcal{I}(t)=\frac{1}{Lt}\int_0^t ds\sum_{j=1}^Lp_j\bra{\psi(s)}\hat{n}_j\ket{\psi(s)},
\end{align}
with $p_j=2\bra{\psi_0}\hat{n}_j\ket{\psi_0}-1$, does not decay to zero (as would be expected in the case of thermalization) but rather settles into a finite plateau persisting indefinitely.

However, also in this case it turns out that a great deal of fine-tuning is required in order to witness DFL, while even perturbative gauge-breaking errors can quickly destroy it \cite{Smith2018}; see red curves in Fig.~\ref{fig:DFL}(a,b), illustrating this for both the $\mathrm{U}(1)$ QLM and $\mathbb{Z}_2$ LGT, respectively. Nevertheless, it has been demonstrated recently that linear gauge protection can stabilize DFL in $\mathrm{U}(1)$ QLMs \cite{Halimeh2021stabilizingDFL}, and even enhance it in the case of $\mathbb{Z}_2$ LGTs \cite{Halimeh2021enhancing}; see Fig.~\ref{fig:DFL}. For both models, the initial state is prepared as a domain wall in the matter fields (left half fully occupied while the right half is empty), and the electric fields are prepared locally as a superposition of both local eigenstates, such that the initial states form a superposition over an extensive number of superselection sectors. Krylov-subspace methods are employed for the time evolution, with $L=8$ lattice sites and periodic boundary conditions in place.

In the case of the $\mathrm{U}(1)$ QLM, the linear gauge protection~\eqref{eq:HG} with $c_j=j(-1)^j$ was utilized to protect against the error term~\eqref{eq:H1}, but with periodic boundary conditions, in Ref.~\cite{lang2022disorder}. This scheme of \textit{Stark gauge protection} (named so due to $c_j$ involving a linear staggered potential) goes beyond the concept of quantum Zeno dynamics, and stabilizes DFL up to all accessible evolution times. The emergent gauge theory still hosts the same $\mathrm{U}(1)$ gauge symmetry as the $\mathrm{U}(1)$ QLM, and DFL is restored to the same plateau of the ideal case, as indicated by the imbalance dynamics in Fig.~\ref{fig:DFL}(b). Note how even a moderate value of $V=10J$ yields almost perfect quantitative agreement with the ideal case. Note that in the case of the $\mathrm{U}(1)$ QLM, $\hat{n}_j=(\hat{\sigma}^z_j+\mathds{1})/2$ in Eq.~\eqref{eq:imbalance}.

For the case of the $\mathbb{Z}_2$ LGT, errors inspired from the implementation of Ref.~\cite{Schweizer2019} are considered. The linear gauge protection~\eqref{eq:HW} based on the LPG~\eqref{eq:LPG} are employed with $c_j=j$. Not only is DFL stabilized in this case, but it is even enhanced, with the imbalance shown in Fig.~\ref{fig:DFL}(b) for $V=10J$ settling into a plateau of significantly larger value than that of the ideal case. This enhancement is due to the local pseudogenerator, which gives rise to an enhanced local symmetry containing the original $\mathbb{Z}_2$ gauge symmetry; cf.~Sec.~\ref{sec:LPG}. This effectively renders the initial state as a superposition over a greater number of local-symmetry sectors when $V$ is sufficiently large. This in turn leads to an effective disorder over a larger variety of background charges associated with all these sectors, and consequently renders the dynamics more localized \cite{Halimeh2021enhancing}.

\section{Experimental applications}\label{sec:experiment}
Our work on linear gauge protection has been used experimentally in large-scale implementations of the spin-$1/2$ $\mathrm{U}(1)$ QLM \cite{Yang2020,Zhou2021}. In these cold-atom experiments, the $\mathrm{U}(1)$ QLM has been mapped onto the Bose--Hubbard model
\begin{align}\nonumber
    \hat{H}_\text{BH}=&-J_\text{BH}\sum_{j=1}^{L_\text{BH}-1}\big(\hat{b}_j^\dagger\hat{b}_{j+1}+\text{H.c.}\big)+\frac{U}{2}\sum_{j=1}^{L_\text{BH}}\hat{n}_j\big(\hat{n}_j-1\big)\\\label{eq:BHM}
    &+\sum_{j=1}^{L_\text{BH}}\bigg[(-1)^j\frac{\delta}{2}+j\Delta\bigg]\hat{n}_j,
\end{align}
where $J_\text{BH}$ is the tunneling strength, $U$ is the on-site interaction strength, $\Delta$ is the strength of a linear potential generated by the gravitational force, and $\delta$ is the strength of the staggered chemical potential. Further, $L_\text{BH}$ is the total number of sites of the bosonic superlattice, corresponding to $L=L_\text{BH}/2$ matter sites and $L_\text{BH}/2-1$ gauge sites of the QLM. The bosonic ladder operators $\hat{b}_j,\hat{b}_j^\dagger$ on site $j$ satisfy the canonical commutation relations $\big[\hat{b}_j,\hat{b}_l\big]=0$ and $\big[\hat{b}_j,\hat{b}_l^\dagger\big]=\delta_{j,l}$, and $\hat{n}_j=\hat{b}_j^\dagger\hat{b}_j$ is the bosonic number operator at site $j$. Figure~\ref{fig:BHM} illustrates the optical superlattice of the model in Eq.~\eqref{eq:BHM}, and the mapping to the spin-$1/2$ $\mathrm{U}(1)$ QLM within the target sector $\hat{G}_j\ket{\phi}=0,\,\forall j$.

The $\mathrm{U}(1)$ QLM can be obtained from second-order perturbation theory in the limit of $U,\delta\gg J_\text{BH}$ and $U\approx2\delta$ \cite{Yang2020}. The parameters of Eq.~\eqref{eq:U1QLM} for $S=1/2$ can further be related to those of Eq.~\eqref{eq:BHM} from degenerate perturbation theory as
\begin{subequations}
\begin{align}\label{eq:kappa}
    \kappa&=2\sqrt{2}J_\text{BH}^2\bigg[\frac{\delta}{\delta^2-\Delta^2}+\frac{U-\delta}{(U-\delta)^2-\Delta^2}\bigg],\\
    \label{eq:mu}
    \mu&=\delta-\frac{U}{2},
\end{align}
\end{subequations}
where $\kappa=J/\sqrt{S(S+1)}$. Note that for $S=1/2$, the term $\propto\eta^2$ in Eq.~\eqref{eq:U1QLM} can be neglected as it is an inconsequential constant energy term, because $\big(\hat{s}^z_{j,j+1}\big)^2=\mathds{1}$ in this case.

One can then show that gauge-breaking processes in this mapping are suppressed by the linear gauge protection term $\sum_\ell c_\ell\hat{\mathcal{G}}_\ell$, where \cite{Halimeh2020e,Halimeh2022Tuning}
\begin{subequations}
\begin{align}
    c_\ell&= (-1)^\ell\bigg[\Delta \ell + \bigg(U-\delta + \frac{\Delta}{2}\bigg)\bigg],\\
    \hat{\mathcal{G}}_\ell&= (-1)^\ell\bigg[\frac{1}{2}\big(\hat{n}_{\ell-1,\ell}+\hat{n}_{\ell,\ell+1}\big) + \hat{n}_\ell -1\bigg],
\end{align}
\end{subequations}
with $\ell$ the index of a matter site and $\hat{n}_{\ell,\ell+1}$ the bosonic number operator at the gauge site between matter sites $\ell$ and $\ell+1$. Additional energy penalties due to $U$ and $\delta$ enforce a maximum of single- and double-occupations on matter and gauge sites, respectively. Within that occupation subspace, the ``proto-Gauss's law'' operators $\hat{\mathcal{G}}_\ell$ stabilize the gauge theory in the quantum simulator, based on the concept of linear gauge protection.

The above mapping has allowed the large-scale quantum simulation of the spin-$1/2$ $\mathrm{U}(1)$ QLM in up to 71 superlattice sites, where in Ref.~\cite{Yang2020} an adiabatic mass ramp was employed to probe Coleman's phase transition and the adherence to Gauss's law was certified for the first time in a quantum-simulation experiment. In Ref.~\cite{Zhou2021}, global quenches were performed to probe the thermalization dynamics of gauge theories. Recently, we have also proposed a modification of this setup that allows introducing a tunable topological $\theta$-angle to study confinement in gauge theories \cite{Halimeh2022Tuning} (see also the related work \cite{Cheng2022tunable}). We emphasize that all these experiments are possible due to the control and suppression of gauge-breaking errors by linear gauge protection.

The linear gauge protection can also be employed fruitfully in other experimental platforms. For example, we have proposed an experiment in Rydberg atoms with optical tweezers for probing DFL, exploiting linear gauge protection in terms of the LPG \cite{Halimeh2021enhancing}. Recently, linear gauge protection has further been used in a superconducting qubit processor within the Early Access Program of Google Quantum AI to tune from a $\mathbb{Z}_2$ gauge symmetry to the larger continuous $\mathrm{U}(1)$ gauge symmetry \cite{Mildenberger2022}. This was achieved by adding a term proportional to the $\mathrm{U}(1)$ Gauss's law generators given in Eq.~\eqref{eq:Gj}, which required only extremely simple single-qubit gates without any relevant experimental overhead. In that experiment, a stabilization of the $\mathrm{U}(1)$ gauge symmetry over 15 Trotter steps has been demonstrated using a very moderate strength of $V=6J$. Further, it was shown how the resulting change from a $\mathbb{Z}_2$ to an effective $\mathrm{U}(1)$ gauge symmetry leads to a drastic modification of the system dynamics, in particular a freezing of the dynamics due to the severe constraints given by the $\mathrm{U}(1)$ Gauss's law generator.

\section{Conclusion}\label{sec:conc}

To conclude, the control and stabilization of gauge symmetry in quantum simulators has made great strides in the recent years, theoretically and experimentally. 
Though important questions for the future will be how the stabilization strategies perform for higher-dimensional systems \cite{Zohar_NewReview} and non-Abelian symmetries \cite{Banerjee2013,Tagliacozzo2013,Zohar2013SU2,Mezzacapo2015,Klco2020,Atas2021,Kasper2021nonabelian,Halimeh2021gauge}, the achieved advances are already laying the foundations for future quantum simulations of effects of importance to high-energy and nuclear physics, including in the near term confinement \cite{Zohar2011,Zohar2012,Tagliacozzo2012,Tagliacozzo2013,Surace2020,Wang2021,Banerjee2021Nematic,Mildenberger2022}, string breaking \cite{Banerjee2012,Hebenstreit2013,Yang2016}, topological terms \cite{Surace2020,Zache2019,Kharzeev2008,Fukushima2008,Kharzeev2016,Koch2017,Kharzeev2020,Halimeh2022Tuning,Cheng2022tunable,Kan2021}, transport coefficients \cite{Cohen2021}, the question how gauge theories reach thermal equilibrium \cite{Berges2019,Zhou2021,Mueller2021}, and the Schwinger effect of particle production \cite{Kasper2017,Zache2018}, to just give some salient examples.

The topics treated in this review also open the horizon to a fresh view on gauge-symmetry violation: while in fundamental models of nature, such as the Standard Model of Particle Physics, gauge symmetry is a postulated exact symmetry, quantum simulators that do not use the approach of integrating out are physically described by an enlarged Hilbert space, where also configurations live that violate Gauss's law. Apart from the important question how gauge-symmetry can emerge in such a setting \cite{Foerster1980,Poppitz2008,Wetterich2017,Witten2018,Barcelo2021,Bass2022,Hastings2005,Sachdev2018,VanDamme2020,Borla2020}, it then becomes natural to investigate violations of gauge symmetry not as an error but as an effect that generates new physical phenomena. An example has been mentioned in this work, where superpositions of different gauge superselection sectors in the initial state induce slow dynamics in the form of disorder-free localization \cite{Smith2017,Brenes2018}. Another example is a phenomenon that we have dubbed \emph{staircase prethermalization}, where a small term that breaks the gauge symmetry drives the system into a series of long-lived plateaus of successively stronger gauge-symmetry violations \cite{Halimeh2020b,Halimeh2020c}.   

As all these examples show, the quantum simulation of lattice gauge theories opens up to an extremely fertile field of rich physical phenomena.

\begin{acknowledgments}
We are grateful to the groups of Fabian Grusdt, Johannes Knolle, Jian-Wei Pan, Zlatko Papi\'c, Bing Yang, Zhen-Sheng Yuan, Zhang Jiang and the Google Quantum AI team, as well as to Debasish Banerjee, Luca Barbiero, Annabelle Bohrdt, Markus Heyl, Haifeng Lang, Julius Mildenberger, Ian P.~McCulloch, Arnab Sen, Maarten Van Damme, and Hongzheng Zhao for collaborations on related work. We are grateful to Lukas Homeier for assistance with Fig.~\ref{fig:schematic}. J.C.H.~acknowledges funding from the European Research Council (ERC) under the European Union’s Horizon 2020 research and innovation programm (Grant Agreement no 948141) — ERC Starting Grant SimUcQuam, and by the Deutsche Forschungsgemeinschaft (DFG, German Research Foundation) under Germany's Excellence Strategy -- EXC-2111 -- 390814868. P.H.~acknowledges support by the ERC Starting Grant StrEnQTh (project ID 804305), the Google Research Scholar Award ProGauge, Provincia Autonoma di Trento, and Q@TN — Quantum Science and Technology in Trento.
\end{acknowledgments}

\bibliography{review_biblio}

\begin{thebibliography}{136}%
\makeatletter
\providecommand \@ifxundefined [1]{%
 \@ifx{#1\undefined}
}%
\providecommand \@ifnum [1]{%
 \ifnum #1\expandafter \@firstoftwo
 \else \expandafter \@secondoftwo
 \fi
}%
\providecommand \@ifx [1]{%
 \ifx #1\expandafter \@firstoftwo
 \else \expandafter \@secondoftwo
 \fi
}%
\providecommand \natexlab [1]{#1}%
\providecommand \enquote  [1]{``#1''}%
\providecommand \bibnamefont  [1]{#1}%
\providecommand \bibfnamefont [1]{#1}%
\providecommand \citenamefont [1]{#1}%
\providecommand \href@noop [0]{\@secondoftwo}%
\providecommand \href [0]{\begingroup \@sanitize@url \@href}%
\providecommand \@href[1]{\@@startlink{#1}\@@href}%
\providecommand \@@href[1]{\endgroup#1\@@endlink}%
\providecommand \@sanitize@url [0]{\catcode `\\12\catcode `\$12\catcode
  `\&12\catcode `\#12\catcode `\^12\catcode `\_12\catcode `\%12\relax}%
\providecommand \@@startlink[1]{}%
\providecommand \@@endlink[0]{}%
\providecommand \url  [0]{\begingroup\@sanitize@url \@url }%
\providecommand \@url [1]{\endgroup\@href {#1}{\urlprefix }}%
\providecommand \urlprefix  [0]{URL }%
\providecommand \Eprint [0]{\href }%
\providecommand \doibase [0]{http://dx.doi.org/}%
\providecommand \selectlanguage [0]{\@gobble}%
\providecommand \bibinfo  [0]{\@secondoftwo}%
\providecommand \bibfield  [0]{\@secondoftwo}%
\providecommand \translation [1]{[#1]}%
\providecommand \BibitemOpen [0]{}%
\providecommand \bibitemStop [0]{}%
\providecommand \bibitemNoStop [0]{.\EOS\space}%
\providecommand \EOS [0]{\spacefactor3000\relax}%
\providecommand \BibitemShut  [1]{\csname bibitem#1\endcsname}%
\let\auto@bib@innerbib\@empty
%</preamble>
\bibitem [{qua()}]{quasiproceedingsSite}%
  \BibitemOpen
  \href@noop {} {}\bibinfo {howpublished} {\href
  {https://iqus.uw.edu/events/quantum-simulation-of-strong-interactions-quasi-workshop-2-implementation-strategies-for-gauge-theories/}
  {\bibinfo {title} {{Proceedings of the 2021 Quantum Simulation for Strong
  Interactions (QuaSi) Workshops}}} (2022), https://iqus.uw.edu/}\BibitemShut
  {NoStop}%
\bibitem [{\citenamefont {Bloch}\ \emph {et~al.}(2008)\citenamefont {Bloch},
  \citenamefont {Dalibard},\ and\ \citenamefont {Zwerger}}]{Bloch2008}%
  \BibitemOpen
  \bibfield  {author} {\bibinfo {author} {\bibfnamefont {Immanuel}\
  \bibnamefont {Bloch}}, \bibinfo {author} {\bibfnamefont {Jean}\ \bibnamefont
  {Dalibard}}, \ and\ \bibinfo {author} {\bibfnamefont {Wilhelm}\ \bibnamefont
  {Zwerger}},\ }\bibfield  {title} {\enquote {\bibinfo {title} {Many-body
  physics with ultracold gases},}\ }\href {\doibase 10.1103/RevModPhys.80.885}
  {\bibfield  {journal} {\bibinfo  {journal} {Rev. Mod. Phys.}\ }\textbf
  {\bibinfo {volume} {80}},\ \bibinfo {pages} {885--964} (\bibinfo {year}
  {2008})}\BibitemShut {NoStop}%
\bibitem [{\citenamefont {Bakr}\ \emph {et~al.}(2009)\citenamefont {Bakr},
  \citenamefont {Gillen}, \citenamefont {Peng}, \citenamefont {F{\"o}lling},\
  and\ \citenamefont {Greiner}}]{Bakr2009}%
  \BibitemOpen
  \bibfield  {author} {\bibinfo {author} {\bibfnamefont {Waseem~S.}\
  \bibnamefont {Bakr}}, \bibinfo {author} {\bibfnamefont {Jonathon~I.}\
  \bibnamefont {Gillen}}, \bibinfo {author} {\bibfnamefont {Amy}\ \bibnamefont
  {Peng}}, \bibinfo {author} {\bibfnamefont {Simon}\ \bibnamefont
  {F{\"o}lling}}, \ and\ \bibinfo {author} {\bibfnamefont {Markus}\
  \bibnamefont {Greiner}},\ }\bibfield  {title} {\enquote {\bibinfo {title} {A
  quantum gas microscope for detecting single atoms in a hubbard-regime optical
  lattice},}\ }\href {\doibase 10.1038/nature08482} {\bibfield  {journal}
  {\bibinfo  {journal} {Nature}\ }\textbf {\bibinfo {volume} {462}},\ \bibinfo
  {pages} {74--77} (\bibinfo {year} {2009})}\BibitemShut {NoStop}%
\bibitem [{\citenamefont {Hauke}\ \emph {et~al.}(2012)\citenamefont {Hauke},
  \citenamefont {Cucchietti}, \citenamefont {Tagliacozzo}, \citenamefont
  {Deutsch},\ and\ \citenamefont {Lewenstein}}]{Hauke2012}%
  \BibitemOpen
  \bibfield  {author} {\bibinfo {author} {\bibfnamefont {Philipp}\ \bibnamefont
  {Hauke}}, \bibinfo {author} {\bibfnamefont {Fernando~M}\ \bibnamefont
  {Cucchietti}}, \bibinfo {author} {\bibfnamefont {Luca}\ \bibnamefont
  {Tagliacozzo}}, \bibinfo {author} {\bibfnamefont {Ivan}\ \bibnamefont
  {Deutsch}}, \ and\ \bibinfo {author} {\bibfnamefont {Maciej}\ \bibnamefont
  {Lewenstein}},\ }\bibfield  {title} {\enquote {\bibinfo {title} {Can one
  trust quantum simulators?}}\ }\href {\doibase 10.1088/0034-4885/75/8/082401}
  {\bibfield  {journal} {\bibinfo  {journal} {Reports on Progress in Physics}\
  }\textbf {\bibinfo {volume} {75}},\ \bibinfo {pages} {082401} (\bibinfo
  {year} {2012})}\BibitemShut {NoStop}%
\bibitem [{\citenamefont {Alexeev}\ \emph {et~al.}(2021)\citenamefont
  {Alexeev}, \citenamefont {Bacon}, \citenamefont {Brown}, \citenamefont
  {Calderbank}, \citenamefont {Carr}, \citenamefont {Chong}, \citenamefont
  {DeMarco}, \citenamefont {Englund}, \citenamefont {Farhi}, \citenamefont
  {Fefferman}, \citenamefont {Gorshkov}, \citenamefont {Houck}, \citenamefont
  {Kim}, \citenamefont {Kimmel}, \citenamefont {Lange}, \citenamefont {Lloyd},
  \citenamefont {Lukin}, \citenamefont {Maslov}, \citenamefont {Maunz},
  \citenamefont {Monroe}, \citenamefont {Preskill}, \citenamefont {Roetteler},
  \citenamefont {Savage},\ and\ \citenamefont {Thompson}}]{Alexeev_review}%
  \BibitemOpen
  \bibfield  {author} {\bibinfo {author} {\bibfnamefont {Yuri}\ \bibnamefont
  {Alexeev}}, \bibinfo {author} {\bibfnamefont {Dave}\ \bibnamefont {Bacon}},
  \bibinfo {author} {\bibfnamefont {Kenneth~R.}\ \bibnamefont {Brown}},
  \bibinfo {author} {\bibfnamefont {Robert}\ \bibnamefont {Calderbank}},
  \bibinfo {author} {\bibfnamefont {Lincoln~D.}\ \bibnamefont {Carr}}, \bibinfo
  {author} {\bibfnamefont {Frederic~T.}\ \bibnamefont {Chong}}, \bibinfo
  {author} {\bibfnamefont {Brian}\ \bibnamefont {DeMarco}}, \bibinfo {author}
  {\bibfnamefont {Dirk}\ \bibnamefont {Englund}}, \bibinfo {author}
  {\bibfnamefont {Edward}\ \bibnamefont {Farhi}}, \bibinfo {author}
  {\bibfnamefont {Bill}\ \bibnamefont {Fefferman}}, \bibinfo {author}
  {\bibfnamefont {Alexey~V.}\ \bibnamefont {Gorshkov}}, \bibinfo {author}
  {\bibfnamefont {Andrew}\ \bibnamefont {Houck}}, \bibinfo {author}
  {\bibfnamefont {Jungsang}\ \bibnamefont {Kim}}, \bibinfo {author}
  {\bibfnamefont {Shelby}\ \bibnamefont {Kimmel}}, \bibinfo {author}
  {\bibfnamefont {Michael}\ \bibnamefont {Lange}}, \bibinfo {author}
  {\bibfnamefont {Seth}\ \bibnamefont {Lloyd}}, \bibinfo {author}
  {\bibfnamefont {Mikhail~D.}\ \bibnamefont {Lukin}}, \bibinfo {author}
  {\bibfnamefont {Dmitri}\ \bibnamefont {Maslov}}, \bibinfo {author}
  {\bibfnamefont {Peter}\ \bibnamefont {Maunz}}, \bibinfo {author}
  {\bibfnamefont {Christopher}\ \bibnamefont {Monroe}}, \bibinfo {author}
  {\bibfnamefont {John}\ \bibnamefont {Preskill}}, \bibinfo {author}
  {\bibfnamefont {Martin}\ \bibnamefont {Roetteler}}, \bibinfo {author}
  {\bibfnamefont {Martin~J.}\ \bibnamefont {Savage}}, \ and\ \bibinfo {author}
  {\bibfnamefont {Jeff}\ \bibnamefont {Thompson}},\ }\bibfield  {title}
  {\enquote {\bibinfo {title} {Quantum computer systems for scientific
  discovery},}\ }\href {\doibase 10.1103/PRXQuantum.2.017001} {\bibfield
  {journal} {\bibinfo  {journal} {PRX Quantum}\ }\textbf {\bibinfo {volume}
  {2}},\ \bibinfo {pages} {017001} (\bibinfo {year} {2021})}\BibitemShut
  {NoStop}%
\bibitem [{\citenamefont {Klco}\ \emph {et~al.}(2021)\citenamefont {Klco},
  \citenamefont {Roggero},\ and\ \citenamefont {Savage}}]{klco2021standard}%
  \BibitemOpen
  \bibfield  {author} {\bibinfo {author} {\bibfnamefont {Natalie}\ \bibnamefont
  {Klco}}, \bibinfo {author} {\bibfnamefont {Alessandro}\ \bibnamefont
  {Roggero}}, \ and\ \bibinfo {author} {\bibfnamefont {Martin~J.}\ \bibnamefont
  {Savage}},\ }\bibfield  {title} {\enquote {\bibinfo {title} {Standard model
  physics and the digital quantum revolution: Thoughts about the interface},}\
  }\href@noop {} {\  (\bibinfo {year} {2021})},\ \Eprint
  {http://arxiv.org/abs/2107.04769} {arXiv:2107.04769 [quant-ph]} \BibitemShut
  {NoStop}%
\bibitem [{\citenamefont {Ba{\~n}uls}\ \emph {et~al.}(2020)\citenamefont
  {Ba{\~n}uls}, \citenamefont {Blatt}, \citenamefont {Catani}, \citenamefont
  {Celi}, \citenamefont {Cirac}, \citenamefont {Dalmonte}, \citenamefont
  {Fallani}, \citenamefont {Jansen}, \citenamefont {Lewenstein}, \citenamefont
  {Montangero}, \citenamefont {Muschik}, \citenamefont {Reznik}, \citenamefont
  {Rico}, \citenamefont {Tagliacozzo}, \citenamefont {Van~Acoleyen},
  \citenamefont {Verstraete}, \citenamefont {Wiese}, \citenamefont {Wingate},
  \citenamefont {Zakrzewski},\ and\ \citenamefont {Zoller}}]{Pasquans_review}%
  \BibitemOpen
  \bibfield  {author} {\bibinfo {author} {\bibfnamefont {Mari~Carmen}\
  \bibnamefont {Ba{\~n}uls}}, \bibinfo {author} {\bibfnamefont {Rainer}\
  \bibnamefont {Blatt}}, \bibinfo {author} {\bibfnamefont {Jacopo}\
  \bibnamefont {Catani}}, \bibinfo {author} {\bibfnamefont {Alessio}\
  \bibnamefont {Celi}}, \bibinfo {author} {\bibfnamefont {Juan~Ignacio}\
  \bibnamefont {Cirac}}, \bibinfo {author} {\bibfnamefont {Marcello}\
  \bibnamefont {Dalmonte}}, \bibinfo {author} {\bibfnamefont {Leonardo}\
  \bibnamefont {Fallani}}, \bibinfo {author} {\bibfnamefont {Karl}\
  \bibnamefont {Jansen}}, \bibinfo {author} {\bibfnamefont {Maciej}\
  \bibnamefont {Lewenstein}}, \bibinfo {author} {\bibfnamefont {Simone}\
  \bibnamefont {Montangero}}, \bibinfo {author} {\bibfnamefont {Christine~A.}\
  \bibnamefont {Muschik}}, \bibinfo {author} {\bibfnamefont {Benni}\
  \bibnamefont {Reznik}}, \bibinfo {author} {\bibfnamefont {Enrique}\
  \bibnamefont {Rico}}, \bibinfo {author} {\bibfnamefont {Luca}\ \bibnamefont
  {Tagliacozzo}}, \bibinfo {author} {\bibfnamefont {Karel}\ \bibnamefont
  {Van~Acoleyen}}, \bibinfo {author} {\bibfnamefont {Frank}\ \bibnamefont
  {Verstraete}}, \bibinfo {author} {\bibfnamefont {Uwe-Jens}\ \bibnamefont
  {Wiese}}, \bibinfo {author} {\bibfnamefont {Matthew}\ \bibnamefont
  {Wingate}}, \bibinfo {author} {\bibfnamefont {Jakub}\ \bibnamefont
  {Zakrzewski}}, \ and\ \bibinfo {author} {\bibfnamefont {Peter}\ \bibnamefont
  {Zoller}},\ }\bibfield  {title} {\enquote {\bibinfo {title} {Simulating
  lattice gauge theories within quantum technologies},}\ }\href {\doibase
  10.1140/epjd/e2020-100571-8} {\bibfield  {journal} {\bibinfo  {journal} {The
  European Physical Journal D}\ }\textbf {\bibinfo {volume} {74}},\ \bibinfo
  {pages} {165} (\bibinfo {year} {2020})}\BibitemShut {NoStop}%
\bibitem [{\citenamefont {Dalmonte}\ and\ \citenamefont
  {Montangero}(2016)}]{Dalmonte_review}%
  \BibitemOpen
  \bibfield  {author} {\bibinfo {author} {\bibfnamefont {M.}~\bibnamefont
  {Dalmonte}}\ and\ \bibinfo {author} {\bibfnamefont {S.}~\bibnamefont
  {Montangero}},\ }\bibfield  {title} {\enquote {\bibinfo {title} {Lattice
  gauge theory simulations in the quantum information era},}\ }\href {\doibase
  10.1080/00107514.2016.1151199} {\bibfield  {journal} {\bibinfo  {journal}
  {Contemporary Physics}\ }\textbf {\bibinfo {volume} {57}},\ \bibinfo {pages}
  {388--412} (\bibinfo {year} {2016})},\ \Eprint
  {http://arxiv.org/abs/https://doi.org/10.1080/00107514.2016.1151199}
  {https://doi.org/10.1080/00107514.2016.1151199} \BibitemShut {NoStop}%
\bibitem [{\citenamefont {Zohar}\ \emph {et~al.}(2015)\citenamefont {Zohar},
  \citenamefont {Cirac},\ and\ \citenamefont {Reznik}}]{Zohar_review}%
  \BibitemOpen
  \bibfield  {author} {\bibinfo {author} {\bibfnamefont {Erez}\ \bibnamefont
  {Zohar}}, \bibinfo {author} {\bibfnamefont {J~Ignacio}\ \bibnamefont
  {Cirac}}, \ and\ \bibinfo {author} {\bibfnamefont {Benni}\ \bibnamefont
  {Reznik}},\ }\bibfield  {title} {\enquote {\bibinfo {title} {Quantum
  simulations of lattice gauge theories using ultracold atoms in optical
  lattices},}\ }\href {\doibase 10.1088/0034-4885/79/1/014401} {\bibfield
  {journal} {\bibinfo  {journal} {Reports on Progress in Physics}\ }\textbf
  {\bibinfo {volume} {79}},\ \bibinfo {pages} {014401} (\bibinfo {year}
  {2015})}\BibitemShut {NoStop}%
\bibitem [{\citenamefont {Aidelsburger}\ \emph {et~al.}(2022)\citenamefont
  {Aidelsburger}, \citenamefont {Barbiero}, \citenamefont {Bermudez},
  \citenamefont {Chanda}, \citenamefont {Dauphin}, \citenamefont
  {González-Cuadra}, \citenamefont {Grzybowski}, \citenamefont {Hands},
  \citenamefont {Jendrzejewski}, \citenamefont {Jünemann}, \citenamefont
  {Juzeliūnas}, \citenamefont {Kasper}, \citenamefont {Piga}, \citenamefont
  {Ran}, \citenamefont {Rizzi}, \citenamefont {Sierra}, \citenamefont
  {Tagliacozzo}, \citenamefont {Tirrito}, \citenamefont {Zache}, \citenamefont
  {Zakrzewski}, \citenamefont {Zohar},\ and\ \citenamefont
  {Lewenstein}}]{aidelsburger2021cold}%
  \BibitemOpen
  \bibfield  {author} {\bibinfo {author} {\bibfnamefont {Monika}\ \bibnamefont
  {Aidelsburger}}, \bibinfo {author} {\bibfnamefont {Luca}\ \bibnamefont
  {Barbiero}}, \bibinfo {author} {\bibfnamefont {Alejandro}\ \bibnamefont
  {Bermudez}}, \bibinfo {author} {\bibfnamefont {Titas}\ \bibnamefont
  {Chanda}}, \bibinfo {author} {\bibfnamefont {Alexandre}\ \bibnamefont
  {Dauphin}}, \bibinfo {author} {\bibfnamefont {Daniel}\ \bibnamefont
  {González-Cuadra}}, \bibinfo {author} {\bibfnamefont {Przemysław~R.}\
  \bibnamefont {Grzybowski}}, \bibinfo {author} {\bibfnamefont {Simon}\
  \bibnamefont {Hands}}, \bibinfo {author} {\bibfnamefont {Fred}\ \bibnamefont
  {Jendrzejewski}}, \bibinfo {author} {\bibfnamefont {Johannes}\ \bibnamefont
  {Jünemann}}, \bibinfo {author} {\bibfnamefont {Gediminas}\ \bibnamefont
  {Juzeliūnas}}, \bibinfo {author} {\bibfnamefont {Valentin}\ \bibnamefont
  {Kasper}}, \bibinfo {author} {\bibfnamefont {Angelo}\ \bibnamefont {Piga}},
  \bibinfo {author} {\bibfnamefont {Shi-Ju}\ \bibnamefont {Ran}}, \bibinfo
  {author} {\bibfnamefont {Matteo}\ \bibnamefont {Rizzi}}, \bibinfo {author}
  {\bibfnamefont {Germán}\ \bibnamefont {Sierra}}, \bibinfo {author}
  {\bibfnamefont {Luca}\ \bibnamefont {Tagliacozzo}}, \bibinfo {author}
  {\bibfnamefont {Emanuele}\ \bibnamefont {Tirrito}}, \bibinfo {author}
  {\bibfnamefont {Torsten~V.}\ \bibnamefont {Zache}}, \bibinfo {author}
  {\bibfnamefont {Jakub}\ \bibnamefont {Zakrzewski}}, \bibinfo {author}
  {\bibfnamefont {Erez}\ \bibnamefont {Zohar}}, \ and\ \bibinfo {author}
  {\bibfnamefont {Maciej}\ \bibnamefont {Lewenstein}},\ }\bibfield  {title}
  {\enquote {\bibinfo {title} {Cold atoms meet lattice gauge theory},}\ }\href
  {\doibase 10.1098/rsta.2021.0064} {\bibfield  {journal} {\bibinfo  {journal}
  {Philosophical Transactions of the Royal Society A: Mathematical, Physical
  and Engineering Sciences}\ }\textbf {\bibinfo {volume} {380}},\ \bibinfo
  {pages} {20210064} (\bibinfo {year} {2022})}\BibitemShut {NoStop}%
\bibitem [{\citenamefont {{Zohar}}(2022)}]{Zohar_NewReview}%
  \BibitemOpen
  \bibfield  {author} {\bibinfo {author} {\bibfnamefont {Erez}\ \bibnamefont
  {{Zohar}}},\ }\bibfield  {title} {\enquote {\bibinfo {title} {{Quantum
  simulation of lattice gauge theories in more than one space
  dimension{\textemdash}requirements, challenges and methods}},}\ }\href
  {\doibase 10.1098/rsta.2021.0069} {\bibfield  {journal} {\bibinfo  {journal}
  {Philosophical Transactions of the Royal Society of London Series A}\
  }\textbf {\bibinfo {volume} {380}},\ \bibinfo {eid} {20210069} (\bibinfo
  {year} {2022})},\ \Eprint {http://arxiv.org/abs/2106.04609} {arXiv:2106.04609
  [quant-ph]} \BibitemShut {NoStop}%
\bibitem [{\citenamefont {Davoudi}\ \emph {et~al.}(2022)\citenamefont
  {Davoudi}, \citenamefont {Balantekin}, \citenamefont {Bhattacharya},
  \citenamefont {Carena}, \citenamefont {de~Jong}, \citenamefont {Draper},
  \citenamefont {El-Khadra}, \citenamefont {Gemelke}, \citenamefont {Hanada},
  \citenamefont {Kharzeev}, \citenamefont {Lamm}, \citenamefont {Li},
  \citenamefont {Liu}, \citenamefont {Lukin}, \citenamefont {Meurice},
  \citenamefont {Monroe}, \citenamefont {Nachman}, \citenamefont {Pagano},
  \citenamefont {Preskill}, \citenamefont {Rinaldi}, \citenamefont {Roggero},
  \citenamefont {Santiago}, \citenamefont {Savage}, \citenamefont {Siddiqi},
  \citenamefont {Siopsis}, \citenamefont {Van~Zanten}, \citenamefont {Wiebe},
  \citenamefont {Yamauchi}, \citenamefont {Yeter-Aydeniz},\ and\ \citenamefont
  {Zorzetti}}]{Bauer_review}%
  \BibitemOpen
  \bibfield  {author} {\bibinfo {author} {\bibfnamefont {Christian W.
  Bauer.~Zohreh}\ \bibnamefont {Davoudi}}, \bibinfo {author} {\bibfnamefont
  {A.~Baha}\ \bibnamefont {Balantekin}}, \bibinfo {author} {\bibfnamefont
  {Tanmoy}\ \bibnamefont {Bhattacharya}}, \bibinfo {author} {\bibfnamefont
  {Marcela}\ \bibnamefont {Carena}}, \bibinfo {author} {\bibfnamefont
  {Wibe~A.}\ \bibnamefont {de~Jong}}, \bibinfo {author} {\bibfnamefont
  {Patrick}\ \bibnamefont {Draper}}, \bibinfo {author} {\bibfnamefont {Aida}\
  \bibnamefont {El-Khadra}}, \bibinfo {author} {\bibfnamefont {Nate}\
  \bibnamefont {Gemelke}}, \bibinfo {author} {\bibfnamefont {Masanori}\
  \bibnamefont {Hanada}}, \bibinfo {author} {\bibfnamefont {Dmitri}\
  \bibnamefont {Kharzeev}}, \bibinfo {author} {\bibfnamefont {Henry}\
  \bibnamefont {Lamm}}, \bibinfo {author} {\bibfnamefont {Ying-Ying}\
  \bibnamefont {Li}}, \bibinfo {author} {\bibfnamefont {Junyu}\ \bibnamefont
  {Liu}}, \bibinfo {author} {\bibfnamefont {Mikhail}\ \bibnamefont {Lukin}},
  \bibinfo {author} {\bibfnamefont {Yannick}\ \bibnamefont {Meurice}}, \bibinfo
  {author} {\bibfnamefont {Christopher}\ \bibnamefont {Monroe}}, \bibinfo
  {author} {\bibfnamefont {Benjamin}\ \bibnamefont {Nachman}}, \bibinfo
  {author} {\bibfnamefont {Guido}\ \bibnamefont {Pagano}}, \bibinfo {author}
  {\bibfnamefont {John}\ \bibnamefont {Preskill}}, \bibinfo {author}
  {\bibfnamefont {Enrico}\ \bibnamefont {Rinaldi}}, \bibinfo {author}
  {\bibfnamefont {Alessandro}\ \bibnamefont {Roggero}}, \bibinfo {author}
  {\bibfnamefont {David~I.}\ \bibnamefont {Santiago}}, \bibinfo {author}
  {\bibfnamefont {Martin~J.}\ \bibnamefont {Savage}}, \bibinfo {author}
  {\bibfnamefont {Irfan}\ \bibnamefont {Siddiqi}}, \bibinfo {author}
  {\bibfnamefont {George}\ \bibnamefont {Siopsis}}, \bibinfo {author}
  {\bibfnamefont {David}\ \bibnamefont {Van~Zanten}}, \bibinfo {author}
  {\bibfnamefont {Nathan}\ \bibnamefont {Wiebe}}, \bibinfo {author}
  {\bibfnamefont {Yukari}\ \bibnamefont {Yamauchi}}, \bibinfo {author}
  {\bibfnamefont {Kübra}\ \bibnamefont {Yeter-Aydeniz}}, \ and\ \bibinfo
  {author} {\bibfnamefont {Silvia}\ \bibnamefont {Zorzetti}},\ }\bibfield
  {title} {\enquote {\bibinfo {title} {Quantum simulation for high energy
  physics},}\ }\href {\doibase 10.48550/ARXIV.2204.03381} {\  (\bibinfo {year}
  {2022}),\ 10.48550/ARXIV.2204.03381}\BibitemShut {NoStop}%
\bibitem [{\citenamefont {Weinberg}(1995)}]{Weinberg_book}%
  \BibitemOpen
  \bibfield  {author} {\bibinfo {author} {\bibfnamefont {S.}~\bibnamefont
  {Weinberg}},\ }\href {https://books.google.de/books?id=doeDB3\_WLvwC} {\emph
  {\bibinfo {title} {The Quantum Theory of Fields}}},\ Vol. 2: Modern
  Applications\ (\bibinfo  {publisher} {Cambridge University Press},\ \bibinfo
  {year} {1995})\BibitemShut {NoStop}%
\bibitem [{\citenamefont {Gattringer}\ and\ \citenamefont
  {Lang}(2009)}]{Gattringer_book}%
  \BibitemOpen
  \bibfield  {author} {\bibinfo {author} {\bibfnamefont {C.}~\bibnamefont
  {Gattringer}}\ and\ \bibinfo {author} {\bibfnamefont {C.}~\bibnamefont
  {Lang}},\ }\href {https://books.google.de/books?id=l2hZKnlYDxoC} {\emph
  {\bibinfo {title} {Quantum Chromodynamics on the Lattice: An Introductory
  Presentation}}},\ Lecture Notes in Physics\ (\bibinfo  {publisher} {Springer
  Berlin Heidelberg},\ \bibinfo {year} {2009})\BibitemShut {NoStop}%
\bibitem [{\citenamefont {Zee}(2003)}]{Zee_book}%
  \BibitemOpen
  \bibfield  {author} {\bibinfo {author} {\bibfnamefont {A.}~\bibnamefont
  {Zee}},\ }\href {https://books.google.de/books?id=85G9QgAACAAJ} {\emph
  {\bibinfo {title} {Quantum Field Theory in a Nutshell}}}\ (\bibinfo
  {publisher} {Princeton University Press},\ \bibinfo {year}
  {2003})\BibitemShut {NoStop}%
\bibitem [{\citenamefont {Hamer}\ \emph {et~al.}(1997)\citenamefont {Hamer},
  \citenamefont {Weihong},\ and\ \citenamefont {Oitmaa}}]{Hamer1997}%
  \BibitemOpen
  \bibfield  {author} {\bibinfo {author} {\bibfnamefont {C.~J.}\ \bibnamefont
  {Hamer}}, \bibinfo {author} {\bibfnamefont {Zheng}\ \bibnamefont {Weihong}},
  \ and\ \bibinfo {author} {\bibfnamefont {J.}~\bibnamefont {Oitmaa}},\
  }\bibfield  {title} {\enquote {\bibinfo {title} {Series expansions for the
  massive schwinger model in hamiltonian lattice theory},}\ }\href {\doibase
  10.1103/PhysRevD.56.55} {\bibfield  {journal} {\bibinfo  {journal} {Phys.
  Rev. D}\ }\textbf {\bibinfo {volume} {56}},\ \bibinfo {pages} {55--67}
  (\bibinfo {year} {1997})}\BibitemShut {NoStop}%
\bibitem [{\citenamefont {Ba{\~n}uls}\ \emph {et~al.}(2013)\citenamefont
  {Ba{\~n}uls}, \citenamefont {Cichy}, \citenamefont {Cirac},\ and\
  \citenamefont {Jansen}}]{Banuls2013}%
  \BibitemOpen
  \bibfield  {author} {\bibinfo {author} {\bibfnamefont {M.~C.}\ \bibnamefont
  {Ba{\~n}uls}}, \bibinfo {author} {\bibfnamefont {K.}~\bibnamefont {Cichy}},
  \bibinfo {author} {\bibfnamefont {J.~I.}\ \bibnamefont {Cirac}}, \ and\
  \bibinfo {author} {\bibfnamefont {K.}~\bibnamefont {Jansen}},\ }\bibfield
  {title} {\enquote {\bibinfo {title} {The mass spectrum of the schwinger model
  with matrix product states},}\ }\href {\doibase 10.1007/JHEP11(2013)158}
  {\bibfield  {journal} {\bibinfo  {journal} {Journal of High Energy Physics}\
  }\textbf {\bibinfo {volume} {2013}},\ \bibinfo {pages} {158} (\bibinfo {year}
  {2013})}\BibitemShut {NoStop}%
\bibitem [{\citenamefont {Bañuls}\ \emph {et~al.}(2013)\citenamefont
  {Bañuls}, \citenamefont {Cichy}, \citenamefont {Cirac}, \citenamefont
  {Jansen},\ and\ \citenamefont {Saito}}]{BanulsPos2013}%
  \BibitemOpen
  \bibfield  {author} {\bibinfo {author} {\bibfnamefont {Mari~Carmen}\
  \bibnamefont {Bañuls}}, \bibinfo {author} {\bibfnamefont {Krzysztof}\
  \bibnamefont {Cichy}}, \bibinfo {author} {\bibfnamefont {J.~Ignacio}\
  \bibnamefont {Cirac}}, \bibinfo {author} {\bibfnamefont {Karl}\ \bibnamefont
  {Jansen}}, \ and\ \bibinfo {author} {\bibfnamefont {Hana}\ \bibnamefont
  {Saito}},\ }\bibfield  {title} {\enquote {\bibinfo {title} {Matrix product
  states for lattice field theories},}\ }\href {\doibase
  10.48550/ARXIV.1310.4118} {\  (\bibinfo {year} {2013}),\
  10.48550/ARXIV.1310.4118}\BibitemShut {NoStop}%
\bibitem [{\citenamefont {Saito}\ \emph {et~al.}(2015)\citenamefont {Saito},
  \citenamefont {Bañuls}, \citenamefont {Cichy}, \citenamefont {Cirac},\ and\
  \citenamefont {Jansen}}]{Saito2015}%
  \BibitemOpen
  \bibfield  {author} {\bibinfo {author} {\bibfnamefont {H.}~\bibnamefont
  {Saito}}, \bibinfo {author} {\bibfnamefont {M.~C.}\ \bibnamefont {Bañuls}},
  \bibinfo {author} {\bibfnamefont {K.}~\bibnamefont {Cichy}}, \bibinfo
  {author} {\bibfnamefont {J.~I.}\ \bibnamefont {Cirac}}, \ and\ \bibinfo
  {author} {\bibfnamefont {K.}~\bibnamefont {Jansen}},\ }\bibfield  {title}
  {\enquote {\bibinfo {title} {Thermal evolution of the one-flavour schwinger
  model using matrix product states},}\ }\href {\doibase
  10.48550/ARXIV.1511.00794} {\  (\bibinfo {year} {2015}),\
  10.48550/ARXIV.1511.00794}\BibitemShut {NoStop}%
\bibitem [{\citenamefont {Ba\~nuls}\ \emph {et~al.}(2016)\citenamefont
  {Ba\~nuls}, \citenamefont {Cichy}, \citenamefont {Jansen},\ and\
  \citenamefont {Saito}}]{Banuls2016}%
  \BibitemOpen
  \bibfield  {author} {\bibinfo {author} {\bibfnamefont {Mari~Carmen}\
  \bibnamefont {Ba\~nuls}}, \bibinfo {author} {\bibfnamefont {Krzysztof}\
  \bibnamefont {Cichy}}, \bibinfo {author} {\bibfnamefont {Karl}\ \bibnamefont
  {Jansen}}, \ and\ \bibinfo {author} {\bibfnamefont {Hana}\ \bibnamefont
  {Saito}},\ }\bibfield  {title} {\enquote {\bibinfo {title} {Chiral condensate
  in the schwinger model with matrix product operators},}\ }\href {\doibase
  10.1103/PhysRevD.93.094512} {\bibfield  {journal} {\bibinfo  {journal} {Phys.
  Rev. D}\ }\textbf {\bibinfo {volume} {93}},\ \bibinfo {pages} {094512}
  (\bibinfo {year} {2016})}\BibitemShut {NoStop}%
\bibitem [{\citenamefont {Martinez}\ \emph {et~al.}(2016)\citenamefont
  {Martinez}, \citenamefont {Muschik}, \citenamefont {Schindler}, \citenamefont
  {Nigg}, \citenamefont {Erhard}, \citenamefont {Heyl}, \citenamefont {Hauke},
  \citenamefont {Dalmonte}, \citenamefont {Monz}, \citenamefont {Zoller},\ and\
  \citenamefont {Blatt}}]{Martinez2016}%
  \BibitemOpen
  \bibfield  {author} {\bibinfo {author} {\bibfnamefont {Esteban~A.}\
  \bibnamefont {Martinez}}, \bibinfo {author} {\bibfnamefont {Christine~A.}\
  \bibnamefont {Muschik}}, \bibinfo {author} {\bibfnamefont {Philipp}\
  \bibnamefont {Schindler}}, \bibinfo {author} {\bibfnamefont {Daniel}\
  \bibnamefont {Nigg}}, \bibinfo {author} {\bibfnamefont {Alexander}\
  \bibnamefont {Erhard}}, \bibinfo {author} {\bibfnamefont {Markus}\
  \bibnamefont {Heyl}}, \bibinfo {author} {\bibfnamefont {Philipp}\
  \bibnamefont {Hauke}}, \bibinfo {author} {\bibfnamefont {Marcello}\
  \bibnamefont {Dalmonte}}, \bibinfo {author} {\bibfnamefont {Thomas}\
  \bibnamefont {Monz}}, \bibinfo {author} {\bibfnamefont {Peter}\ \bibnamefont
  {Zoller}}, \ and\ \bibinfo {author} {\bibfnamefont {Rainer}\ \bibnamefont
  {Blatt}},\ }\bibfield  {title} {\enquote {\bibinfo {title} {Real-time
  dynamics of lattice gauge theories with a few-qubit quantum computer},}\
  }\href {\doibase 10.1038/nature18318} {\bibfield  {journal} {\bibinfo
  {journal} {Nature}\ }\textbf {\bibinfo {volume} {534}},\ \bibinfo {pages}
  {516--519} (\bibinfo {year} {2016})}\BibitemShut {NoStop}%
\bibitem [{\citenamefont {Bernien}\ \emph {et~al.}(2017)\citenamefont
  {Bernien}, \citenamefont {Schwartz}, \citenamefont {Keesling}, \citenamefont
  {Levine}, \citenamefont {Omran}, \citenamefont {Pichler}, \citenamefont
  {Choi}, \citenamefont {Zibrov}, \citenamefont {Endres}, \citenamefont
  {Greiner}, \citenamefont {Vuleti{\'c}},\ and\ \citenamefont
  {Lukin}}]{Bernien2017}%
  \BibitemOpen
  \bibfield  {author} {\bibinfo {author} {\bibfnamefont {Hannes}\ \bibnamefont
  {Bernien}}, \bibinfo {author} {\bibfnamefont {Sylvain}\ \bibnamefont
  {Schwartz}}, \bibinfo {author} {\bibfnamefont {Alexander}\ \bibnamefont
  {Keesling}}, \bibinfo {author} {\bibfnamefont {Harry}\ \bibnamefont
  {Levine}}, \bibinfo {author} {\bibfnamefont {Ahmed}\ \bibnamefont {Omran}},
  \bibinfo {author} {\bibfnamefont {Hannes}\ \bibnamefont {Pichler}}, \bibinfo
  {author} {\bibfnamefont {Soonwon}\ \bibnamefont {Choi}}, \bibinfo {author}
  {\bibfnamefont {Alexander~S.}\ \bibnamefont {Zibrov}}, \bibinfo {author}
  {\bibfnamefont {Manuel}\ \bibnamefont {Endres}}, \bibinfo {author}
  {\bibfnamefont {Markus}\ \bibnamefont {Greiner}}, \bibinfo {author}
  {\bibfnamefont {Vladan}\ \bibnamefont {Vuleti{\'c}}}, \ and\ \bibinfo
  {author} {\bibfnamefont {Mikhail~D.}\ \bibnamefont {Lukin}},\ }\bibfield
  {title} {\enquote {\bibinfo {title} {Probing many-body dynamics on a 51-atom
  quantum simulator},}\ }\href {\doibase 10.1038/nature24622} {\bibfield
  {journal} {\bibinfo  {journal} {Nature}\ }\textbf {\bibinfo {volume} {551}},\
  \bibinfo {pages} {579--584} (\bibinfo {year} {2017})}\BibitemShut {NoStop}%
\bibitem [{\citenamefont {Surace}\ \emph {et~al.}(2020)\citenamefont {Surace},
  \citenamefont {Mazza}, \citenamefont {Giudici}, \citenamefont {Lerose},
  \citenamefont {Gambassi},\ and\ \citenamefont {Dalmonte}}]{Surace2020}%
  \BibitemOpen
  \bibfield  {author} {\bibinfo {author} {\bibfnamefont {Federica~M.}\
  \bibnamefont {Surace}}, \bibinfo {author} {\bibfnamefont {Paolo~P.}\
  \bibnamefont {Mazza}}, \bibinfo {author} {\bibfnamefont {Giuliano}\
  \bibnamefont {Giudici}}, \bibinfo {author} {\bibfnamefont {Alessio}\
  \bibnamefont {Lerose}}, \bibinfo {author} {\bibfnamefont {Andrea}\
  \bibnamefont {Gambassi}}, \ and\ \bibinfo {author} {\bibfnamefont {Marcello}\
  \bibnamefont {Dalmonte}},\ }\bibfield  {title} {\enquote {\bibinfo {title}
  {Lattice gauge theories and string dynamics in rydberg atom quantum
  simulators},}\ }\href {\doibase 10.1103/PhysRevX.10.021041} {\bibfield
  {journal} {\bibinfo  {journal} {Phys. Rev. X}\ }\textbf {\bibinfo {volume}
  {10}},\ \bibinfo {pages} {021041} (\bibinfo {year} {2020})}\BibitemShut
  {NoStop}%
\bibitem [{\citenamefont {Klco}\ \emph {et~al.}(2018)\citenamefont {Klco},
  \citenamefont {Dumitrescu}, \citenamefont {McCaskey}, \citenamefont {Morris},
  \citenamefont {Pooser}, \citenamefont {Sanz}, \citenamefont {Solano},
  \citenamefont {Lougovski},\ and\ \citenamefont {Savage}}]{Klco2018}%
  \BibitemOpen
  \bibfield  {author} {\bibinfo {author} {\bibfnamefont {N.}~\bibnamefont
  {Klco}}, \bibinfo {author} {\bibfnamefont {E.~F.}\ \bibnamefont
  {Dumitrescu}}, \bibinfo {author} {\bibfnamefont {A.~J.}\ \bibnamefont
  {McCaskey}}, \bibinfo {author} {\bibfnamefont {T.~D.}\ \bibnamefont
  {Morris}}, \bibinfo {author} {\bibfnamefont {R.~C.}\ \bibnamefont {Pooser}},
  \bibinfo {author} {\bibfnamefont {M.}~\bibnamefont {Sanz}}, \bibinfo {author}
  {\bibfnamefont {E.}~\bibnamefont {Solano}}, \bibinfo {author} {\bibfnamefont
  {P.}~\bibnamefont {Lougovski}}, \ and\ \bibinfo {author} {\bibfnamefont
  {M.~J.}\ \bibnamefont {Savage}},\ }\bibfield  {title} {\enquote {\bibinfo
  {title} {Quantum-classical computation of schwinger model dynamics using
  quantum computers},}\ }\href {\doibase 10.1103/PhysRevA.98.032331} {\bibfield
   {journal} {\bibinfo  {journal} {Phys. Rev. A}\ }\textbf {\bibinfo {volume}
  {98}},\ \bibinfo {pages} {032331} (\bibinfo {year} {2018})}\BibitemShut
  {NoStop}%
\bibitem [{\citenamefont {Kokail}\ \emph {et~al.}(2019)\citenamefont {Kokail},
  \citenamefont {Maier}, \citenamefont {van Bijnen}, \citenamefont {Brydges},
  \citenamefont {Joshi}, \citenamefont {Jurcevic}, \citenamefont {Muschik},
  \citenamefont {Silvi}, \citenamefont {Blatt}, \citenamefont {Roos},\ and\
  \citenamefont {Zoller}}]{Kokail2019}%
  \BibitemOpen
  \bibfield  {author} {\bibinfo {author} {\bibfnamefont {C.}~\bibnamefont
  {Kokail}}, \bibinfo {author} {\bibfnamefont {C.}~\bibnamefont {Maier}},
  \bibinfo {author} {\bibfnamefont {R.}~\bibnamefont {van Bijnen}}, \bibinfo
  {author} {\bibfnamefont {T.}~\bibnamefont {Brydges}}, \bibinfo {author}
  {\bibfnamefont {M.~K.}\ \bibnamefont {Joshi}}, \bibinfo {author}
  {\bibfnamefont {P.}~\bibnamefont {Jurcevic}}, \bibinfo {author}
  {\bibfnamefont {C.~A.}\ \bibnamefont {Muschik}}, \bibinfo {author}
  {\bibfnamefont {P.}~\bibnamefont {Silvi}}, \bibinfo {author} {\bibfnamefont
  {R.}~\bibnamefont {Blatt}}, \bibinfo {author} {\bibfnamefont {C.~F.}\
  \bibnamefont {Roos}}, \ and\ \bibinfo {author} {\bibfnamefont
  {P.}~\bibnamefont {Zoller}},\ }\bibfield  {title} {\enquote {\bibinfo {title}
  {Self-verifying variational quantum simulation of lattice models},}\ }\href
  {\doibase 10.1038/s41586-019-1177-4} {\bibfield  {journal} {\bibinfo
  {journal} {Nature}\ }\textbf {\bibinfo {volume} {569}},\ \bibinfo {pages}
  {355--360} (\bibinfo {year} {2019})}\BibitemShut {NoStop}%
\bibitem [{\citenamefont {Zohar}\ and\ \citenamefont
  {Cirac}(2019)}]{Zohar2019}%
  \BibitemOpen
  \bibfield  {author} {\bibinfo {author} {\bibfnamefont {Erez}\ \bibnamefont
  {Zohar}}\ and\ \bibinfo {author} {\bibfnamefont {J.~Ignacio}\ \bibnamefont
  {Cirac}},\ }\bibfield  {title} {\enquote {\bibinfo {title} {Removing
  staggered fermionic matter in $u(n)$ and $su(n)$ lattice gauge theories},}\
  }\href {\doibase 10.1103/PhysRevD.99.114511} {\bibfield  {journal} {\bibinfo
  {journal} {Phys. Rev. D}\ }\textbf {\bibinfo {volume} {99}},\ \bibinfo
  {pages} {114511} (\bibinfo {year} {2019})}\BibitemShut {NoStop}%
\bibitem [{\citenamefont {Klco}\ \emph {et~al.}(2020)\citenamefont {Klco},
  \citenamefont {Savage},\ and\ \citenamefont {Stryker}}]{Klco2020}%
  \BibitemOpen
  \bibfield  {author} {\bibinfo {author} {\bibfnamefont {Natalie}\ \bibnamefont
  {Klco}}, \bibinfo {author} {\bibfnamefont {Martin~J.}\ \bibnamefont
  {Savage}}, \ and\ \bibinfo {author} {\bibfnamefont {Jesse~R.}\ \bibnamefont
  {Stryker}},\ }\bibfield  {title} {\enquote {\bibinfo {title} {Su(2)
  non-abelian gauge field theory in one dimension on digital quantum
  computers},}\ }\href {\doibase 10.1103/PhysRevD.101.074512} {\bibfield
  {journal} {\bibinfo  {journal} {Phys. Rev. D}\ }\textbf {\bibinfo {volume}
  {101}},\ \bibinfo {pages} {074512} (\bibinfo {year} {2020})}\BibitemShut
  {NoStop}%
\bibitem [{\citenamefont {Ciavarella}\ \emph {et~al.}(2021)\citenamefont
  {Ciavarella}, \citenamefont {Klco},\ and\ \citenamefont
  {Savage}}]{Ciavarella2021}%
  \BibitemOpen
  \bibfield  {author} {\bibinfo {author} {\bibfnamefont {Anthony}\ \bibnamefont
  {Ciavarella}}, \bibinfo {author} {\bibfnamefont {Natalie}\ \bibnamefont
  {Klco}}, \ and\ \bibinfo {author} {\bibfnamefont {Martin~J.}\ \bibnamefont
  {Savage}},\ }\bibfield  {title} {\enquote {\bibinfo {title} {Trailhead for
  quantum simulation of su(3) yang-mills lattice gauge theory in the local
  multiplet basis},}\ }\href {\doibase 10.1103/PhysRevD.103.094501} {\bibfield
  {journal} {\bibinfo  {journal} {Phys. Rev. D}\ }\textbf {\bibinfo {volume}
  {103}},\ \bibinfo {pages} {094501} (\bibinfo {year} {2021})}\BibitemShut
  {NoStop}%
\bibitem [{\citenamefont {Atas}\ \emph {et~al.}(2021)\citenamefont {Atas},
  \citenamefont {Zhang}, \citenamefont {Lewis}, \citenamefont {Jahanpour},
  \citenamefont {Haase},\ and\ \citenamefont {Muschik}}]{Atas2021}%
  \BibitemOpen
  \bibfield  {author} {\bibinfo {author} {\bibfnamefont {Yasar~Y.}\
  \bibnamefont {Atas}}, \bibinfo {author} {\bibfnamefont {Jinglei}\
  \bibnamefont {Zhang}}, \bibinfo {author} {\bibfnamefont {Randy}\ \bibnamefont
  {Lewis}}, \bibinfo {author} {\bibfnamefont {Amin}\ \bibnamefont {Jahanpour}},
  \bibinfo {author} {\bibfnamefont {Jan~F.}\ \bibnamefont {Haase}}, \ and\
  \bibinfo {author} {\bibfnamefont {Christine~A.}\ \bibnamefont {Muschik}},\
  }\bibfield  {title} {\enquote {\bibinfo {title} {Su(2) hadrons on a quantum
  computer via a variational approach},}\ }\href {\doibase
  10.1038/s41467-021-26825-4} {\bibfield  {journal} {\bibinfo  {journal}
  {Nature Communications}\ }\textbf {\bibinfo {volume} {12}},\ \bibinfo {pages}
  {6499} (\bibinfo {year} {2021})}\BibitemShut {NoStop}%
\bibitem [{\citenamefont {Nguyen}\ \emph {et~al.}(2021)\citenamefont {Nguyen},
  \citenamefont {Tran}, \citenamefont {Zhu}, \citenamefont {Green},
  \citenamefont {Alderete}, \citenamefont {Davoudi},\ and\ \citenamefont
  {Linke}}]{Nguyen2021}%
  \BibitemOpen
  \bibfield  {author} {\bibinfo {author} {\bibfnamefont {Nhung~H.}\
  \bibnamefont {Nguyen}}, \bibinfo {author} {\bibfnamefont {Minh~C.}\
  \bibnamefont {Tran}}, \bibinfo {author} {\bibfnamefont {Yingyue}\
  \bibnamefont {Zhu}}, \bibinfo {author} {\bibfnamefont {Alaina~M.}\
  \bibnamefont {Green}}, \bibinfo {author} {\bibfnamefont {C.~Huerta}\
  \bibnamefont {Alderete}}, \bibinfo {author} {\bibfnamefont {Zohreh}\
  \bibnamefont {Davoudi}}, \ and\ \bibinfo {author} {\bibfnamefont
  {Norbert~M.}\ \bibnamefont {Linke}},\ }\bibfield  {title} {\enquote {\bibinfo
  {title} {Digital quantum simulation of the schwinger model and symmetry
  protection with trapped ions},}\ }\href {\doibase 10.48550/ARXIV.2112.14262}
  {\  (\bibinfo {year} {2021}),\ 10.48550/ARXIV.2112.14262}\BibitemShut
  {NoStop}%
\bibitem [{\citenamefont {Muschik}\ \emph {et~al.}(2017)\citenamefont
  {Muschik}, \citenamefont {Heyl}, \citenamefont {Martinez}, \citenamefont
  {Monz}, \citenamefont {Schindler}, \citenamefont {Vogell}, \citenamefont
  {Dalmonte}, \citenamefont {Hauke}, \citenamefont {Blatt},\ and\ \citenamefont
  {Zoller}}]{Muschik2017}%
  \BibitemOpen
  \bibfield  {author} {\bibinfo {author} {\bibfnamefont {Christine}\
  \bibnamefont {Muschik}}, \bibinfo {author} {\bibfnamefont {Markus}\
  \bibnamefont {Heyl}}, \bibinfo {author} {\bibfnamefont {Esteban}\
  \bibnamefont {Martinez}}, \bibinfo {author} {\bibfnamefont {Thomas}\
  \bibnamefont {Monz}}, \bibinfo {author} {\bibfnamefont {Philipp}\
  \bibnamefont {Schindler}}, \bibinfo {author} {\bibfnamefont {Berit}\
  \bibnamefont {Vogell}}, \bibinfo {author} {\bibfnamefont {Marcello}\
  \bibnamefont {Dalmonte}}, \bibinfo {author} {\bibfnamefont {Philipp}\
  \bibnamefont {Hauke}}, \bibinfo {author} {\bibfnamefont {Rainer}\
  \bibnamefont {Blatt}}, \ and\ \bibinfo {author} {\bibfnamefont {Peter}\
  \bibnamefont {Zoller}},\ }\bibfield  {title} {\enquote {\bibinfo {title}
  {U(1) wilson lattice gauge theories in digital quantum simulators},}\ }\href
  {\doibase 10.1088/1367-2630/aa89ab} {\bibfield  {journal} {\bibinfo
  {journal} {New Journal of Physics}\ }\textbf {\bibinfo {volume} {19}},\
  \bibinfo {pages} {103020} (\bibinfo {year} {2017})}\BibitemShut {NoStop}%
\bibitem [{\citenamefont {Zohar}\ and\ \citenamefont
  {Reznik}(2011)}]{Zohar2011}%
  \BibitemOpen
  \bibfield  {author} {\bibinfo {author} {\bibfnamefont {Erez}\ \bibnamefont
  {Zohar}}\ and\ \bibinfo {author} {\bibfnamefont {Benni}\ \bibnamefont
  {Reznik}},\ }\bibfield  {title} {\enquote {\bibinfo {title} {Confinement and
  lattice quantum-electrodynamic electric flux tubes simulated with ultracold
  atoms},}\ }\href {\doibase 10.1103/PhysRevLett.107.275301} {\bibfield
  {journal} {\bibinfo  {journal} {Phys. Rev. Lett.}\ }\textbf {\bibinfo
  {volume} {107}},\ \bibinfo {pages} {275301} (\bibinfo {year}
  {2011})}\BibitemShut {NoStop}%
\bibitem [{\citenamefont {Zohar}\ \emph {et~al.}(2012)\citenamefont {Zohar},
  \citenamefont {Cirac},\ and\ \citenamefont {Reznik}}]{Zohar2012}%
  \BibitemOpen
  \bibfield  {author} {\bibinfo {author} {\bibfnamefont {Erez}\ \bibnamefont
  {Zohar}}, \bibinfo {author} {\bibfnamefont {J.~Ignacio}\ \bibnamefont
  {Cirac}}, \ and\ \bibinfo {author} {\bibfnamefont {Benni}\ \bibnamefont
  {Reznik}},\ }\bibfield  {title} {\enquote {\bibinfo {title} {Simulating
  compact quantum electrodynamics with ultracold atoms: Probing confinement and
  nonperturbative effects},}\ }\href {\doibase 10.1103/PhysRevLett.109.125302}
  {\bibfield  {journal} {\bibinfo  {journal} {Phys. Rev. Lett.}\ }\textbf
  {\bibinfo {volume} {109}},\ \bibinfo {pages} {125302} (\bibinfo {year}
  {2012})}\BibitemShut {NoStop}%
\bibitem [{\citenamefont {Banerjee}\ \emph {et~al.}(2012)\citenamefont
  {Banerjee}, \citenamefont {Dalmonte}, \citenamefont {Müller}, \citenamefont
  {Rico}, \citenamefont {Stebler}, \citenamefont {Wiese},\ and\ \citenamefont
  {Zoller}}]{Banerjee2012}%
  \BibitemOpen
  \bibfield  {author} {\bibinfo {author} {\bibfnamefont {D.}~\bibnamefont
  {Banerjee}}, \bibinfo {author} {\bibfnamefont {M.}~\bibnamefont {Dalmonte}},
  \bibinfo {author} {\bibfnamefont {M.}~\bibnamefont {Müller}}, \bibinfo
  {author} {\bibfnamefont {E.}~\bibnamefont {Rico}}, \bibinfo {author}
  {\bibfnamefont {P.}~\bibnamefont {Stebler}}, \bibinfo {author} {\bibfnamefont
  {U.-J.}\ \bibnamefont {Wiese}}, \ and\ \bibinfo {author} {\bibfnamefont
  {P.}~\bibnamefont {Zoller}},\ }\bibfield  {title} {\enquote {\bibinfo {title}
  {Atomic quantum simulation of dynamical gauge fields coupled to fermionic
  matter: From string breaking to evolution after a quench},}\ }\href {\doibase
  10.1103/physrevlett.109.175302} {\bibfield  {journal} {\bibinfo  {journal}
  {Physical Review Letters}\ }\textbf {\bibinfo {volume} {109}} (\bibinfo
  {year} {2012}),\ 10.1103/physrevlett.109.175302}\BibitemShut {NoStop}%
\bibitem [{\citenamefont {Tagliacozzo}\ \emph
  {et~al.}(2013{\natexlab{a}})\citenamefont {Tagliacozzo}, \citenamefont
  {Celi}, \citenamefont {Zamora},\ and\ \citenamefont
  {Lewenstein}}]{Tagliacozzo2012}%
  \BibitemOpen
  \bibfield  {author} {\bibinfo {author} {\bibfnamefont {L.}~\bibnamefont
  {Tagliacozzo}}, \bibinfo {author} {\bibfnamefont {A.}~\bibnamefont {Celi}},
  \bibinfo {author} {\bibfnamefont {A.}~\bibnamefont {Zamora}}, \ and\ \bibinfo
  {author} {\bibfnamefont {M.}~\bibnamefont {Lewenstein}},\ }\bibfield  {title}
  {\enquote {\bibinfo {title} {Optical abelian lattice gauge theories},}\
  }\href {\doibase https://doi.org/10.1016/j.aop.2012.11.009} {\bibfield
  {journal} {\bibinfo  {journal} {Annals of Physics}\ }\textbf {\bibinfo
  {volume} {330}},\ \bibinfo {pages} {160--191} (\bibinfo {year}
  {2013}{\natexlab{a}})}\BibitemShut {NoStop}%
\bibitem [{\citenamefont {Zohar}\ \emph
  {et~al.}(2013{\natexlab{a}})\citenamefont {Zohar}, \citenamefont {Cirac},\
  and\ \citenamefont {Reznik}}]{Zohar2013}%
  \BibitemOpen
  \bibfield  {author} {\bibinfo {author} {\bibfnamefont {Erez}\ \bibnamefont
  {Zohar}}, \bibinfo {author} {\bibfnamefont {J.~Ignacio}\ \bibnamefont
  {Cirac}}, \ and\ \bibinfo {author} {\bibfnamefont {Benni}\ \bibnamefont
  {Reznik}},\ }\bibfield  {title} {\enquote {\bibinfo {title} {Simulating
  ($2+1$)-dimensional lattice qed with dynamical matter using ultracold
  atoms},}\ }\href {\doibase 10.1103/PhysRevLett.110.055302} {\bibfield
  {journal} {\bibinfo  {journal} {Phys. Rev. Lett.}\ }\textbf {\bibinfo
  {volume} {110}},\ \bibinfo {pages} {055302} (\bibinfo {year}
  {2013}{\natexlab{a}})}\BibitemShut {NoStop}%
\bibitem [{\citenamefont {Banerjee}\ \emph {et~al.}(2013)\citenamefont
  {Banerjee}, \citenamefont {B\"ogli}, \citenamefont {Dalmonte}, \citenamefont
  {Rico}, \citenamefont {Stebler}, \citenamefont {Wiese},\ and\ \citenamefont
  {Zoller}}]{Banerjee2013}%
  \BibitemOpen
  \bibfield  {author} {\bibinfo {author} {\bibfnamefont {D.}~\bibnamefont
  {Banerjee}}, \bibinfo {author} {\bibfnamefont {M.}~\bibnamefont {B\"ogli}},
  \bibinfo {author} {\bibfnamefont {M.}~\bibnamefont {Dalmonte}}, \bibinfo
  {author} {\bibfnamefont {E.}~\bibnamefont {Rico}}, \bibinfo {author}
  {\bibfnamefont {P.}~\bibnamefont {Stebler}}, \bibinfo {author} {\bibfnamefont
  {U.-J.}\ \bibnamefont {Wiese}}, \ and\ \bibinfo {author} {\bibfnamefont
  {P.}~\bibnamefont {Zoller}},\ }\bibfield  {title} {\enquote {\bibinfo {title}
  {Atomic quantum simulation of $\mathbf{U}(n)$ and $\mathrm{SU}(n)$
  non-abelian lattice gauge theories},}\ }\href {\doibase
  10.1103/PhysRevLett.110.125303} {\bibfield  {journal} {\bibinfo  {journal}
  {Phys. Rev. Lett.}\ }\textbf {\bibinfo {volume} {110}},\ \bibinfo {pages}
  {125303} (\bibinfo {year} {2013})}\BibitemShut {NoStop}%
\bibitem [{\citenamefont {Tagliacozzo}\ \emph
  {et~al.}(2013{\natexlab{b}})\citenamefont {Tagliacozzo}, \citenamefont
  {Celi}, \citenamefont {Orland}, \citenamefont {Mitchell},\ and\ \citenamefont
  {Lewenstein}}]{Tagliacozzo2013}%
  \BibitemOpen
  \bibfield  {author} {\bibinfo {author} {\bibfnamefont {L.}~\bibnamefont
  {Tagliacozzo}}, \bibinfo {author} {\bibfnamefont {A.}~\bibnamefont {Celi}},
  \bibinfo {author} {\bibfnamefont {P.}~\bibnamefont {Orland}}, \bibinfo
  {author} {\bibfnamefont {M.~W.}\ \bibnamefont {Mitchell}}, \ and\ \bibinfo
  {author} {\bibfnamefont {M.}~\bibnamefont {Lewenstein}},\ }\bibfield  {title}
  {\enquote {\bibinfo {title} {Simulation of non-abelian gauge theories with
  optical lattices},}\ }\href {\doibase 10.1038/ncomms3615} {\bibfield
  {journal} {\bibinfo  {journal} {Nature Communications}\ }\textbf {\bibinfo
  {volume} {4}},\ \bibinfo {pages} {2615} (\bibinfo {year}
  {2013}{\natexlab{b}})}\BibitemShut {NoStop}%
\bibitem [{\citenamefont {Hauke}\ \emph {et~al.}(2013)\citenamefont {Hauke},
  \citenamefont {Marcos}, \citenamefont {Dalmonte},\ and\ \citenamefont
  {Zoller}}]{Hauke2013}%
  \BibitemOpen
  \bibfield  {author} {\bibinfo {author} {\bibfnamefont {P.}~\bibnamefont
  {Hauke}}, \bibinfo {author} {\bibfnamefont {D.}~\bibnamefont {Marcos}},
  \bibinfo {author} {\bibfnamefont {M.}~\bibnamefont {Dalmonte}}, \ and\
  \bibinfo {author} {\bibfnamefont {P.}~\bibnamefont {Zoller}},\ }\bibfield
  {title} {\enquote {\bibinfo {title} {Quantum simulation of a lattice
  schwinger model in a chain of trapped ions},}\ }\href {\doibase
  10.1103/PhysRevX.3.041018} {\bibfield  {journal} {\bibinfo  {journal} {Phys.
  Rev. X}\ }\textbf {\bibinfo {volume} {3}},\ \bibinfo {pages} {041018}
  (\bibinfo {year} {2013})}\BibitemShut {NoStop}%
\bibitem [{\citenamefont {Schweizer}\ \emph {et~al.}(2019)\citenamefont
  {Schweizer}, \citenamefont {Grusdt}, \citenamefont {Berngruber},
  \citenamefont {Barbiero}, \citenamefont {Demler}, \citenamefont {Goldman},
  \citenamefont {Bloch},\ and\ \citenamefont {Aidelsburger}}]{Schweizer2019}%
  \BibitemOpen
  \bibfield  {author} {\bibinfo {author} {\bibfnamefont {Christian}\
  \bibnamefont {Schweizer}}, \bibinfo {author} {\bibfnamefont {Fabian}\
  \bibnamefont {Grusdt}}, \bibinfo {author} {\bibfnamefont {Moritz}\
  \bibnamefont {Berngruber}}, \bibinfo {author} {\bibfnamefont {Luca}\
  \bibnamefont {Barbiero}}, \bibinfo {author} {\bibfnamefont {Eugene}\
  \bibnamefont {Demler}}, \bibinfo {author} {\bibfnamefont {Nathan}\
  \bibnamefont {Goldman}}, \bibinfo {author} {\bibfnamefont {Immanuel}\
  \bibnamefont {Bloch}}, \ and\ \bibinfo {author} {\bibfnamefont {Monika}\
  \bibnamefont {Aidelsburger}},\ }\bibfield  {title} {\enquote {\bibinfo
  {title} {Floquet approach to $\mathbb{Z}$2 lattice gauge theories with
  ultracold atoms in optical lattices},}\ }\href {\doibase
  10.1038/s41567-019-0649-7} {\bibfield  {journal} {\bibinfo  {journal} {Nature
  Physics}\ }\textbf {\bibinfo {volume} {15}},\ \bibinfo {pages} {1168--1173}
  (\bibinfo {year} {2019})}\BibitemShut {NoStop}%
\bibitem [{\citenamefont {G{\"o}rg}\ \emph {et~al.}(2019)\citenamefont
  {G{\"o}rg}, \citenamefont {Sandholzer}, \citenamefont {Minguzzi},
  \citenamefont {Desbuquois}, \citenamefont {Messer},\ and\ \citenamefont
  {Esslinger}}]{Goerg2019}%
  \BibitemOpen
  \bibfield  {author} {\bibinfo {author} {\bibfnamefont {Frederik}\
  \bibnamefont {G{\"o}rg}}, \bibinfo {author} {\bibfnamefont {Kilian}\
  \bibnamefont {Sandholzer}}, \bibinfo {author} {\bibfnamefont {Joaqu{\'\i}n}\
  \bibnamefont {Minguzzi}}, \bibinfo {author} {\bibfnamefont {R{\'e}mi}\
  \bibnamefont {Desbuquois}}, \bibinfo {author} {\bibfnamefont {Michael}\
  \bibnamefont {Messer}}, \ and\ \bibinfo {author} {\bibfnamefont {Tilman}\
  \bibnamefont {Esslinger}},\ }\bibfield  {title} {\enquote {\bibinfo {title}
  {Realization of density-dependent peierls phases to engineer quantized gauge
  fields coupled to ultracold matter},}\ }\href {\doibase
  10.1038/s41567-019-0615-4} {\bibfield  {journal} {\bibinfo  {journal} {Nature
  Physics}\ }\textbf {\bibinfo {volume} {15}},\ \bibinfo {pages} {1161--1167}
  (\bibinfo {year} {2019})}\BibitemShut {NoStop}%
\bibitem [{\citenamefont {Mil}\ \emph {et~al.}(2020)\citenamefont {Mil},
  \citenamefont {Zache}, \citenamefont {Hegde}, \citenamefont {Xia},
  \citenamefont {Bhatt}, \citenamefont {Oberthaler}, \citenamefont {Hauke},
  \citenamefont {Berges},\ and\ \citenamefont {Jendrzejewski}}]{Mil2020}%
  \BibitemOpen
  \bibfield  {author} {\bibinfo {author} {\bibfnamefont {Alexander}\
  \bibnamefont {Mil}}, \bibinfo {author} {\bibfnamefont {Torsten~V.}\
  \bibnamefont {Zache}}, \bibinfo {author} {\bibfnamefont {Apoorva}\
  \bibnamefont {Hegde}}, \bibinfo {author} {\bibfnamefont {Andy}\ \bibnamefont
  {Xia}}, \bibinfo {author} {\bibfnamefont {Rohit~P.}\ \bibnamefont {Bhatt}},
  \bibinfo {author} {\bibfnamefont {Markus~K.}\ \bibnamefont {Oberthaler}},
  \bibinfo {author} {\bibfnamefont {Philipp}\ \bibnamefont {Hauke}}, \bibinfo
  {author} {\bibfnamefont {J{\"u}rgen}\ \bibnamefont {Berges}}, \ and\ \bibinfo
  {author} {\bibfnamefont {Fred}\ \bibnamefont {Jendrzejewski}},\ }\bibfield
  {title} {\enquote {\bibinfo {title} {A scalable realization of local u(1)
  gauge invariance in cold atomic mixtures},}\ }\href {\doibase
  10.1126/science.aaz5312} {\bibfield  {journal} {\bibinfo  {journal}
  {Science}\ }\textbf {\bibinfo {volume} {367}},\ \bibinfo {pages} {1128--1130}
  (\bibinfo {year} {2020})}\BibitemShut {NoStop}%
\bibitem [{\citenamefont {Yang}\ \emph
  {et~al.}(2020{\natexlab{a}})\citenamefont {Yang}, \citenamefont {Sun},
  \citenamefont {Ott}, \citenamefont {Wang}, \citenamefont {Zache},
  \citenamefont {Halimeh}, \citenamefont {Yuan}, \citenamefont {Hauke},\ and\
  \citenamefont {Pan}}]{Yang2020}%
  \BibitemOpen
  \bibfield  {author} {\bibinfo {author} {\bibfnamefont {Bing}\ \bibnamefont
  {Yang}}, \bibinfo {author} {\bibfnamefont {Hui}\ \bibnamefont {Sun}},
  \bibinfo {author} {\bibfnamefont {Robert}\ \bibnamefont {Ott}}, \bibinfo
  {author} {\bibfnamefont {Han-Yi}\ \bibnamefont {Wang}}, \bibinfo {author}
  {\bibfnamefont {Torsten~V.}\ \bibnamefont {Zache}}, \bibinfo {author}
  {\bibfnamefont {Jad~C.}\ \bibnamefont {Halimeh}}, \bibinfo {author}
  {\bibfnamefont {Zhen-Sheng}\ \bibnamefont {Yuan}}, \bibinfo {author}
  {\bibfnamefont {Philipp}\ \bibnamefont {Hauke}}, \ and\ \bibinfo {author}
  {\bibfnamefont {Jian-Wei}\ \bibnamefont {Pan}},\ }\bibfield  {title}
  {\enquote {\bibinfo {title} {Observation of gauge invariance in a 71-site
  bose--hubbard quantum simulator},}\ }\href {\doibase
  10.1038/s41586-020-2910-8} {\bibfield  {journal} {\bibinfo  {journal}
  {Nature}\ }\textbf {\bibinfo {volume} {587}},\ \bibinfo {pages} {392--396}
  (\bibinfo {year} {2020}{\natexlab{a}})}\BibitemShut {NoStop}%
\bibitem [{\citenamefont {Zhou}\ \emph {et~al.}(2021)\citenamefont {Zhou},
  \citenamefont {Su}, \citenamefont {Halimeh}, \citenamefont {Ott},
  \citenamefont {Sun}, \citenamefont {Hauke}, \citenamefont {Yang},
  \citenamefont {Yuan}, \citenamefont {Berges},\ and\ \citenamefont
  {Pan}}]{Zhou2021}%
  \BibitemOpen
  \bibfield  {author} {\bibinfo {author} {\bibfnamefont {Zhao-Yu}\ \bibnamefont
  {Zhou}}, \bibinfo {author} {\bibfnamefont {Guo-Xian}\ \bibnamefont {Su}},
  \bibinfo {author} {\bibfnamefont {Jad~C.}\ \bibnamefont {Halimeh}}, \bibinfo
  {author} {\bibfnamefont {Robert}\ \bibnamefont {Ott}}, \bibinfo {author}
  {\bibfnamefont {Hui}\ \bibnamefont {Sun}}, \bibinfo {author} {\bibfnamefont
  {Philipp}\ \bibnamefont {Hauke}}, \bibinfo {author} {\bibfnamefont {Bing}\
  \bibnamefont {Yang}}, \bibinfo {author} {\bibfnamefont {Zhen-Sheng}\
  \bibnamefont {Yuan}}, \bibinfo {author} {\bibfnamefont {Jürgen}\
  \bibnamefont {Berges}}, \ and\ \bibinfo {author} {\bibfnamefont {Jian-Wei}\
  \bibnamefont {Pan}},\ }\bibfield  {title} {\enquote {\bibinfo {title}
  {Thermalization dynamics of a gauge theory on a quantum simulator},}\
  }\href@noop {} {\  (\bibinfo {year} {2021})},\ \Eprint
  {http://arxiv.org/abs/2107.13563} {arXiv:2107.13563 [cond-mat.quant-gas]}
  \BibitemShut {NoStop}%
\bibitem [{\citenamefont {Wang}\ \emph {et~al.}(2021)\citenamefont {Wang},
  \citenamefont {Ge}, \citenamefont {Xiang}, \citenamefont {Song},
  \citenamefont {Huang}, \citenamefont {Song}, \citenamefont {Guo},
  \citenamefont {Su}, \citenamefont {Xu}, \citenamefont {Zheng},\ and\
  \citenamefont {Fan}}]{Wang2021}%
  \BibitemOpen
  \bibfield  {author} {\bibinfo {author} {\bibfnamefont {Zhan}\ \bibnamefont
  {Wang}}, \bibinfo {author} {\bibfnamefont {Zi-Yong}\ \bibnamefont {Ge}},
  \bibinfo {author} {\bibfnamefont {Zhongcheng}\ \bibnamefont {Xiang}},
  \bibinfo {author} {\bibfnamefont {Xiaohui}\ \bibnamefont {Song}}, \bibinfo
  {author} {\bibfnamefont {Rui-Zhen}\ \bibnamefont {Huang}}, \bibinfo {author}
  {\bibfnamefont {Pengtao}\ \bibnamefont {Song}}, \bibinfo {author}
  {\bibfnamefont {Xue-Yi}\ \bibnamefont {Guo}}, \bibinfo {author}
  {\bibfnamefont {Luhong}\ \bibnamefont {Su}}, \bibinfo {author} {\bibfnamefont
  {Kai}\ \bibnamefont {Xu}}, \bibinfo {author} {\bibfnamefont {Dongning}\
  \bibnamefont {Zheng}}, \ and\ \bibinfo {author} {\bibfnamefont {Heng}\
  \bibnamefont {Fan}},\ }\bibfield  {title} {\enquote {\bibinfo {title}
  {Observation of emergent $\mathbb{Z}_2$ gauge invariance in a superconducting
  circuit},}\ }\href {\doibase 10.48550/ARXIV.2111.05048} {\  (\bibinfo {year}
  {2021}),\ 10.48550/ARXIV.2111.05048}\BibitemShut {NoStop}%
\bibitem [{\citenamefont {Mildenberger}\ \emph {et~al.}(2022)\citenamefont
  {Mildenberger}, \citenamefont {Mruczkiewicz}, \citenamefont {Halimeh},
  \citenamefont {Jiang},\ and\ \citenamefont {Hauke}}]{Mildenberger2022}%
  \BibitemOpen
  \bibfield  {author} {\bibinfo {author} {\bibfnamefont {Julius}\ \bibnamefont
  {Mildenberger}}, \bibinfo {author} {\bibfnamefont {Wojciech}\ \bibnamefont
  {Mruczkiewicz}}, \bibinfo {author} {\bibfnamefont {Jad~C.}\ \bibnamefont
  {Halimeh}}, \bibinfo {author} {\bibfnamefont {Zhang}\ \bibnamefont {Jiang}},
  \ and\ \bibinfo {author} {\bibfnamefont {Philipp}\ \bibnamefont {Hauke}},\
  }\bibfield  {title} {\enquote {\bibinfo {title} {Probing confinement in a
  $\mathbb{Z}_2$ lattice gauge theory on a quantum computer},}\ }\href
  {\doibase 10.48550/ARXIV.2203.08905} {\  (\bibinfo {year} {2022}),\
  10.48550/ARXIV.2203.08905}\BibitemShut {NoStop}%
\bibitem [{\citenamefont {Foerster}\ \emph {et~al.}(1980)\citenamefont
  {Foerster}, \citenamefont {Nielsen},\ and\ \citenamefont
  {Ninomiya}}]{Foerster1980}%
  \BibitemOpen
  \bibfield  {author} {\bibinfo {author} {\bibfnamefont {D.}~\bibnamefont
  {Foerster}}, \bibinfo {author} {\bibfnamefont {H.B.}\ \bibnamefont
  {Nielsen}}, \ and\ \bibinfo {author} {\bibfnamefont {M.}~\bibnamefont
  {Ninomiya}},\ }\bibfield  {title} {\enquote {\bibinfo {title} {Dynamical
  stability of local gauge symmetry creation of light from chaos},}\ }\href
  {\doibase https://doi.org/10.1016/0370-2693(80)90842-4} {\bibfield  {journal}
  {\bibinfo  {journal} {Physics Letters B}\ }\textbf {\bibinfo {volume} {94}},\
  \bibinfo {pages} {135 -- 140} (\bibinfo {year} {1980})}\BibitemShut {NoStop}%
\bibitem [{\citenamefont {Poppitz}\ and\ \citenamefont
  {Shang}(2008)}]{Poppitz2008}%
  \BibitemOpen
  \bibfield  {author} {\bibinfo {author} {\bibfnamefont {E.}~\bibnamefont
  {Poppitz}}\ and\ \bibinfo {author} {\bibfnamefont {Y.}~\bibnamefont
  {Shang}},\ }\bibfield  {title} {\enquote {\bibinfo {title} {"light from
  chaos" in two dimensions},}\ }\href {\doibase 10.1142/S0217751X08041281}
  {\bibfield  {journal} {\bibinfo  {journal} {International Journal of Modern
  Physics A}\ }\textbf {\bibinfo {volume} {23}},\ \bibinfo {pages} {4545--4556}
  (\bibinfo {year} {2008})},\ \Eprint
  {http://arxiv.org/abs/https://doi.org/10.1142/S0217751X08041281}
  {https://doi.org/10.1142/S0217751X08041281} \BibitemShut {NoStop}%
\bibitem [{\citenamefont {Wetterich}(2017)}]{Wetterich2017}%
  \BibitemOpen
  \bibfield  {author} {\bibinfo {author} {\bibfnamefont {C.}~\bibnamefont
  {Wetterich}},\ }\bibfield  {title} {\enquote {\bibinfo {title} {Gauge
  symmetry from decoupling},}\ }\href {\doibase
  https://doi.org/10.1016/j.nuclphysb.2016.12.008} {\bibfield  {journal}
  {\bibinfo  {journal} {Nuclear Physics B}\ }\textbf {\bibinfo {volume}
  {915}},\ \bibinfo {pages} {135 -- 167} (\bibinfo {year} {2017})}\BibitemShut
  {NoStop}%
\bibitem [{\citenamefont {Witten}(2018)}]{Witten2018}%
  \BibitemOpen
  \bibfield  {author} {\bibinfo {author} {\bibfnamefont {Edward}\ \bibnamefont
  {Witten}},\ }\bibfield  {title} {\enquote {\bibinfo {title} {Symmetry and
  emergence},}\ }\href {\doibase 10.1038/nphys4348} {\bibfield  {journal}
  {\bibinfo  {journal} {Nature Physics}\ }\textbf {\bibinfo {volume} {14}},\
  \bibinfo {pages} {116--119} (\bibinfo {year} {2018})}\BibitemShut {NoStop}%
\bibitem [{\citenamefont {Barcel\'o}\ \emph {et~al.}(2021)\citenamefont
  {Barcel\'o}, \citenamefont {Carballo-Rubio}, \citenamefont {Garay},\ and\
  \citenamefont {Garc\'{\i}a-Moreno}}]{Barcelo2021}%
  \BibitemOpen
  \bibfield  {author} {\bibinfo {author} {\bibfnamefont {Carlos}\ \bibnamefont
  {Barcel\'o}}, \bibinfo {author} {\bibfnamefont {Ra\'ul}\ \bibnamefont
  {Carballo-Rubio}}, \bibinfo {author} {\bibfnamefont {Luis~J.}\ \bibnamefont
  {Garay}}, \ and\ \bibinfo {author} {\bibfnamefont {Gerardo}\ \bibnamefont
  {Garc\'{\i}a-Moreno}},\ }\bibfield  {title} {\enquote {\bibinfo {title}
  {Emergent gauge symmetries: Yang-mills theory},}\ }\href {\doibase
  10.1103/PhysRevD.104.025017} {\bibfield  {journal} {\bibinfo  {journal}
  {Phys. Rev. D}\ }\textbf {\bibinfo {volume} {104}},\ \bibinfo {pages}
  {025017} (\bibinfo {year} {2021})}\BibitemShut {NoStop}%
\bibitem [{\citenamefont {Bass}(2022)}]{Bass2022}%
  \BibitemOpen
  \bibfield  {author} {\bibinfo {author} {\bibfnamefont {Steven~D.}\
  \bibnamefont {Bass}},\ }\bibfield  {title} {\enquote {\bibinfo {title}
  {Emergent gauge symmetries: making symmetry as well as breaking it},}\ }\href
  {\doibase 10.1098/rsta.2021.0059} {\bibfield  {journal} {\bibinfo  {journal}
  {Philosophical Transactions of the Royal Society A: Mathematical, Physical
  and Engineering Sciences}\ }\textbf {\bibinfo {volume} {380}},\ \bibinfo
  {pages} {20210059} (\bibinfo {year} {2022})},\ \Eprint
  {http://arxiv.org/abs/https://royalsocietypublishing.org/doi/pdf/10.1098/rsta.2021.0059}
  {https://royalsocietypublishing.org/doi/pdf/10.1098/rsta.2021.0059}
  \BibitemShut {NoStop}%
\bibitem [{\citenamefont {Stannigel}\ \emph {et~al.}(2014)\citenamefont
  {Stannigel}, \citenamefont {Hauke}, \citenamefont {Marcos}, \citenamefont
  {Hafezi}, \citenamefont {Diehl}, \citenamefont {Dalmonte},\ and\
  \citenamefont {Zoller}}]{Stannigel2014}%
  \BibitemOpen
  \bibfield  {author} {\bibinfo {author} {\bibfnamefont {K.}~\bibnamefont
  {Stannigel}}, \bibinfo {author} {\bibfnamefont {P.}~\bibnamefont {Hauke}},
  \bibinfo {author} {\bibfnamefont {D.}~\bibnamefont {Marcos}}, \bibinfo
  {author} {\bibfnamefont {M.}~\bibnamefont {Hafezi}}, \bibinfo {author}
  {\bibfnamefont {S.}~\bibnamefont {Diehl}}, \bibinfo {author} {\bibfnamefont
  {M.}~\bibnamefont {Dalmonte}}, \ and\ \bibinfo {author} {\bibfnamefont
  {P.}~\bibnamefont {Zoller}},\ }\bibfield  {title} {\enquote {\bibinfo {title}
  {Constrained dynamics via the zeno effect in quantum simulation: Implementing
  non-abelian lattice gauge theories with cold atoms},}\ }\href {\doibase
  10.1103/PhysRevLett.112.120406} {\bibfield  {journal} {\bibinfo  {journal}
  {Phys. Rev. Lett.}\ }\textbf {\bibinfo {volume} {112}},\ \bibinfo {pages}
  {120406} (\bibinfo {year} {2014})}\BibitemShut {NoStop}%
\bibitem [{\citenamefont {K\"uhn}\ \emph {et~al.}(2014)\citenamefont {K\"uhn},
  \citenamefont {Cirac},\ and\ \citenamefont {Ba\~nuls}}]{Kuehn2014}%
  \BibitemOpen
  \bibfield  {author} {\bibinfo {author} {\bibfnamefont {Stefan}\ \bibnamefont
  {K\"uhn}}, \bibinfo {author} {\bibfnamefont {J.~Ignacio}\ \bibnamefont
  {Cirac}}, \ and\ \bibinfo {author} {\bibfnamefont {Mari-Carmen}\ \bibnamefont
  {Ba\~nuls}},\ }\bibfield  {title} {\enquote {\bibinfo {title} {Quantum
  simulation of the schwinger model: A study of feasibility},}\ }\href
  {\doibase 10.1103/PhysRevA.90.042305} {\bibfield  {journal} {\bibinfo
  {journal} {Phys. Rev. A}\ }\textbf {\bibinfo {volume} {90}},\ \bibinfo
  {pages} {042305} (\bibinfo {year} {2014})}\BibitemShut {NoStop}%
\bibitem [{\citenamefont {Kuno}\ \emph {et~al.}(2015)\citenamefont {Kuno},
  \citenamefont {Kasamatsu}, \citenamefont {Takahashi}, \citenamefont
  {Ichinose},\ and\ \citenamefont {Matsui}}]{Kuno2015}%
  \BibitemOpen
  \bibfield  {author} {\bibinfo {author} {\bibfnamefont {Yoshihito}\
  \bibnamefont {Kuno}}, \bibinfo {author} {\bibfnamefont {Kenichi}\
  \bibnamefont {Kasamatsu}}, \bibinfo {author} {\bibfnamefont {Yoshiro}\
  \bibnamefont {Takahashi}}, \bibinfo {author} {\bibfnamefont {Ikuo}\
  \bibnamefont {Ichinose}}, \ and\ \bibinfo {author} {\bibfnamefont {Tetsuo}\
  \bibnamefont {Matsui}},\ }\bibfield  {title} {\enquote {\bibinfo {title}
  {Real-time dynamics and proposal for feasible experiments of lattice
  gauge{\textendash}higgs model simulated by cold atoms},}\ }\href {\doibase
  10.1088/1367-2630/17/6/063005} {\bibfield  {journal} {\bibinfo  {journal}
  {New Journal of Physics}\ }\textbf {\bibinfo {volume} {17}},\ \bibinfo
  {pages} {063005} (\bibinfo {year} {2015})}\BibitemShut {NoStop}%
\bibitem [{\citenamefont {Yang}\ \emph {et~al.}(2016)\citenamefont {Yang},
  \citenamefont {Giri}, \citenamefont {Johanning}, \citenamefont {Wunderlich},
  \citenamefont {Zoller},\ and\ \citenamefont {Hauke}}]{Yang2016}%
  \BibitemOpen
  \bibfield  {author} {\bibinfo {author} {\bibfnamefont {Dayou}\ \bibnamefont
  {Yang}}, \bibinfo {author} {\bibfnamefont {Gouri~Shankar}\ \bibnamefont
  {Giri}}, \bibinfo {author} {\bibfnamefont {Michael}\ \bibnamefont
  {Johanning}}, \bibinfo {author} {\bibfnamefont {Christof}\ \bibnamefont
  {Wunderlich}}, \bibinfo {author} {\bibfnamefont {Peter}\ \bibnamefont
  {Zoller}}, \ and\ \bibinfo {author} {\bibfnamefont {Philipp}\ \bibnamefont
  {Hauke}},\ }\bibfield  {title} {\enquote {\bibinfo {title} {Analog quantum
  simulation of $(1+1)$-dimensional lattice qed with trapped ions},}\ }\href
  {\doibase 10.1103/PhysRevA.94.052321} {\bibfield  {journal} {\bibinfo
  {journal} {Phys. Rev. A}\ }\textbf {\bibinfo {volume} {94}},\ \bibinfo
  {pages} {052321} (\bibinfo {year} {2016})}\BibitemShut {NoStop}%
\bibitem [{\citenamefont {Kuno}\ \emph {et~al.}(2017)\citenamefont {Kuno},
  \citenamefont {Sakane}, \citenamefont {Kasamatsu}, \citenamefont {Ichinose},\
  and\ \citenamefont {Matsui}}]{Kuno2017}%
  \BibitemOpen
  \bibfield  {author} {\bibinfo {author} {\bibfnamefont {Yoshihito}\
  \bibnamefont {Kuno}}, \bibinfo {author} {\bibfnamefont {Shinya}\ \bibnamefont
  {Sakane}}, \bibinfo {author} {\bibfnamefont {Kenichi}\ \bibnamefont
  {Kasamatsu}}, \bibinfo {author} {\bibfnamefont {Ikuo}\ \bibnamefont
  {Ichinose}}, \ and\ \bibinfo {author} {\bibfnamefont {Tetsuo}\ \bibnamefont
  {Matsui}},\ }\bibfield  {title} {\enquote {\bibinfo {title} {Quantum
  simulation of ($1+1$)-dimensional u(1) gauge-higgs model on a lattice by cold
  bose gases},}\ }\href {\doibase 10.1103/PhysRevD.95.094507} {\bibfield
  {journal} {\bibinfo  {journal} {Phys. Rev. D}\ }\textbf {\bibinfo {volume}
  {95}},\ \bibinfo {pages} {094507} (\bibinfo {year} {2017})}\BibitemShut
  {NoStop}%
\bibitem [{\citenamefont {Dehkharghani}\ \emph {et~al.}(2017)\citenamefont
  {Dehkharghani}, \citenamefont {Rico}, \citenamefont {Zinner},\ and\
  \citenamefont {Negretti}}]{Negretti2017}%
  \BibitemOpen
  \bibfield  {author} {\bibinfo {author} {\bibfnamefont {A.~S.}\ \bibnamefont
  {Dehkharghani}}, \bibinfo {author} {\bibfnamefont {E.}~\bibnamefont {Rico}},
  \bibinfo {author} {\bibfnamefont {N.~T.}\ \bibnamefont {Zinner}}, \ and\
  \bibinfo {author} {\bibfnamefont {A.}~\bibnamefont {Negretti}},\ }\bibfield
  {title} {\enquote {\bibinfo {title} {Quantum simulation of abelian lattice
  gauge theories via state-dependent hopping},}\ }\href {\doibase
  10.1103/PhysRevA.96.043611} {\bibfield  {journal} {\bibinfo  {journal} {Phys.
  Rev. A}\ }\textbf {\bibinfo {volume} {96}},\ \bibinfo {pages} {043611}
  (\bibinfo {year} {2017})}\BibitemShut {NoStop}%
\bibitem [{\citenamefont {Dutta}\ \emph {et~al.}(2017)\citenamefont {Dutta},
  \citenamefont {Tagliacozzo}, \citenamefont {Lewenstein},\ and\ \citenamefont
  {Zakrzewski}}]{Dutta2017}%
  \BibitemOpen
  \bibfield  {author} {\bibinfo {author} {\bibfnamefont {Omjyoti}\ \bibnamefont
  {Dutta}}, \bibinfo {author} {\bibfnamefont {Luca}\ \bibnamefont
  {Tagliacozzo}}, \bibinfo {author} {\bibfnamefont {Maciej}\ \bibnamefont
  {Lewenstein}}, \ and\ \bibinfo {author} {\bibfnamefont {Jakub}\ \bibnamefont
  {Zakrzewski}},\ }\bibfield  {title} {\enquote {\bibinfo {title} {Toolbox for
  abelian lattice gauge theories with synthetic matter},}\ }\href {\doibase
  10.1103/PhysRevA.95.053608} {\bibfield  {journal} {\bibinfo  {journal} {Phys.
  Rev. A}\ }\textbf {\bibinfo {volume} {95}},\ \bibinfo {pages} {053608}
  (\bibinfo {year} {2017})}\BibitemShut {NoStop}%
\bibitem [{\citenamefont {Pinto~Barros}\ \emph {et~al.}(2020)\citenamefont
  {Pinto~Barros}, \citenamefont {Burrello},\ and\ \citenamefont
  {Trombettoni}}]{Barros2019}%
  \BibitemOpen
  \bibfield  {author} {\bibinfo {author} {\bibfnamefont {Jo{\~a}o~C.}\
  \bibnamefont {Pinto~Barros}}, \bibinfo {author} {\bibfnamefont {Michele}\
  \bibnamefont {Burrello}}, \ and\ \bibinfo {author} {\bibfnamefont {Andrea}\
  \bibnamefont {Trombettoni}},\ }\bibfield  {title} {\enquote {\bibinfo {title}
  {Gauge theories with ultracold atoms},}\ }in\ \href@noop {} {\emph {\bibinfo
  {booktitle} {Strongly Coupled Field Theories for Condensed Matter and Quantum
  Information Theory}}},\ \bibinfo {editor} {edited by\ \bibinfo {editor}
  {\bibfnamefont {Alvaro}\ \bibnamefont {Ferraz}}, \bibinfo {editor}
  {\bibfnamefont {Kumar~S.}\ \bibnamefont {Gupta}}, \bibinfo {editor}
  {\bibfnamefont {Gordon~Walter}\ \bibnamefont {Semenoff}}, \ and\ \bibinfo
  {editor} {\bibfnamefont {Pasquale}\ \bibnamefont {Sodano}}}\ (\bibinfo
  {publisher} {Springer International Publishing},\ \bibinfo {address} {Cham},\
  \bibinfo {year} {2020})\ pp.\ \bibinfo {pages} {217--245}\BibitemShut
  {NoStop}%
\bibitem [{\citenamefont {Halimeh}\ and\ \citenamefont
  {Hauke}(2020{\natexlab{a}})}]{Halimeh2020a}%
  \BibitemOpen
  \bibfield  {author} {\bibinfo {author} {\bibfnamefont {Jad~C.}\ \bibnamefont
  {Halimeh}}\ and\ \bibinfo {author} {\bibfnamefont {Philipp}\ \bibnamefont
  {Hauke}},\ }\bibfield  {title} {\enquote {\bibinfo {title} {Reliability of
  lattice gauge theories},}\ }\href {\doibase 10.1103/PhysRevLett.125.030503}
  {\bibfield  {journal} {\bibinfo  {journal} {Phys. Rev. Lett.}\ }\textbf
  {\bibinfo {volume} {125}},\ \bibinfo {pages} {030503} (\bibinfo {year}
  {2020}{\natexlab{a}})}\BibitemShut {NoStop}%
\bibitem [{\citenamefont {Kasper}\ \emph {et~al.}(2021)\citenamefont {Kasper},
  \citenamefont {Zache}, \citenamefont {Jendrzejewski}, \citenamefont
  {Lewenstein},\ and\ \citenamefont {Zohar}}]{Kasper2021nonabelian}%
  \BibitemOpen
  \bibfield  {author} {\bibinfo {author} {\bibfnamefont {Valentin}\
  \bibnamefont {Kasper}}, \bibinfo {author} {\bibfnamefont {Torsten~V.}\
  \bibnamefont {Zache}}, \bibinfo {author} {\bibfnamefont {Fred}\ \bibnamefont
  {Jendrzejewski}}, \bibinfo {author} {\bibfnamefont {Maciej}\ \bibnamefont
  {Lewenstein}}, \ and\ \bibinfo {author} {\bibfnamefont {Erez}\ \bibnamefont
  {Zohar}},\ }\bibfield  {title} {\enquote {\bibinfo {title} {Non-abelian gauge
  invariance from dynamical decoupling},}\ }\href@noop {} {\  (\bibinfo {year}
  {2021})},\ \Eprint {http://arxiv.org/abs/2012.08620} {arXiv:2012.08620
  [quant-ph]} \BibitemShut {NoStop}%
\bibitem [{\citenamefont {Lamm}\ \emph {et~al.}(2020)\citenamefont {Lamm},
  \citenamefont {Lawrence},\ and\ \citenamefont {Yamauchi}}]{Lamm2020}%
  \BibitemOpen
  \bibfield  {author} {\bibinfo {author} {\bibfnamefont {Henry}\ \bibnamefont
  {Lamm}}, \bibinfo {author} {\bibfnamefont {Scott}\ \bibnamefont {Lawrence}},
  \ and\ \bibinfo {author} {\bibfnamefont {Yukari}\ \bibnamefont {Yamauchi}},\
  }\bibfield  {title} {\enquote {\bibinfo {title} {Suppressing coherent gauge
  drift in quantum simulations},}\ }\href {https://arxiv.org/abs/2005.12688} {\
   (\bibinfo {year} {2020})},\ \Eprint {http://arxiv.org/abs/2005.12688}
  {arXiv:2005.12688 [quant-ph]} \BibitemShut {NoStop}%
\bibitem [{\citenamefont {Chubb}\ and\ \citenamefont
  {Flammia}(2017)}]{Chubb2017}%
  \BibitemOpen
  \bibfield  {author} {\bibinfo {author} {\bibfnamefont {Christopher~T.}\
  \bibnamefont {Chubb}}\ and\ \bibinfo {author} {\bibfnamefont {Steven~T.}\
  \bibnamefont {Flammia}},\ }\bibfield  {title} {\enquote {\bibinfo {title}
  {Approximate symmetries of hamiltonians},}\ }\href {\doibase
  10.1063/1.4998921} {\bibfield  {journal} {\bibinfo  {journal} {Journal of
  Mathematical Physics}\ }\textbf {\bibinfo {volume} {58}},\ \bibinfo {pages}
  {082202} (\bibinfo {year} {2017})},\ \Eprint
  {http://arxiv.org/abs/https://doi.org/10.1063/1.4998921}
  {https://doi.org/10.1063/1.4998921} \BibitemShut {NoStop}%
\bibitem [{\citenamefont {Tran}\ \emph {et~al.}(2021)\citenamefont {Tran},
  \citenamefont {Su}, \citenamefont {Carney},\ and\ \citenamefont
  {Taylor}}]{Tran2021}%
  \BibitemOpen
  \bibfield  {author} {\bibinfo {author} {\bibfnamefont {Minh~C.}\ \bibnamefont
  {Tran}}, \bibinfo {author} {\bibfnamefont {Yuan}\ \bibnamefont {Su}},
  \bibinfo {author} {\bibfnamefont {Daniel}\ \bibnamefont {Carney}}, \ and\
  \bibinfo {author} {\bibfnamefont {Jacob~M.}\ \bibnamefont {Taylor}},\
  }\bibfield  {title} {\enquote {\bibinfo {title} {Faster digital quantum
  simulation by symmetry protection},}\ }\href {\doibase
  10.1103/PRXQuantum.2.010323} {\bibfield  {journal} {\bibinfo  {journal} {PRX
  Quantum}\ }\textbf {\bibinfo {volume} {2}},\ \bibinfo {pages} {010323}
  (\bibinfo {year} {2021})}\BibitemShut {NoStop}%
\bibitem [{\citenamefont {Gonz{\'a}lez-Cuadra}\ \emph
  {et~al.}(2019)\citenamefont {Gonz{\'a}lez-Cuadra}, \citenamefont {Bermudez},
  \citenamefont {Grzybowski}, \citenamefont {Lewenstein},\ and\ \citenamefont
  {Dauphin}}]{GonzalezCuadra2019}%
  \BibitemOpen
  \bibfield  {author} {\bibinfo {author} {\bibfnamefont {Daniel}\ \bibnamefont
  {Gonz{\'a}lez-Cuadra}}, \bibinfo {author} {\bibfnamefont {Alejandro}\
  \bibnamefont {Bermudez}}, \bibinfo {author} {\bibfnamefont
  {Przemys{\l}aw~R.}\ \bibnamefont {Grzybowski}}, \bibinfo {author}
  {\bibfnamefont {Maciej}\ \bibnamefont {Lewenstein}}, \ and\ \bibinfo {author}
  {\bibfnamefont {Alexandre}\ \bibnamefont {Dauphin}},\ }\bibfield  {title}
  {\enquote {\bibinfo {title} {Intertwined topological phases induced by
  emergent symmetry protection},}\ }\href {\doibase 10.1038/s41467-019-10796-8}
  {\bibfield  {journal} {\bibinfo  {journal} {Nature Communications}\ }\textbf
  {\bibinfo {volume} {10}},\ \bibinfo {pages} {2694} (\bibinfo {year}
  {2019})}\BibitemShut {NoStop}%
\bibitem [{\citenamefont {Gonz\'alez-Cuadra}\ \emph {et~al.}(2020)\citenamefont
  {Gonz\'alez-Cuadra}, \citenamefont {Dauphin}, \citenamefont {Grzybowski},
  \citenamefont {Lewenstein},\ and\ \citenamefont
  {Bermudez}}]{GonzalezCuadra2020}%
  \BibitemOpen
  \bibfield  {author} {\bibinfo {author} {\bibfnamefont {D.}~\bibnamefont
  {Gonz\'alez-Cuadra}}, \bibinfo {author} {\bibfnamefont {A.}~\bibnamefont
  {Dauphin}}, \bibinfo {author} {\bibfnamefont {P.~R.}\ \bibnamefont
  {Grzybowski}}, \bibinfo {author} {\bibfnamefont {M.}~\bibnamefont
  {Lewenstein}}, \ and\ \bibinfo {author} {\bibfnamefont {A.}~\bibnamefont
  {Bermudez}},\ }\bibfield  {title} {\enquote {\bibinfo {title} {Dynamical
  solitons and boson fractionalization in cold-atom topological insulators},}\
  }\href {\doibase 10.1103/PhysRevLett.125.265301} {\bibfield  {journal}
  {\bibinfo  {journal} {Phys. Rev. Lett.}\ }\textbf {\bibinfo {volume} {125}},\
  \bibinfo {pages} {265301} (\bibinfo {year} {2020})}\BibitemShut {NoStop}%
\bibitem [{\citenamefont {Damme}\ \emph {et~al.}(2020)\citenamefont {Damme},
  \citenamefont {Halimeh},\ and\ \citenamefont {Hauke}}]{VanDamme2020}%
  \BibitemOpen
  \bibfield  {author} {\bibinfo {author} {\bibfnamefont {Maarten~Van}\
  \bibnamefont {Damme}}, \bibinfo {author} {\bibfnamefont {Jad~C.}\
  \bibnamefont {Halimeh}}, \ and\ \bibinfo {author} {\bibfnamefont {Philipp}\
  \bibnamefont {Hauke}},\ }\bibfield  {title} {\enquote {\bibinfo {title}
  {Gauge-symmetry violation quantum phase transition in lattice gauge
  theories},}\ }\href@noop {} {\  (\bibinfo {year} {2020})},\ \Eprint
  {http://arxiv.org/abs/2010.07338} {arXiv:2010.07338 [cond-mat.quant-gas]}
  \BibitemShut {NoStop}%
\bibitem [{\citenamefont {Borla}\ \emph {et~al.}(2021)\citenamefont {Borla},
  \citenamefont {Verresen}, \citenamefont {Shah},\ and\ \citenamefont
  {Moroz}}]{Borla2020}%
  \BibitemOpen
  \bibfield  {author} {\bibinfo {author} {\bibfnamefont {Umberto}\ \bibnamefont
  {Borla}}, \bibinfo {author} {\bibfnamefont {Ruben}\ \bibnamefont {Verresen}},
  \bibinfo {author} {\bibfnamefont {Jeet}\ \bibnamefont {Shah}}, \ and\
  \bibinfo {author} {\bibfnamefont {Sergej}\ \bibnamefont {Moroz}},\ }\bibfield
   {title} {\enquote {\bibinfo {title} {{Gauging the Kitaev chain}},}\ }\href
  {\doibase 10.21468/SciPostPhys.10.6.148} {\bibfield  {journal} {\bibinfo
  {journal} {SciPost Phys.}\ }\textbf {\bibinfo {volume} {10}},\ \bibinfo
  {pages} {148} (\bibinfo {year} {2021})}\BibitemShut {NoStop}%
\bibitem [{\citenamefont {Halimeh}\ \emph
  {et~al.}(2021{\natexlab{a}})\citenamefont {Halimeh}, \citenamefont {Homeier},
  \citenamefont {Zhao}, \citenamefont {Bohrdt}, \citenamefont {Grusdt},
  \citenamefont {Hauke},\ and\ \citenamefont {Knolle}}]{Halimeh2021enhancing}%
  \BibitemOpen
  \bibfield  {author} {\bibinfo {author} {\bibfnamefont {Jad~C.}\ \bibnamefont
  {Halimeh}}, \bibinfo {author} {\bibfnamefont {Lukas}\ \bibnamefont
  {Homeier}}, \bibinfo {author} {\bibfnamefont {Hongzheng}\ \bibnamefont
  {Zhao}}, \bibinfo {author} {\bibfnamefont {Annabelle}\ \bibnamefont
  {Bohrdt}}, \bibinfo {author} {\bibfnamefont {Fabian}\ \bibnamefont {Grusdt}},
  \bibinfo {author} {\bibfnamefont {Philipp}\ \bibnamefont {Hauke}}, \ and\
  \bibinfo {author} {\bibfnamefont {Johannes}\ \bibnamefont {Knolle}},\
  }\bibfield  {title} {\enquote {\bibinfo {title} {Enhancing disorder-free
  localization through dynamically emergent local symmetries},}\ }\href@noop {}
  {\  (\bibinfo {year} {2021}{\natexlab{a}})},\ \Eprint
  {http://arxiv.org/abs/2111.08715} {arXiv:2111.08715 [cond-mat.quant-gas]}
  \BibitemShut {NoStop}%
\bibitem [{\citenamefont {Bazavov}\ \emph {et~al.}(2015)\citenamefont
  {Bazavov}, \citenamefont {Meurice}, \citenamefont {Tsai}, \citenamefont
  {Unmuth-Yockey},\ and\ \citenamefont {Zhang}}]{Bazavov2015}%
  \BibitemOpen
  \bibfield  {author} {\bibinfo {author} {\bibfnamefont {A.}~\bibnamefont
  {Bazavov}}, \bibinfo {author} {\bibfnamefont {Y.}~\bibnamefont {Meurice}},
  \bibinfo {author} {\bibfnamefont {S.-W.}\ \bibnamefont {Tsai}}, \bibinfo
  {author} {\bibfnamefont {J.}~\bibnamefont {Unmuth-Yockey}}, \ and\ \bibinfo
  {author} {\bibfnamefont {Jin}\ \bibnamefont {Zhang}},\ }\bibfield  {title}
  {\enquote {\bibinfo {title} {Gauge-invariant implementation of the
  abelian-higgs model on optical lattices},}\ }\href {\doibase
  10.1103/PhysRevD.92.076003} {\bibfield  {journal} {\bibinfo  {journal} {Phys.
  Rev. D}\ }\textbf {\bibinfo {volume} {92}},\ \bibinfo {pages} {076003}
  (\bibinfo {year} {2015})}\BibitemShut {NoStop}%
\bibitem [{\citenamefont {Heitger}(1997)}]{HeitgerPhDThesis}%
  \BibitemOpen
  \bibfield  {author} {\bibinfo {author} {\bibfnamefont {Jochen}\ \bibnamefont
  {Heitger}},\ }\href@noop {} {\enquote {\bibinfo {title} {Numerical
  simulations of gauge-higgs models on the lattice},}\ }\bibinfo {howpublished}
  {\href{https://www.uni-muenster.de/imperia/md/content/physik_tp/theses/muenster/heitger_dr.pdf}{PhD
  Thesis}} (\bibinfo {year} {1997})\BibitemShut {NoStop}%
\bibitem [{\citenamefont {Hastings}\ and\ \citenamefont
  {Wen}(2005)}]{Hastings2005}%
  \BibitemOpen
  \bibfield  {author} {\bibinfo {author} {\bibfnamefont {M.~B.}\ \bibnamefont
  {Hastings}}\ and\ \bibinfo {author} {\bibfnamefont {Xiao-Gang}\ \bibnamefont
  {Wen}},\ }\bibfield  {title} {\enquote {\bibinfo {title} {Quasiadiabatic
  continuation of quantum states: The stability of topological ground-state
  degeneracy and emergent gauge invariance},}\ }\href {\doibase
  10.1103/PhysRevB.72.045141} {\bibfield  {journal} {\bibinfo  {journal} {Phys.
  Rev. B}\ }\textbf {\bibinfo {volume} {72}},\ \bibinfo {pages} {045141}
  (\bibinfo {year} {2005})}\BibitemShut {NoStop}%
\bibitem [{\citenamefont {Sachdev}(2018)}]{Sachdev2018}%
  \BibitemOpen
  \bibfield  {author} {\bibinfo {author} {\bibfnamefont {Subir}\ \bibnamefont
  {Sachdev}},\ }\bibfield  {title} {\enquote {\bibinfo {title} {Topological
  order, emergent gauge fields, and fermi surface reconstruction},}\ }\href
  {\doibase 10.1088/1361-6633/aae110} {\bibfield  {journal} {\bibinfo
  {journal} {Reports on Progress in Physics}\ }\textbf {\bibinfo {volume}
  {82}},\ \bibinfo {pages} {014001} (\bibinfo {year} {2018})}\BibitemShut
  {NoStop}%
\bibitem [{\citenamefont {Damme}\ \emph
  {et~al.}(2021{\natexlab{a}})\citenamefont {Damme}, \citenamefont {Lang},
  \citenamefont {Hauke},\ and\ \citenamefont
  {Halimeh}}]{vandamme2021reliability}%
  \BibitemOpen
  \bibfield  {author} {\bibinfo {author} {\bibfnamefont {Maarten~Van}\
  \bibnamefont {Damme}}, \bibinfo {author} {\bibfnamefont {Haifeng}\
  \bibnamefont {Lang}}, \bibinfo {author} {\bibfnamefont {Philipp}\
  \bibnamefont {Hauke}}, \ and\ \bibinfo {author} {\bibfnamefont {Jad~C.}\
  \bibnamefont {Halimeh}},\ }\bibfield  {title} {\enquote {\bibinfo {title}
  {Reliability of lattice gauge theories in the thermodynamic limit},}\
  }\href@noop {} {\  (\bibinfo {year} {2021}{\natexlab{a}})},\ \Eprint
  {http://arxiv.org/abs/2104.07040} {arXiv:2104.07040 [cond-mat.quant-gas]}
  \BibitemShut {NoStop}%
\bibitem [{\citenamefont {Halimeh}\ \emph {et~al.}(2022)\citenamefont
  {Halimeh}, \citenamefont {Lang},\ and\ \citenamefont
  {Hauke}}]{Halimeh2021gauge}%
  \BibitemOpen
  \bibfield  {author} {\bibinfo {author} {\bibfnamefont {Jad~C.}\ \bibnamefont
  {Halimeh}}, \bibinfo {author} {\bibfnamefont {Haifeng}\ \bibnamefont {Lang}},
  \ and\ \bibinfo {author} {\bibfnamefont {Philipp}\ \bibnamefont {Hauke}},\
  }\bibfield  {title} {\enquote {\bibinfo {title} {Gauge protection in
  non-abelian lattice gauge theories},}\ }\href
  {http://iopscience.iop.org/article/10.1088/1367-2630/ac5564} {\bibfield
  {journal} {\bibinfo  {journal} {New Journal of Physics}\ } (\bibinfo {year}
  {2022})}\BibitemShut {NoStop}%
\bibitem [{\citenamefont {Halimeh}\ \emph
  {et~al.}(2021{\natexlab{b}})\citenamefont {Halimeh}, \citenamefont {Lang},
  \citenamefont {Mildenberger}, \citenamefont {Jiang},\ and\ \citenamefont
  {Hauke}}]{Halimeh2020e}%
  \BibitemOpen
  \bibfield  {author} {\bibinfo {author} {\bibfnamefont {Jad~C.}\ \bibnamefont
  {Halimeh}}, \bibinfo {author} {\bibfnamefont {Haifeng}\ \bibnamefont {Lang}},
  \bibinfo {author} {\bibfnamefont {Julius}\ \bibnamefont {Mildenberger}},
  \bibinfo {author} {\bibfnamefont {Zhang}\ \bibnamefont {Jiang}}, \ and\
  \bibinfo {author} {\bibfnamefont {Philipp}\ \bibnamefont {Hauke}},\
  }\bibfield  {title} {\enquote {\bibinfo {title} {Gauge-symmetry protection
  using single-body terms},}\ }\href {\doibase 10.1103/PRXQuantum.2.040311}
  {\bibfield  {journal} {\bibinfo  {journal} {PRX Quantum}\ }\textbf {\bibinfo
  {volume} {2}},\ \bibinfo {pages} {040311} (\bibinfo {year}
  {2021}{\natexlab{b}})}\BibitemShut {NoStop}%
\bibitem [{\citenamefont {Halimeh}\ \emph
  {et~al.}(2021{\natexlab{c}})\citenamefont {Halimeh}, \citenamefont {Homeier},
  \citenamefont {Schweizer}, \citenamefont {Aidelsburger}, \citenamefont
  {Hauke},\ and\ \citenamefont {Grusdt}}]{Halimeh2021stabilizing}%
  \BibitemOpen
  \bibfield  {author} {\bibinfo {author} {\bibfnamefont {Jad~C.}\ \bibnamefont
  {Halimeh}}, \bibinfo {author} {\bibfnamefont {Lukas}\ \bibnamefont
  {Homeier}}, \bibinfo {author} {\bibfnamefont {Christian}\ \bibnamefont
  {Schweizer}}, \bibinfo {author} {\bibfnamefont {Monika}\ \bibnamefont
  {Aidelsburger}}, \bibinfo {author} {\bibfnamefont {Philipp}\ \bibnamefont
  {Hauke}}, \ and\ \bibinfo {author} {\bibfnamefont {Fabian}\ \bibnamefont
  {Grusdt}},\ }\bibfield  {title} {\enquote {\bibinfo {title} {Stabilizing
  lattice gauge theories through simplified local pseudo generators},}\
  }\href@noop {} {\  (\bibinfo {year} {2021}{\natexlab{c}})},\ \Eprint
  {http://arxiv.org/abs/2108.02203} {arXiv:2108.02203 [cond-mat.quant-gas]}
  \BibitemShut {NoStop}%
\bibitem [{\citenamefont {Chandrasekharan}\ and\ \citenamefont
  {Wiese}(1997)}]{Chandrasekharan1997}%
  \BibitemOpen
  \bibfield  {author} {\bibinfo {author} {\bibfnamefont {S}~\bibnamefont
  {Chandrasekharan}}\ and\ \bibinfo {author} {\bibfnamefont {U.-J}\
  \bibnamefont {Wiese}},\ }\bibfield  {title} {\enquote {\bibinfo {title}
  {Quantum link models: A discrete approach to gauge theories},}\ }\href
  {\doibase https://doi.org/10.1016/S0550-3213(97)80041-7} {\bibfield
  {journal} {\bibinfo  {journal} {Nuclear Physics B}\ }\textbf {\bibinfo
  {volume} {492}},\ \bibinfo {pages} {455 -- 471} (\bibinfo {year}
  {1997})}\BibitemShut {NoStop}%
\bibitem [{\citenamefont {Wiese}(2013)}]{Wiese_review}%
  \BibitemOpen
  \bibfield  {author} {\bibinfo {author} {\bibfnamefont {U.-J.}\ \bibnamefont
  {Wiese}},\ }\bibfield  {title} {\enquote {\bibinfo {title} {Ultracold quantum
  gases and lattice systems: quantum simulation of lattice gauge theories},}\
  }\href {\doibase 10.1002/andp.201300104} {\bibfield  {journal} {\bibinfo
  {journal} {Annalen der Physik}\ }\textbf {\bibinfo {volume} {525}},\ \bibinfo
  {pages} {777--796} (\bibinfo {year} {2013})}\BibitemShut {NoStop}%
\bibitem [{\citenamefont {Kasper}\ \emph {et~al.}(2017)\citenamefont {Kasper},
  \citenamefont {Hebenstreit}, \citenamefont {Jendrzejewski}, \citenamefont
  {Oberthaler},\ and\ \citenamefont {Berges}}]{Kasper2017}%
  \BibitemOpen
  \bibfield  {author} {\bibinfo {author} {\bibfnamefont {V}~\bibnamefont
  {Kasper}}, \bibinfo {author} {\bibfnamefont {F}~\bibnamefont {Hebenstreit}},
  \bibinfo {author} {\bibfnamefont {F}~\bibnamefont {Jendrzejewski}}, \bibinfo
  {author} {\bibfnamefont {M~K}\ \bibnamefont {Oberthaler}}, \ and\ \bibinfo
  {author} {\bibfnamefont {J}~\bibnamefont {Berges}},\ }\bibfield  {title}
  {\enquote {\bibinfo {title} {Implementing quantum electrodynamics with
  ultracold atomic systems},}\ }\href {\doibase 10.1088/1367-2630/aa54e0}
  {\bibfield  {journal} {\bibinfo  {journal} {New Journal of Physics}\ }\textbf
  {\bibinfo {volume} {19}},\ \bibinfo {pages} {023030} (\bibinfo {year}
  {2017})}\BibitemShut {NoStop}%
\bibitem [{\citenamefont {Zache}\ \emph {et~al.}(2021)\citenamefont {Zache},
  \citenamefont {Damme}, \citenamefont {Halimeh}, \citenamefont {Hauke},\ and\
  \citenamefont {Banerjee}}]{Zache2021achieving}%
  \BibitemOpen
  \bibfield  {author} {\bibinfo {author} {\bibfnamefont {Torsten~V.}\
  \bibnamefont {Zache}}, \bibinfo {author} {\bibfnamefont {Maarten~Van}\
  \bibnamefont {Damme}}, \bibinfo {author} {\bibfnamefont {Jad~C.}\
  \bibnamefont {Halimeh}}, \bibinfo {author} {\bibfnamefont {Philipp}\
  \bibnamefont {Hauke}}, \ and\ \bibinfo {author} {\bibfnamefont {Debasish}\
  \bibnamefont {Banerjee}},\ }\bibfield  {title} {\enquote {\bibinfo {title}
  {Achieving the continuum limit of quantum link lattice gauge theories on
  quantum devices},}\ }\href@noop {} {\  (\bibinfo {year} {2021})},\ \Eprint
  {http://arxiv.org/abs/2104.00025} {arXiv:2104.00025 [hep-lat]} \BibitemShut
  {NoStop}%
\bibitem [{\citenamefont {Halimeh}\ \emph
  {et~al.}(2021{\natexlab{d}})\citenamefont {Halimeh}, \citenamefont {Damme},
  \citenamefont {Zache}, \citenamefont {Banerjee},\ and\ \citenamefont
  {Hauke}}]{Halimeh2021achieving}%
  \BibitemOpen
  \bibfield  {author} {\bibinfo {author} {\bibfnamefont {Jad~C.}\ \bibnamefont
  {Halimeh}}, \bibinfo {author} {\bibfnamefont {Maarten~Van}\ \bibnamefont
  {Damme}}, \bibinfo {author} {\bibfnamefont {Torsten~V.}\ \bibnamefont
  {Zache}}, \bibinfo {author} {\bibfnamefont {Debasish}\ \bibnamefont
  {Banerjee}}, \ and\ \bibinfo {author} {\bibfnamefont {Philipp}\ \bibnamefont
  {Hauke}},\ }\bibfield  {title} {\enquote {\bibinfo {title} {Achieving the
  quantum field theory limit in far-from-equilibrium quantum link models},}\
  }\href@noop {} {\  (\bibinfo {year} {2021}{\natexlab{d}})},\ \Eprint
  {http://arxiv.org/abs/2112.04501} {arXiv:2112.04501 [cond-mat.quant-gas]}
  \BibitemShut {NoStop}%
\bibitem [{\citenamefont {Buyens}\ \emph {et~al.}(2017)\citenamefont {Buyens},
  \citenamefont {Montangero}, \citenamefont {Haegeman}, \citenamefont
  {Verstraete},\ and\ \citenamefont {Van~Acoleyen}}]{Buyens2017}%
  \BibitemOpen
  \bibfield  {author} {\bibinfo {author} {\bibfnamefont {Boye}\ \bibnamefont
  {Buyens}}, \bibinfo {author} {\bibfnamefont {Simone}\ \bibnamefont
  {Montangero}}, \bibinfo {author} {\bibfnamefont {Jutho}\ \bibnamefont
  {Haegeman}}, \bibinfo {author} {\bibfnamefont {Frank}\ \bibnamefont
  {Verstraete}}, \ and\ \bibinfo {author} {\bibfnamefont {Karel}\ \bibnamefont
  {Van~Acoleyen}},\ }\bibfield  {title} {\enquote {\bibinfo {title}
  {Finite-representation approximation of lattice gauge theories at the
  continuum limit with tensor networks},}\ }\href {\doibase
  10.1103/PhysRevD.95.094509} {\bibfield  {journal} {\bibinfo  {journal} {Phys.
  Rev. D}\ }\textbf {\bibinfo {volume} {95}},\ \bibinfo {pages} {094509}
  (\bibinfo {year} {2017})}\BibitemShut {NoStop}%
\bibitem [{\citenamefont {Banuls}\ \emph {et~al.}(2019)\citenamefont {Banuls},
  \citenamefont {Cichy}, \citenamefont {Cirac}, \citenamefont {Jansen},\ and\
  \citenamefont {Kühn}}]{Banuls2018}%
  \BibitemOpen
  \bibfield  {author} {\bibinfo {author} {\bibfnamefont {Mari~Carmen}\
  \bibnamefont {Banuls}}, \bibinfo {author} {\bibfnamefont {Krzysztof}\
  \bibnamefont {Cichy}}, \bibinfo {author} {\bibfnamefont {J.~Ignacio}\
  \bibnamefont {Cirac}}, \bibinfo {author} {\bibfnamefont {Karl}\ \bibnamefont
  {Jansen}}, \ and\ \bibinfo {author} {\bibfnamefont {Stefan}\ \bibnamefont
  {Kühn}},\ }\bibfield  {title} {\enquote {\bibinfo {title} {{Tensor Networks
  and their use for Lattice Gauge Theories}},}\ }\href {\doibase
  10.22323/1.334.0022} {\bibfield  {journal} {\bibinfo  {journal} {PoS}\
  }\textbf {\bibinfo {volume} {LATTICE2018}},\ \bibinfo {pages} {022} (\bibinfo
  {year} {2019})}\BibitemShut {NoStop}%
\bibitem [{\citenamefont {Ba{\~{n}}uls}\ and\ \citenamefont
  {Cichy}(2020)}]{Banuls2020}%
  \BibitemOpen
  \bibfield  {author} {\bibinfo {author} {\bibfnamefont {Mari~Carmen}\
  \bibnamefont {Ba{\~{n}}uls}}\ and\ \bibinfo {author} {\bibfnamefont
  {Krzysztof}\ \bibnamefont {Cichy}},\ }\bibfield  {title} {\enquote {\bibinfo
  {title} {Review on novel methods for lattice gauge theories},}\ }\href
  {\doibase 10.1088/1361-6633/ab6311} {\bibfield  {journal} {\bibinfo
  {journal} {Reports on Progress in Physics}\ }\textbf {\bibinfo {volume}
  {83}},\ \bibinfo {pages} {024401} (\bibinfo {year} {2020})}\BibitemShut
  {NoStop}%
\bibitem [{\citenamefont {Abanin}\ \emph {et~al.}(2017)\citenamefont {Abanin},
  \citenamefont {De~Roeck}, \citenamefont {Ho},\ and\ \citenamefont
  {Huveneers}}]{abanin2017rigorous}%
  \BibitemOpen
  \bibfield  {author} {\bibinfo {author} {\bibfnamefont {Dmitry}\ \bibnamefont
  {Abanin}}, \bibinfo {author} {\bibfnamefont {Wojciech}\ \bibnamefont
  {De~Roeck}}, \bibinfo {author} {\bibfnamefont {Wen~Wei}\ \bibnamefont {Ho}},
  \ and\ \bibinfo {author} {\bibfnamefont {Fran{\c c}ois}\ \bibnamefont
  {Huveneers}},\ }\bibfield  {title} {\enquote {\bibinfo {title} {A rigorous
  theory of many-body prethermalization for periodically driven and closed
  quantum systems},}\ }\href {\doibase 10.1007/s00220-017-2930-x} {\bibfield
  {journal} {\bibinfo  {journal} {Communications in Mathematical Physics}\
  }\textbf {\bibinfo {volume} {354}},\ \bibinfo {pages} {809--827} (\bibinfo
  {year} {2017})}\BibitemShut {NoStop}%
\bibitem [{\citenamefont {Facchi}\ and\ \citenamefont
  {Pascazio}(2002)}]{facchi2002quantum}%
  \BibitemOpen
  \bibfield  {author} {\bibinfo {author} {\bibfnamefont {P.}~\bibnamefont
  {Facchi}}\ and\ \bibinfo {author} {\bibfnamefont {S.}~\bibnamefont
  {Pascazio}},\ }\bibfield  {title} {\enquote {\bibinfo {title} {Quantum zeno
  subspaces},}\ }\href {\doibase 10.1103/PhysRevLett.89.080401} {\bibfield
  {journal} {\bibinfo  {journal} {Phys. Rev. Lett.}\ }\textbf {\bibinfo
  {volume} {89}},\ \bibinfo {pages} {080401} (\bibinfo {year}
  {2002})}\BibitemShut {NoStop}%
\bibitem [{\citenamefont {Facchi}\ \emph {et~al.}(2004)\citenamefont {Facchi},
  \citenamefont {Lidar},\ and\ \citenamefont
  {Pascazio}}]{facchi2004unification}%
  \BibitemOpen
  \bibfield  {author} {\bibinfo {author} {\bibfnamefont {P.}~\bibnamefont
  {Facchi}}, \bibinfo {author} {\bibfnamefont {D.~A.}\ \bibnamefont {Lidar}}, \
  and\ \bibinfo {author} {\bibfnamefont {S.}~\bibnamefont {Pascazio}},\
  }\bibfield  {title} {\enquote {\bibinfo {title} {Unification of dynamical
  decoupling and the quantum zeno effect},}\ }\href {\doibase
  10.1103/PhysRevA.69.032314} {\bibfield  {journal} {\bibinfo  {journal} {Phys.
  Rev. A}\ }\textbf {\bibinfo {volume} {69}},\ \bibinfo {pages} {032314}
  (\bibinfo {year} {2004})}\BibitemShut {NoStop}%
\bibitem [{\citenamefont {Facchi}\ \emph {et~al.}(2009)\citenamefont {Facchi},
  \citenamefont {Marmo},\ and\ \citenamefont {Pascazio}}]{facchi2009quantum}%
  \BibitemOpen
  \bibfield  {author} {\bibinfo {author} {\bibfnamefont {Paolo}\ \bibnamefont
  {Facchi}}, \bibinfo {author} {\bibfnamefont {Giuseppe}\ \bibnamefont
  {Marmo}}, \ and\ \bibinfo {author} {\bibfnamefont {Saverio}\ \bibnamefont
  {Pascazio}},\ }\bibfield  {title} {\enquote {\bibinfo {title} {Quantum zeno
  dynamics and quantum zeno subspaces},}\ }\href {\doibase
  10.1088/1742-6596/196/1/012017} {\ \textbf {\bibinfo {volume} {196}},\
  \bibinfo {pages} {012017} (\bibinfo {year} {2009})}\BibitemShut {NoStop}%
\bibitem [{\citenamefont {Burgarth}\ \emph {et~al.}(2019)\citenamefont
  {Burgarth}, \citenamefont {Facchi}, \citenamefont {Nakazato}, \citenamefont
  {Pascazio},\ and\ \citenamefont {Yuasa}}]{burgarth2019generalized}%
  \BibitemOpen
  \bibfield  {author} {\bibinfo {author} {\bibfnamefont {Daniel}\ \bibnamefont
  {Burgarth}}, \bibinfo {author} {\bibfnamefont {Paolo}\ \bibnamefont
  {Facchi}}, \bibinfo {author} {\bibfnamefont {Hiromichi}\ \bibnamefont
  {Nakazato}}, \bibinfo {author} {\bibfnamefont {Saverio}\ \bibnamefont
  {Pascazio}}, \ and\ \bibinfo {author} {\bibfnamefont {Kazuya}\ \bibnamefont
  {Yuasa}},\ }\bibfield  {title} {\enquote {\bibinfo {title} {Generalized
  {A}diabatic {T}heorem and {S}trong-{C}oupling {L}imits},}\ }\href {\doibase
  10.22331/q-2019-06-12-152} {\bibfield  {journal} {\bibinfo  {journal}
  {{Quantum}}\ }\textbf {\bibinfo {volume} {3}},\ \bibinfo {pages} {152}
  (\bibinfo {year} {2019})}\BibitemShut {NoStop}%
\bibitem [{\citenamefont {Smith}\ \emph {et~al.}(2017)\citenamefont {Smith},
  \citenamefont {Knolle}, \citenamefont {Kovrizhin},\ and\ \citenamefont
  {Moessner}}]{Smith2017}%
  \BibitemOpen
  \bibfield  {author} {\bibinfo {author} {\bibfnamefont {A.}~\bibnamefont
  {Smith}}, \bibinfo {author} {\bibfnamefont {J.}~\bibnamefont {Knolle}},
  \bibinfo {author} {\bibfnamefont {D.~L.}\ \bibnamefont {Kovrizhin}}, \ and\
  \bibinfo {author} {\bibfnamefont {R.}~\bibnamefont {Moessner}},\ }\bibfield
  {title} {\enquote {\bibinfo {title} {Disorder-free localization},}\ }\href
  {\doibase 10.1103/PhysRevLett.118.266601} {\bibfield  {journal} {\bibinfo
  {journal} {Phys. Rev. Lett.}\ }\textbf {\bibinfo {volume} {118}},\ \bibinfo
  {pages} {266601} (\bibinfo {year} {2017})}\BibitemShut {NoStop}%
\bibitem [{\citenamefont {Brenes}\ \emph {et~al.}(2018)\citenamefont {Brenes},
  \citenamefont {Dalmonte}, \citenamefont {Heyl},\ and\ \citenamefont
  {Scardicchio}}]{Brenes2018}%
  \BibitemOpen
  \bibfield  {author} {\bibinfo {author} {\bibfnamefont {Marlon}\ \bibnamefont
  {Brenes}}, \bibinfo {author} {\bibfnamefont {Marcello}\ \bibnamefont
  {Dalmonte}}, \bibinfo {author} {\bibfnamefont {Markus}\ \bibnamefont {Heyl}},
  \ and\ \bibinfo {author} {\bibfnamefont {Antonello}\ \bibnamefont
  {Scardicchio}},\ }\bibfield  {title} {\enquote {\bibinfo {title} {Many-body
  localization dynamics from gauge invariance},}\ }\href {\doibase
  10.1103/PhysRevLett.120.030601} {\bibfield  {journal} {\bibinfo  {journal}
  {Phys. Rev. Lett.}\ }\textbf {\bibinfo {volume} {120}},\ \bibinfo {pages}
  {030601} (\bibinfo {year} {2018})}\BibitemShut {NoStop}%
\bibitem [{\citenamefont {Halimeh}\ \emph
  {et~al.}(2021{\natexlab{e}})\citenamefont {Halimeh}, \citenamefont {Zhao},
  \citenamefont {Hauke},\ and\ \citenamefont
  {Knolle}}]{Halimeh2021stabilizingDFL}%
  \BibitemOpen
  \bibfield  {author} {\bibinfo {author} {\bibfnamefont {Jad~C.}\ \bibnamefont
  {Halimeh}}, \bibinfo {author} {\bibfnamefont {Hongzheng}\ \bibnamefont
  {Zhao}}, \bibinfo {author} {\bibfnamefont {Philipp}\ \bibnamefont {Hauke}}, \
  and\ \bibinfo {author} {\bibfnamefont {Johannes}\ \bibnamefont {Knolle}},\
  }\bibfield  {title} {\enquote {\bibinfo {title} {Stabilizing disorder-free
  localization},}\ }\href@noop {} {\  (\bibinfo {year} {2021}{\natexlab{e}})},\
  \Eprint {http://arxiv.org/abs/2111.02427} {arXiv:2111.02427
  [cond-mat.dis-nn]} \BibitemShut {NoStop}%
\bibitem [{\citenamefont {Barbiero}\ \emph {et~al.}(2019)\citenamefont
  {Barbiero}, \citenamefont {Schweizer}, \citenamefont {Aidelsburger},
  \citenamefont {Demler}, \citenamefont {Goldman},\ and\ \citenamefont
  {Grusdt}}]{Barbiero2019}%
  \BibitemOpen
  \bibfield  {author} {\bibinfo {author} {\bibfnamefont {Luca}\ \bibnamefont
  {Barbiero}}, \bibinfo {author} {\bibfnamefont {Christian}\ \bibnamefont
  {Schweizer}}, \bibinfo {author} {\bibfnamefont {Monika}\ \bibnamefont
  {Aidelsburger}}, \bibinfo {author} {\bibfnamefont {Eugene}\ \bibnamefont
  {Demler}}, \bibinfo {author} {\bibfnamefont {Nathan}\ \bibnamefont
  {Goldman}}, \ and\ \bibinfo {author} {\bibfnamefont {Fabian}\ \bibnamefont
  {Grusdt}},\ }\bibfield  {title} {\enquote {\bibinfo {title} {Coupling
  ultracold matter to dynamical gauge fields in optical lattices: From flux
  attachment to $\mathbb{Z}_2$ lattice gauge theories},}\ }\href {\doibase
  10.1126/sciadv.aav7444} {\bibfield  {journal} {\bibinfo  {journal} {Science
  Advances}\ }\textbf {\bibinfo {volume} {5}} (\bibinfo {year} {2019}),\
  10.1126/sciadv.aav7444}\BibitemShut {NoStop}%
\bibitem [{\citenamefont {Zohar}\ \emph {et~al.}(2017)\citenamefont {Zohar},
  \citenamefont {Farace}, \citenamefont {Reznik},\ and\ \citenamefont
  {Cirac}}]{Zohar2017}%
  \BibitemOpen
  \bibfield  {author} {\bibinfo {author} {\bibfnamefont {Erez}\ \bibnamefont
  {Zohar}}, \bibinfo {author} {\bibfnamefont {Alessandro}\ \bibnamefont
  {Farace}}, \bibinfo {author} {\bibfnamefont {Benni}\ \bibnamefont {Reznik}},
  \ and\ \bibinfo {author} {\bibfnamefont {J.~Ignacio}\ \bibnamefont {Cirac}},\
  }\bibfield  {title} {\enquote {\bibinfo {title} {Digital quantum simulation
  of ${\mathbb{z}}_{2}$ lattice gauge theories with dynamical fermionic
  matter},}\ }\href {\doibase 10.1103/PhysRevLett.118.070501} {\bibfield
  {journal} {\bibinfo  {journal} {Phys. Rev. Lett.}\ }\textbf {\bibinfo
  {volume} {118}},\ \bibinfo {pages} {070501} (\bibinfo {year}
  {2017})}\BibitemShut {NoStop}%
\bibitem [{\citenamefont {Borla}\ \emph {et~al.}(2020)\citenamefont {Borla},
  \citenamefont {Verresen}, \citenamefont {Grusdt},\ and\ \citenamefont
  {Moroz}}]{Borla2019}%
  \BibitemOpen
  \bibfield  {author} {\bibinfo {author} {\bibfnamefont {Umberto}\ \bibnamefont
  {Borla}}, \bibinfo {author} {\bibfnamefont {Ruben}\ \bibnamefont {Verresen}},
  \bibinfo {author} {\bibfnamefont {Fabian}\ \bibnamefont {Grusdt}}, \ and\
  \bibinfo {author} {\bibfnamefont {Sergej}\ \bibnamefont {Moroz}},\ }\bibfield
   {title} {\enquote {\bibinfo {title} {Confined phases of one-dimensional
  spinless fermions coupled to ${Z}_{2}$ gauge theory},}\ }\href {\doibase
  10.1103/PhysRevLett.124.120503} {\bibfield  {journal} {\bibinfo  {journal}
  {Phys. Rev. Lett.}\ }\textbf {\bibinfo {volume} {124}},\ \bibinfo {pages}
  {120503} (\bibinfo {year} {2020})}\BibitemShut {NoStop}%
\bibitem [{\citenamefont {Yang}\ \emph
  {et~al.}(2020{\natexlab{b}})\citenamefont {Yang}, \citenamefont {Liu},
  \citenamefont {Gorshkov},\ and\ \citenamefont
  {Iadecola}}]{Yang2020fragmentation}%
  \BibitemOpen
  \bibfield  {author} {\bibinfo {author} {\bibfnamefont {Zhi-Cheng}\
  \bibnamefont {Yang}}, \bibinfo {author} {\bibfnamefont {Fangli}\ \bibnamefont
  {Liu}}, \bibinfo {author} {\bibfnamefont {Alexey~V.}\ \bibnamefont
  {Gorshkov}}, \ and\ \bibinfo {author} {\bibfnamefont {Thomas}\ \bibnamefont
  {Iadecola}},\ }\bibfield  {title} {\enquote {\bibinfo {title} {Hilbert-space
  fragmentation from strict confinement},}\ }\href {\doibase
  10.1103/PhysRevLett.124.207602} {\bibfield  {journal} {\bibinfo  {journal}
  {Phys. Rev. Lett.}\ }\textbf {\bibinfo {volume} {124}},\ \bibinfo {pages}
  {207602} (\bibinfo {year} {2020}{\natexlab{b}})}\BibitemShut {NoStop}%
\bibitem [{\citenamefont {Kebri\ifmmode~\check{c}\else \v{c}\fi{}}\ \emph
  {et~al.}(2021)\citenamefont {Kebri\ifmmode~\check{c}\else \v{c}\fi{}},
  \citenamefont {Barbiero}, \citenamefont {Reinmoser}, \citenamefont
  {Schollw\"ock},\ and\ \citenamefont {Grusdt}}]{kebric2021confinement}%
  \BibitemOpen
  \bibfield  {author} {\bibinfo {author} {\bibfnamefont {Matja\ifmmode
  \check{z}\else~\v{z}\fi{}}\ \bibnamefont {Kebri\ifmmode~\check{c}\else
  \v{c}\fi{}}}, \bibinfo {author} {\bibfnamefont {Luca}\ \bibnamefont
  {Barbiero}}, \bibinfo {author} {\bibfnamefont {Christian}\ \bibnamefont
  {Reinmoser}}, \bibinfo {author} {\bibfnamefont {Ulrich}\ \bibnamefont
  {Schollw\"ock}}, \ and\ \bibinfo {author} {\bibfnamefont {Fabian}\
  \bibnamefont {Grusdt}},\ }\bibfield  {title} {\enquote {\bibinfo {title}
  {Confinement and mott transitions of dynamical charges in one-dimensional
  lattice gauge theories},}\ }\href {\doibase 10.1103/PhysRevLett.127.167203}
  {\bibfield  {journal} {\bibinfo  {journal} {Phys. Rev. Lett.}\ }\textbf
  {\bibinfo {volume} {127}},\ \bibinfo {pages} {167203} (\bibinfo {year}
  {2021})}\BibitemShut {NoStop}%
\bibitem [{\citenamefont {{Halimeh}}\ \emph {et~al.}(2022)\citenamefont
  {{Halimeh}}, \citenamefont {{Barbiero}}, \citenamefont {{Hauke}},
  \citenamefont {{Grusdt}},\ and\ \citenamefont
  {{Bohrdt}}}]{Halimeh2022robust}%
  \BibitemOpen
  \bibfield  {author} {\bibinfo {author} {\bibfnamefont {Jad~C.}\ \bibnamefont
  {{Halimeh}}}, \bibinfo {author} {\bibfnamefont {Luca}\ \bibnamefont
  {{Barbiero}}}, \bibinfo {author} {\bibfnamefont {Philipp}\ \bibnamefont
  {{Hauke}}}, \bibinfo {author} {\bibfnamefont {Fabian}\ \bibnamefont
  {{Grusdt}}}, \ and\ \bibinfo {author} {\bibfnamefont {Annabelle}\
  \bibnamefont {{Bohrdt}}},\ }\bibfield  {title} {\enquote {\bibinfo {title}
  {{Robust quantum many-body scars in lattice gauge theories}},}\ }\href@noop
  {} {\bibfield  {journal} {\bibinfo  {journal} {arXiv e-prints}\ ,\ \bibinfo
  {eid} {arXiv:2203.08828}} (\bibinfo {year} {2022})},\ \Eprint
  {http://arxiv.org/abs/2203.08828} {arXiv:2203.08828 [cond-mat.quant-gas]}
  \BibitemShut {NoStop}%
\bibitem [{\citenamefont {Damme}\ \emph
  {et~al.}(2021{\natexlab{b}})\citenamefont {Damme}, \citenamefont
  {Mildenberger}, \citenamefont {Grusdt}, \citenamefont {Hauke},\ and\
  \citenamefont {Halimeh}}]{vandamme2021suppressing}%
  \BibitemOpen
  \bibfield  {author} {\bibinfo {author} {\bibfnamefont {Maarten~Van}\
  \bibnamefont {Damme}}, \bibinfo {author} {\bibfnamefont {Julius}\
  \bibnamefont {Mildenberger}}, \bibinfo {author} {\bibfnamefont {Fabian}\
  \bibnamefont {Grusdt}}, \bibinfo {author} {\bibfnamefont {Philipp}\
  \bibnamefont {Hauke}}, \ and\ \bibinfo {author} {\bibfnamefont {Jad~C.}\
  \bibnamefont {Halimeh}},\ }\bibfield  {title} {\enquote {\bibinfo {title}
  {Suppressing nonperturbative gauge errors in the thermodynamic limit using
  local pseudogenerators},}\ }\href@noop {} {\  (\bibinfo {year}
  {2021}{\natexlab{b}})},\ \Eprint {http://arxiv.org/abs/2110.08041}
  {arXiv:2110.08041 [quant-ph]} \BibitemShut {NoStop}%
\bibitem [{\citenamefont {{Desaules}}\ \emph {et~al.}(2022)\citenamefont
  {{Desaules}}, \citenamefont {{Banerjee}}, \citenamefont {{Hudomal}},
  \citenamefont {{Papi{\'c}}}, \citenamefont {{Sen}},\ and\ \citenamefont
  {{Halimeh}}}]{Desaules2022weak}%
  \BibitemOpen
  \bibfield  {author} {\bibinfo {author} {\bibfnamefont {Jean-Yves}\
  \bibnamefont {{Desaules}}}, \bibinfo {author} {\bibfnamefont {Debasish}\
  \bibnamefont {{Banerjee}}}, \bibinfo {author} {\bibfnamefont {Ana}\
  \bibnamefont {{Hudomal}}}, \bibinfo {author} {\bibfnamefont {Zlatko}\
  \bibnamefont {{Papi{\'c}}}}, \bibinfo {author} {\bibfnamefont {Arnab}\
  \bibnamefont {{Sen}}}, \ and\ \bibinfo {author} {\bibfnamefont {Jad~C.}\
  \bibnamefont {{Halimeh}}},\ }\bibfield  {title} {\enquote {\bibinfo {title}
  {{Weak Ergodicity Breaking in the Schwinger Model}},}\ }\href@noop {}
  {\bibfield  {journal} {\bibinfo  {journal} {arXiv e-prints}\ ,\ \bibinfo
  {eid} {arXiv:2203.08830}} (\bibinfo {year} {2022})},\ \Eprint
  {http://arxiv.org/abs/2203.08830} {arXiv:2203.08830 [cond-mat.str-el]}
  \BibitemShut {NoStop}%
\bibitem [{\citenamefont {Desaules}\ \emph {et~al.}(2022)\citenamefont
  {Desaules}, \citenamefont {Hudomal}, \citenamefont {Banerjee}, \citenamefont
  {Sen}, \citenamefont {Papić},\ and\ \citenamefont
  {Halimeh}}]{Desaules2022prominent}%
  \BibitemOpen
  \bibfield  {author} {\bibinfo {author} {\bibfnamefont {Jean-Yves}\
  \bibnamefont {Desaules}}, \bibinfo {author} {\bibfnamefont {Ana}\
  \bibnamefont {Hudomal}}, \bibinfo {author} {\bibfnamefont {Debasish}\
  \bibnamefont {Banerjee}}, \bibinfo {author} {\bibfnamefont {Arnab}\
  \bibnamefont {Sen}}, \bibinfo {author} {\bibfnamefont {Zlatko}\ \bibnamefont
  {Papić}}, \ and\ \bibinfo {author} {\bibfnamefont {Jad~C.}\ \bibnamefont
  {Halimeh}},\ }\bibfield  {title} {\enquote {\bibinfo {title} {Prominent
  quantum many-body scars in a truncated schwinger model},}\ }\href {\doibase
  10.48550/ARXIV.2204.01745} {\  (\bibinfo {year} {2022}),\
  10.48550/ARXIV.2204.01745}\BibitemShut {NoStop}%
\bibitem [{\citenamefont {Iadecola}\ and\ \citenamefont
  {Schecter}(2020)}]{Iadecola2020}%
  \BibitemOpen
  \bibfield  {author} {\bibinfo {author} {\bibfnamefont {Thomas}\ \bibnamefont
  {Iadecola}}\ and\ \bibinfo {author} {\bibfnamefont {Michael}\ \bibnamefont
  {Schecter}},\ }\bibfield  {title} {\enquote {\bibinfo {title} {Quantum
  many-body scar states with emergent kinetic constraints and
  finite-entanglement revivals},}\ }\href {\doibase
  10.1103/PhysRevB.101.024306} {\bibfield  {journal} {\bibinfo  {journal}
  {Phys. Rev. B}\ }\textbf {\bibinfo {volume} {101}},\ \bibinfo {pages}
  {024306} (\bibinfo {year} {2020})}\BibitemShut {NoStop}%
\bibitem [{\citenamefont {Aramthottil}\ \emph {et~al.}(2022)\citenamefont
  {Aramthottil}, \citenamefont {Bhattacharya}, \citenamefont
  {González-Cuadra}, \citenamefont {Lewenstein}, \citenamefont {Barbiero},\
  and\ \citenamefont {Zakrzewski}}]{aramthottil2022scar}%
  \BibitemOpen
  \bibfield  {author} {\bibinfo {author} {\bibfnamefont {Adith~Sai}\
  \bibnamefont {Aramthottil}}, \bibinfo {author} {\bibfnamefont {Utso}\
  \bibnamefont {Bhattacharya}}, \bibinfo {author} {\bibfnamefont {Daniel}\
  \bibnamefont {González-Cuadra}}, \bibinfo {author} {\bibfnamefont {Maciej}\
  \bibnamefont {Lewenstein}}, \bibinfo {author} {\bibfnamefont {Luca}\
  \bibnamefont {Barbiero}}, \ and\ \bibinfo {author} {\bibfnamefont {Jakub}\
  \bibnamefont {Zakrzewski}},\ }\bibfield  {title} {\enquote {\bibinfo {title}
  {Scar states in deconfined $\mathbb{Z}_2$ lattice gauge theories},}\ }\href
  {\doibase 10.48550/ARXIV.2201.10260} {\  (\bibinfo {year} {2022}),\
  10.48550/ARXIV.2201.10260}\BibitemShut {NoStop}%
\bibitem [{\citenamefont {Shiraishi}\ and\ \citenamefont
  {Mori}(2017)}]{ShiraishiMori}%
  \BibitemOpen
  \bibfield  {author} {\bibinfo {author} {\bibfnamefont {Naoto}\ \bibnamefont
  {Shiraishi}}\ and\ \bibinfo {author} {\bibfnamefont {Takashi}\ \bibnamefont
  {Mori}},\ }\bibfield  {title} {\enquote {\bibinfo {title} {Systematic
  construction of counterexamples to the eigenstate thermalization
  hypothesis},}\ }\href {\doibase 10.1103/PhysRevLett.119.030601} {\bibfield
  {journal} {\bibinfo  {journal} {Phys. Rev. Lett.}\ }\textbf {\bibinfo
  {volume} {119}},\ \bibinfo {pages} {030601} (\bibinfo {year}
  {2017})}\BibitemShut {NoStop}%
\bibitem [{\citenamefont {Moudgalya}\ \emph
  {et~al.}(2018{\natexlab{a}})\citenamefont {Moudgalya}, \citenamefont
  {Rachel}, \citenamefont {Bernevig},\ and\ \citenamefont
  {Regnault}}]{Moudgalya2018}%
  \BibitemOpen
  \bibfield  {author} {\bibinfo {author} {\bibfnamefont {Sanjay}\ \bibnamefont
  {Moudgalya}}, \bibinfo {author} {\bibfnamefont {Stephan}\ \bibnamefont
  {Rachel}}, \bibinfo {author} {\bibfnamefont {B.~Andrei}\ \bibnamefont
  {Bernevig}}, \ and\ \bibinfo {author} {\bibfnamefont {Nicolas}\ \bibnamefont
  {Regnault}},\ }\bibfield  {title} {\enquote {\bibinfo {title} {Exact excited
  states of nonintegrable models},}\ }\href {\doibase
  10.1103/PhysRevB.98.235155} {\bibfield  {journal} {\bibinfo  {journal} {Phys.
  Rev. B}\ }\textbf {\bibinfo {volume} {98}},\ \bibinfo {pages} {235155}
  (\bibinfo {year} {2018}{\natexlab{a}})}\BibitemShut {NoStop}%
\bibitem [{\citenamefont {Moudgalya}\ \emph
  {et~al.}(2018{\natexlab{b}})\citenamefont {Moudgalya}, \citenamefont
  {Regnault},\ and\ \citenamefont {Bernevig}}]{BernevigEnt}%
  \BibitemOpen
  \bibfield  {author} {\bibinfo {author} {\bibfnamefont {Sanjay}\ \bibnamefont
  {Moudgalya}}, \bibinfo {author} {\bibfnamefont {Nicolas}\ \bibnamefont
  {Regnault}}, \ and\ \bibinfo {author} {\bibfnamefont {B.~Andrei}\
  \bibnamefont {Bernevig}},\ }\bibfield  {title} {\enquote {\bibinfo {title}
  {Entanglement of exact excited states of {Affleck-Kennedy-Lieb-Tasaki}
  models: Exact results, many-body scars, and violation of the strong
  eigenstate thermalization hypothesis},}\ }\href {\doibase
  10.1103/PhysRevB.98.235156} {\bibfield  {journal} {\bibinfo  {journal} {Phys.
  Rev. B}\ }\textbf {\bibinfo {volume} {98}},\ \bibinfo {pages} {235156}
  (\bibinfo {year} {2018}{\natexlab{b}})}\BibitemShut {NoStop}%
\bibitem [{\citenamefont {Lin}\ and\ \citenamefont
  {Motrunich}(2019)}]{lin2018exact}%
  \BibitemOpen
  \bibfield  {author} {\bibinfo {author} {\bibfnamefont {Cheng-Ju}\
  \bibnamefont {Lin}}\ and\ \bibinfo {author} {\bibfnamefont {Olexei~I.}\
  \bibnamefont {Motrunich}},\ }\bibfield  {title} {\enquote {\bibinfo {title}
  {Exact quantum many-body scar states in the {Rydberg}-blockaded atom
  chain},}\ }\href {\doibase 10.1103/PhysRevLett.122.173401} {\bibfield
  {journal} {\bibinfo  {journal} {Phys. Rev. Lett.}\ }\textbf {\bibinfo
  {volume} {122}},\ \bibinfo {pages} {173401} (\bibinfo {year}
  {2019})}\BibitemShut {NoStop}%
\bibitem [{\citenamefont {Schecter}\ and\ \citenamefont
  {Iadecola}(2019)}]{Iadecola2019_2}%
  \BibitemOpen
  \bibfield  {author} {\bibinfo {author} {\bibfnamefont {Michael}\ \bibnamefont
  {Schecter}}\ and\ \bibinfo {author} {\bibfnamefont {Thomas}\ \bibnamefont
  {Iadecola}},\ }\bibfield  {title} {\enquote {\bibinfo {title} {Weak
  ergodicity breaking and quantum many-body scars in spin-1 {XY} magnets},}\
  }\href {\doibase 10.1103/PhysRevLett.123.147201} {\bibfield  {journal}
  {\bibinfo  {journal} {Phys. Rev. Lett.}\ }\textbf {\bibinfo {volume} {123}},\
  \bibinfo {pages} {147201} (\bibinfo {year} {2019})}\BibitemShut {NoStop}%
\bibitem [{\citenamefont {Mark}\ \emph {et~al.}(2020)\citenamefont {Mark},
  \citenamefont {Lin},\ and\ \citenamefont {Motrunich}}]{MotrunichTowers}%
  \BibitemOpen
  \bibfield  {author} {\bibinfo {author} {\bibfnamefont {Daniel~K.}\
  \bibnamefont {Mark}}, \bibinfo {author} {\bibfnamefont {Cheng-Ju}\
  \bibnamefont {Lin}}, \ and\ \bibinfo {author} {\bibfnamefont {Olexei~I.}\
  \bibnamefont {Motrunich}},\ }\bibfield  {title} {\enquote {\bibinfo {title}
  {Unified structure for exact towers of scar states in the
  {Affleck}-{Kennedy}-{Lieb}-{Tasaki} and other models},}\ }\href {\doibase
  10.1103/PhysRevB.101.195131} {\bibfield  {journal} {\bibinfo  {journal}
  {Phys. Rev. B}\ }\textbf {\bibinfo {volume} {101}},\ \bibinfo {pages}
  {195131} (\bibinfo {year} {2020})}\BibitemShut {NoStop}%
\bibitem [{\citenamefont {Turner}\ \emph {et~al.}(2018)\citenamefont {Turner},
  \citenamefont {Michailidis}, \citenamefont {Abanin}, \citenamefont {Serbyn},\
  and\ \citenamefont {Papi{\'c}}}]{Turner2018}%
  \BibitemOpen
  \bibfield  {author} {\bibinfo {author} {\bibfnamefont {C.~J.}\ \bibnamefont
  {Turner}}, \bibinfo {author} {\bibfnamefont {A.~A.}\ \bibnamefont
  {Michailidis}}, \bibinfo {author} {\bibfnamefont {D.~A.}\ \bibnamefont
  {Abanin}}, \bibinfo {author} {\bibfnamefont {M.}~\bibnamefont {Serbyn}}, \
  and\ \bibinfo {author} {\bibfnamefont {Z.}~\bibnamefont {Papi{\'c}}},\
  }\bibfield  {title} {\enquote {\bibinfo {title} {Weak ergodicity breaking
  from quantum many-body scars},}\ }\href {\doibase 10.1038/s41567-018-0137-5}
  {\bibfield  {journal} {\bibinfo  {journal} {Nature Physics}\ }\textbf
  {\bibinfo {volume} {14}},\ \bibinfo {pages} {745--749} (\bibinfo {year}
  {2018})}\BibitemShut {NoStop}%
\bibitem [{\citenamefont {Surace}\ \emph {et~al.}(2021)\citenamefont {Surace},
  \citenamefont {Votto}, \citenamefont {Lazo}, \citenamefont {Silva},
  \citenamefont {Dalmonte},\ and\ \citenamefont {Giudici}}]{Surace2021}%
  \BibitemOpen
  \bibfield  {author} {\bibinfo {author} {\bibfnamefont {Federica~Maria}\
  \bibnamefont {Surace}}, \bibinfo {author} {\bibfnamefont {Matteo}\
  \bibnamefont {Votto}}, \bibinfo {author} {\bibfnamefont {Eduardo~Gonzalez}\
  \bibnamefont {Lazo}}, \bibinfo {author} {\bibfnamefont {Alessandro}\
  \bibnamefont {Silva}}, \bibinfo {author} {\bibfnamefont {Marcello}\
  \bibnamefont {Dalmonte}}, \ and\ \bibinfo {author} {\bibfnamefont {Giuliano}\
  \bibnamefont {Giudici}},\ }\bibfield  {title} {\enquote {\bibinfo {title}
  {Exact many-body scars and their stability in constrained quantum chains},}\
  }\href {\doibase 10.1103/PhysRevB.103.104302} {\bibfield  {journal} {\bibinfo
   {journal} {Phys. Rev. B}\ }\textbf {\bibinfo {volume} {103}},\ \bibinfo
  {pages} {104302} (\bibinfo {year} {2021})}\BibitemShut {NoStop}%
\bibitem [{\citenamefont {Moler}\ and\ \citenamefont
  {Van~Loan}(2003)}]{Moler2003}%
  \BibitemOpen
  \bibfield  {author} {\bibinfo {author} {\bibfnamefont {Cleve}\ \bibnamefont
  {Moler}}\ and\ \bibinfo {author} {\bibfnamefont {Charles}\ \bibnamefont
  {Van~Loan}},\ }\bibfield  {title} {\enquote {\bibinfo {title} {Nineteen
  dubious ways to compute the exponential of a matrix, twenty-five years
  later},}\ }\href {\doibase 10.1137/S00361445024180} {\bibfield  {journal}
  {\bibinfo  {journal} {SIAM Review}\ }\textbf {\bibinfo {volume} {45}},\
  \bibinfo {pages} {3--49} (\bibinfo {year} {2003})},\ \Eprint
  {http://arxiv.org/abs/https://doi.org/10.1137/S00361445024180}
  {https://doi.org/10.1137/S00361445024180} \BibitemShut {NoStop}%
\bibitem [{\citenamefont {Sidje}(1998)}]{EXPOKIT}%
  \BibitemOpen
  \bibfield  {author} {\bibinfo {author} {\bibfnamefont {R.~B.}\ \bibnamefont
  {Sidje}},\ }\bibfield  {title} {\enquote {\bibinfo {title} {{\sc Expokit.}
  {A} software package for computing matrix exponentials},}\ }\href@noop {}
  {\bibfield  {journal} {\bibinfo  {journal} {ACM Trans. Math. Softw.}\
  }\textbf {\bibinfo {volume} {24}},\ \bibinfo {pages} {130--156} (\bibinfo
  {year} {1998})}\BibitemShut {NoStop}%
\bibitem [{\citenamefont {Lang}\ \emph {et~al.}(2022)\citenamefont {Lang},
  \citenamefont {Hauke}, \citenamefont {Knolle}, \citenamefont {Grusdt},\ and\
  \citenamefont {Halimeh}}]{lang2022disorder}%
  \BibitemOpen
  \bibfield  {author} {\bibinfo {author} {\bibfnamefont {Haifeng}\ \bibnamefont
  {Lang}}, \bibinfo {author} {\bibfnamefont {Philipp}\ \bibnamefont {Hauke}},
  \bibinfo {author} {\bibfnamefont {Johannes}\ \bibnamefont {Knolle}}, \bibinfo
  {author} {\bibfnamefont {Fabian}\ \bibnamefont {Grusdt}}, \ and\ \bibinfo
  {author} {\bibfnamefont {Jad~C}\ \bibnamefont {Halimeh}},\ }\bibfield
  {title} {\enquote {\bibinfo {title} {Disorder-free localization with stark
  gauge protection},}\ }\href@noop {} {\bibfield  {journal} {\bibinfo
  {journal} {arXiv preprint arXiv:2203.01338}\ } (\bibinfo {year}
  {2022})}\BibitemShut {NoStop}%
\bibitem [{\citenamefont {Smith}\ \emph {et~al.}(2018)\citenamefont {Smith},
  \citenamefont {Knolle}, \citenamefont {Moessner},\ and\ \citenamefont
  {Kovrizhin}}]{Smith2018}%
  \BibitemOpen
  \bibfield  {author} {\bibinfo {author} {\bibfnamefont {Adam}\ \bibnamefont
  {Smith}}, \bibinfo {author} {\bibfnamefont {Johannes}\ \bibnamefont
  {Knolle}}, \bibinfo {author} {\bibfnamefont {Roderich}\ \bibnamefont
  {Moessner}}, \ and\ \bibinfo {author} {\bibfnamefont {Dmitry~L.}\
  \bibnamefont {Kovrizhin}},\ }\bibfield  {title} {\enquote {\bibinfo {title}
  {Dynamical localization in ${\ensuremath{\mathbb{z}}}_{2}$ lattice gauge
  theories},}\ }\href {\doibase 10.1103/PhysRevB.97.245137} {\bibfield
  {journal} {\bibinfo  {journal} {Phys. Rev. B}\ }\textbf {\bibinfo {volume}
  {97}},\ \bibinfo {pages} {245137} (\bibinfo {year} {2018})}\BibitemShut
  {NoStop}%
\bibitem [{\citenamefont {Halimeh}\ \emph {et~al.}(2022)\citenamefont
  {Halimeh}, \citenamefont {McCulloch}, \citenamefont {Yang},\ and\
  \citenamefont {Hauke}}]{Halimeh2022Tuning}%
  \BibitemOpen
  \bibfield  {author} {\bibinfo {author} {\bibfnamefont {Jad~C.}\ \bibnamefont
  {Halimeh}}, \bibinfo {author} {\bibfnamefont {Ian~P.}\ \bibnamefont
  {McCulloch}}, \bibinfo {author} {\bibfnamefont {Bing}\ \bibnamefont {Yang}},
  \ and\ \bibinfo {author} {\bibfnamefont {Philipp}\ \bibnamefont {Hauke}},\
  }\bibfield  {title} {\enquote {\bibinfo {title} {Tuning the topological
  $\theta$-angle in cold-atom quantum simulators of gauge theories},}\ }\href
  {\doibase 10.48550/ARXIV.2204.06570} {\  (\bibinfo {year} {2022}),\
  10.48550/ARXIV.2204.06570}\BibitemShut {NoStop}%
\bibitem [{\citenamefont {Cheng}\ \emph {et~al.}(2022)\citenamefont {Cheng},
  \citenamefont {Liu}, \citenamefont {Zheng}, \citenamefont {Zhang},\ and\
  \citenamefont {Zhai}}]{Cheng2022tunable}%
  \BibitemOpen
  \bibfield  {author} {\bibinfo {author} {\bibfnamefont {Yanting}\ \bibnamefont
  {Cheng}}, \bibinfo {author} {\bibfnamefont {Shang}\ \bibnamefont {Liu}},
  \bibinfo {author} {\bibfnamefont {Wei}\ \bibnamefont {Zheng}}, \bibinfo
  {author} {\bibfnamefont {Pengfei}\ \bibnamefont {Zhang}}, \ and\ \bibinfo
  {author} {\bibfnamefont {Hui}\ \bibnamefont {Zhai}},\ }\bibfield  {title}
  {\enquote {\bibinfo {title} {Tunable confinement-deconfinement transition in
  an ultracold atom quantum simulator},}\ }\href {\doibase
  10.48550/ARXIV.2204.06586} {\  (\bibinfo {year} {2022}),\
  10.48550/ARXIV.2204.06586}\BibitemShut {NoStop}%
\bibitem [{\citenamefont {Zohar}\ \emph
  {et~al.}(2013{\natexlab{b}})\citenamefont {Zohar}, \citenamefont {Cirac},\
  and\ \citenamefont {Reznik}}]{Zohar2013SU2}%
  \BibitemOpen
  \bibfield  {author} {\bibinfo {author} {\bibfnamefont {Erez}\ \bibnamefont
  {Zohar}}, \bibinfo {author} {\bibfnamefont {J.~Ignacio}\ \bibnamefont
  {Cirac}}, \ and\ \bibinfo {author} {\bibfnamefont {Benni}\ \bibnamefont
  {Reznik}},\ }\bibfield  {title} {\enquote {\bibinfo {title} {Cold-atom
  quantum simulator for su(2) yang-mills lattice gauge theory},}\ }\href
  {\doibase 10.1103/PhysRevLett.110.125304} {\bibfield  {journal} {\bibinfo
  {journal} {Phys. Rev. Lett.}\ }\textbf {\bibinfo {volume} {110}},\ \bibinfo
  {pages} {125304} (\bibinfo {year} {2013}{\natexlab{b}})}\BibitemShut
  {NoStop}%
\bibitem [{\citenamefont {Mezzacapo}\ \emph {et~al.}(2015)\citenamefont
  {Mezzacapo}, \citenamefont {Rico}, \citenamefont {Sab\'{\i}n}, \citenamefont
  {Egusquiza}, \citenamefont {Lamata},\ and\ \citenamefont
  {Solano}}]{Mezzacapo2015}%
  \BibitemOpen
  \bibfield  {author} {\bibinfo {author} {\bibfnamefont {A.}~\bibnamefont
  {Mezzacapo}}, \bibinfo {author} {\bibfnamefont {E.}~\bibnamefont {Rico}},
  \bibinfo {author} {\bibfnamefont {C.}~\bibnamefont {Sab\'{\i}n}}, \bibinfo
  {author} {\bibfnamefont {I.~L.}\ \bibnamefont {Egusquiza}}, \bibinfo {author}
  {\bibfnamefont {L.}~\bibnamefont {Lamata}}, \ and\ \bibinfo {author}
  {\bibfnamefont {E.}~\bibnamefont {Solano}},\ }\bibfield  {title} {\enquote
  {\bibinfo {title} {Non-abelian su(2) lattice gauge theories in
  superconducting circuits},}\ }\href {\doibase 10.1103/PhysRevLett.115.240502}
  {\bibfield  {journal} {\bibinfo  {journal} {Phys. Rev. Lett.}\ }\textbf
  {\bibinfo {volume} {115}},\ \bibinfo {pages} {240502} (\bibinfo {year}
  {2015})}\BibitemShut {NoStop}%
\bibitem [{\citenamefont {Banerjee}\ \emph {et~al.}(2021)\citenamefont
  {Banerjee}, \citenamefont {Caspar}, \citenamefont {Jiang}, \citenamefont
  {Peng},\ and\ \citenamefont {Wiese}}]{Banerjee2021Nematic}%
  \BibitemOpen
  \bibfield  {author} {\bibinfo {author} {\bibfnamefont {D.}~\bibnamefont
  {Banerjee}}, \bibinfo {author} {\bibfnamefont {S.}~\bibnamefont {Caspar}},
  \bibinfo {author} {\bibfnamefont {F.~J.}\ \bibnamefont {Jiang}}, \bibinfo
  {author} {\bibfnamefont {J.~H.}\ \bibnamefont {Peng}}, \ and\ \bibinfo
  {author} {\bibfnamefont {U.~J.}\ \bibnamefont {Wiese}},\ }\bibfield  {title}
  {\enquote {\bibinfo {title} {Nematic confined phases in the $u(1)$ quantum
  link model on a triangular lattice: An opportunity for near-term quantum
  computations of string dynamics on a chip},}\ }\href {\doibase
  10.48550/ARXIV.2107.01283} {\  (\bibinfo {year} {2021}),\
  10.48550/ARXIV.2107.01283}\BibitemShut {NoStop}%
\bibitem [{\citenamefont {Hebenstreit}\ \emph {et~al.}(2013)\citenamefont
  {Hebenstreit}, \citenamefont {Berges},\ and\ \citenamefont
  {Gelfand}}]{Hebenstreit2013}%
  \BibitemOpen
  \bibfield  {author} {\bibinfo {author} {\bibfnamefont {F.}~\bibnamefont
  {Hebenstreit}}, \bibinfo {author} {\bibfnamefont {J.}~\bibnamefont {Berges}},
  \ and\ \bibinfo {author} {\bibfnamefont {D.}~\bibnamefont {Gelfand}},\
  }\bibfield  {title} {\enquote {\bibinfo {title} {Real-time dynamics of string
  breaking},}\ }\href {\doibase 10.1103/PhysRevLett.111.201601} {\bibfield
  {journal} {\bibinfo  {journal} {Phys. Rev. Lett.}\ }\textbf {\bibinfo
  {volume} {111}},\ \bibinfo {pages} {201601} (\bibinfo {year}
  {2013})}\BibitemShut {NoStop}%
\bibitem [{\citenamefont {Zache}\ \emph {et~al.}(2019)\citenamefont {Zache},
  \citenamefont {Mueller}, \citenamefont {Schneider}, \citenamefont
  {Jendrzejewski}, \citenamefont {Berges},\ and\ \citenamefont
  {Hauke}}]{Zache2019}%
  \BibitemOpen
  \bibfield  {author} {\bibinfo {author} {\bibfnamefont {T.~V.}\ \bibnamefont
  {Zache}}, \bibinfo {author} {\bibfnamefont {N.}~\bibnamefont {Mueller}},
  \bibinfo {author} {\bibfnamefont {J.~T.}\ \bibnamefont {Schneider}}, \bibinfo
  {author} {\bibfnamefont {F.}~\bibnamefont {Jendrzejewski}}, \bibinfo {author}
  {\bibfnamefont {J.}~\bibnamefont {Berges}}, \ and\ \bibinfo {author}
  {\bibfnamefont {P.}~\bibnamefont {Hauke}},\ }\bibfield  {title} {\enquote
  {\bibinfo {title} {Dynamical topological transitions in the massive schwinger
  model with a $\ensuremath{\theta}$ term},}\ }\href {\doibase
  10.1103/PhysRevLett.122.050403} {\bibfield  {journal} {\bibinfo  {journal}
  {Phys. Rev. Lett.}\ }\textbf {\bibinfo {volume} {122}},\ \bibinfo {pages}
  {050403} (\bibinfo {year} {2019})}\BibitemShut {NoStop}%
\bibitem [{\citenamefont {Kharzeev}\ \emph {et~al.}(2008)\citenamefont
  {Kharzeev}, \citenamefont {McLerran},\ and\ \citenamefont
  {Warringa}}]{Kharzeev2008}%
  \BibitemOpen
  \bibfield  {author} {\bibinfo {author} {\bibfnamefont {Dmitri~E.}\
  \bibnamefont {Kharzeev}}, \bibinfo {author} {\bibfnamefont {Larry~D.}\
  \bibnamefont {McLerran}}, \ and\ \bibinfo {author} {\bibfnamefont
  {Harmen~J.}\ \bibnamefont {Warringa}},\ }\bibfield  {title} {\enquote
  {\bibinfo {title} {The effects of topological charge change in heavy ion
  collisions: “event by event p and cp violation”},}\ }\href {\doibase
  https://doi.org/10.1016/j.nuclphysa.2008.02.298} {\bibfield  {journal}
  {\bibinfo  {journal} {Nuclear Physics A}\ }\textbf {\bibinfo {volume}
  {803}},\ \bibinfo {pages} {227--253} (\bibinfo {year} {2008})}\BibitemShut
  {NoStop}%
\bibitem [{\citenamefont {Fukushima}\ \emph {et~al.}(2008)\citenamefont
  {Fukushima}, \citenamefont {Kharzeev},\ and\ \citenamefont
  {Warringa}}]{Fukushima2008}%
  \BibitemOpen
  \bibfield  {author} {\bibinfo {author} {\bibfnamefont {Kenji}\ \bibnamefont
  {Fukushima}}, \bibinfo {author} {\bibfnamefont {Dmitri~E.}\ \bibnamefont
  {Kharzeev}}, \ and\ \bibinfo {author} {\bibfnamefont {Harmen~J.}\
  \bibnamefont {Warringa}},\ }\bibfield  {title} {\enquote {\bibinfo {title}
  {Chiral magnetic effect},}\ }\href {\doibase 10.1103/PhysRevD.78.074033}
  {\bibfield  {journal} {\bibinfo  {journal} {Phys. Rev. D}\ }\textbf {\bibinfo
  {volume} {78}},\ \bibinfo {pages} {074033} (\bibinfo {year}
  {2008})}\BibitemShut {NoStop}%
\bibitem [{\citenamefont {Kharzeev}\ \emph {et~al.}(2016)\citenamefont
  {Kharzeev}, \citenamefont {Liao}, \citenamefont {Voloshin},\ and\
  \citenamefont {Wang}}]{Kharzeev2016}%
  \BibitemOpen
  \bibfield  {author} {\bibinfo {author} {\bibfnamefont {D.E.}\ \bibnamefont
  {Kharzeev}}, \bibinfo {author} {\bibfnamefont {J.}~\bibnamefont {Liao}},
  \bibinfo {author} {\bibfnamefont {S.A.}\ \bibnamefont {Voloshin}}, \ and\
  \bibinfo {author} {\bibfnamefont {G.}~\bibnamefont {Wang}},\ }\bibfield
  {title} {\enquote {\bibinfo {title} {Chiral magnetic and vortical effects in
  high-energy nuclear collisions—a status report},}\ }\href {\doibase
  https://doi.org/10.1016/j.ppnp.2016.01.001} {\bibfield  {journal} {\bibinfo
  {journal} {Progress in Particle and Nuclear Physics}\ }\textbf {\bibinfo
  {volume} {88}},\ \bibinfo {pages} {1--28} (\bibinfo {year}
  {2016})}\BibitemShut {NoStop}%
\bibitem [{\citenamefont {Koch}\ \emph {et~al.}(2017)\citenamefont {Koch},
  \citenamefont {Schlichting}, \citenamefont {Skokov}, \citenamefont
  {Sorensen}, \citenamefont {Thomas}, \citenamefont {Voloshin}, \citenamefont
  {Wang},\ and\ \citenamefont {Yee}}]{Koch2017}%
  \BibitemOpen
  \bibfield  {author} {\bibinfo {author} {\bibfnamefont {Volker}\ \bibnamefont
  {Koch}}, \bibinfo {author} {\bibfnamefont {Soeren}\ \bibnamefont
  {Schlichting}}, \bibinfo {author} {\bibfnamefont {Vladimir}\ \bibnamefont
  {Skokov}}, \bibinfo {author} {\bibfnamefont {Paul}\ \bibnamefont {Sorensen}},
  \bibinfo {author} {\bibfnamefont {Jim}\ \bibnamefont {Thomas}}, \bibinfo
  {author} {\bibfnamefont {Sergei}\ \bibnamefont {Voloshin}}, \bibinfo {author}
  {\bibfnamefont {Gang}\ \bibnamefont {Wang}}, \ and\ \bibinfo {author}
  {\bibfnamefont {Ho-Ung}\ \bibnamefont {Yee}},\ }\bibfield  {title} {\enquote
  {\bibinfo {title} {Status of the chiral magnetic effect and collisions of
  isobars},}\ }\href {\doibase 10.1088/1674-1137/41/7/072001} {\bibfield
  {journal} {\bibinfo  {journal} {Chinese Physics C}\ }\textbf {\bibinfo
  {volume} {41}},\ \bibinfo {pages} {072001} (\bibinfo {year}
  {2017})}\BibitemShut {NoStop}%
\bibitem [{\citenamefont {Kharzeev}\ and\ \citenamefont
  {Kikuchi}(2020)}]{Kharzeev2020}%
  \BibitemOpen
  \bibfield  {author} {\bibinfo {author} {\bibfnamefont {Dmitri~E.}\
  \bibnamefont {Kharzeev}}\ and\ \bibinfo {author} {\bibfnamefont {Yuta}\
  \bibnamefont {Kikuchi}},\ }\bibfield  {title} {\enquote {\bibinfo {title}
  {Real-time chiral dynamics from a digital quantum simulation},}\ }\href
  {\doibase 10.1103/PhysRevResearch.2.023342} {\bibfield  {journal} {\bibinfo
  {journal} {Phys. Rev. Research}\ }\textbf {\bibinfo {volume} {2}},\ \bibinfo
  {pages} {023342} (\bibinfo {year} {2020})}\BibitemShut {NoStop}%
\bibitem [{\citenamefont {Kan}\ \emph {et~al.}(2021)\citenamefont {Kan},
  \citenamefont {Funcke}, \citenamefont {Kühn}, \citenamefont {Dellantonio},
  \citenamefont {Zhang}, \citenamefont {Haase}, \citenamefont {Muschik},\ and\
  \citenamefont {Jansen}}]{Kan2021}%
  \BibitemOpen
  \bibfield  {author} {\bibinfo {author} {\bibfnamefont {Angus}\ \bibnamefont
  {Kan}}, \bibinfo {author} {\bibfnamefont {Lena}\ \bibnamefont {Funcke}},
  \bibinfo {author} {\bibfnamefont {Stefan}\ \bibnamefont {Kühn}}, \bibinfo
  {author} {\bibfnamefont {Luca}\ \bibnamefont {Dellantonio}}, \bibinfo
  {author} {\bibfnamefont {Jinglei}\ \bibnamefont {Zhang}}, \bibinfo {author}
  {\bibfnamefont {Jan~F.}\ \bibnamefont {Haase}}, \bibinfo {author}
  {\bibfnamefont {Christine~A.}\ \bibnamefont {Muschik}}, \ and\ \bibinfo
  {author} {\bibfnamefont {Karl}\ \bibnamefont {Jansen}},\ }\bibfield  {title}
  {\enquote {\bibinfo {title} {3+1d $\theta$-term on the lattice from the
  hamiltonian perspective},}\ }\href {\doibase 10.48550/ARXIV.2111.02238} {\
  (\bibinfo {year} {2021}),\ 10.48550/ARXIV.2111.02238}\BibitemShut {NoStop}%
\bibitem [{\citenamefont {Cohen}\ \emph {et~al.}(2021)\citenamefont {Cohen},
  \citenamefont {Lamm}, \citenamefont {Lawrence},\ and\ \citenamefont
  {Yamauchi}}]{Cohen2021}%
  \BibitemOpen
  \bibfield  {author} {\bibinfo {author} {\bibfnamefont {Thomas~D.}\
  \bibnamefont {Cohen}}, \bibinfo {author} {\bibfnamefont {Henry}\ \bibnamefont
  {Lamm}}, \bibinfo {author} {\bibfnamefont {Scott}\ \bibnamefont {Lawrence}},
  \ and\ \bibinfo {author} {\bibfnamefont {Yukari}\ \bibnamefont {Yamauchi}}
  (\bibinfo {collaboration} {NuQS Collaboration}),\ }\bibfield  {title}
  {\enquote {\bibinfo {title} {Quantum algorithms for transport coefficients in
  gauge theories},}\ }\href {\doibase 10.1103/PhysRevD.104.094514} {\bibfield
  {journal} {\bibinfo  {journal} {Phys. Rev. D}\ }\textbf {\bibinfo {volume}
  {104}},\ \bibinfo {pages} {094514} (\bibinfo {year} {2021})}\BibitemShut
  {NoStop}%
\bibitem [{\citenamefont {Berges}(2019)}]{Berges2019}%
  \BibitemOpen
  \bibfield  {author} {\bibinfo {author} {\bibfnamefont {J.}~\bibnamefont
  {Berges}},\ }\bibfield  {title} {\enquote {\bibinfo {title} {Scaling up
  quantum simulations},}\ }\href {\doibase 10.1038/d41586-019-01483-1}
  {\bibfield  {journal} {\bibinfo  {journal} {Nature}\ }\textbf {\bibinfo
  {volume} {569}},\ \bibinfo {pages} {339--340} (\bibinfo {year}
  {2019})}\BibitemShut {NoStop}%
\bibitem [{\citenamefont {Mueller}\ \emph {et~al.}(2021)\citenamefont
  {Mueller}, \citenamefont {Zache},\ and\ \citenamefont {Ott}}]{Mueller2021}%
  \BibitemOpen
  \bibfield  {author} {\bibinfo {author} {\bibfnamefont {Niklas}\ \bibnamefont
  {Mueller}}, \bibinfo {author} {\bibfnamefont {Torsten~V.}\ \bibnamefont
  {Zache}}, \ and\ \bibinfo {author} {\bibfnamefont {Robert}\ \bibnamefont
  {Ott}},\ }\bibfield  {title} {\enquote {\bibinfo {title} {Quantum
  thermalization of gauge theories: chaos, turbulence and universality},}\
  }\href {\doibase 10.48550/ARXIV.2111.01155} {\  (\bibinfo {year} {2021}),\
  10.48550/ARXIV.2111.01155}\BibitemShut {NoStop}%
\bibitem [{\citenamefont {Zache}\ \emph {et~al.}(2018)\citenamefont {Zache},
  \citenamefont {Hebenstreit}, \citenamefont {Jendrzejewski}, \citenamefont
  {Oberthaler}, \citenamefont {Berges},\ and\ \citenamefont
  {Hauke}}]{Zache2018}%
  \BibitemOpen
  \bibfield  {author} {\bibinfo {author} {\bibfnamefont {T~V}\ \bibnamefont
  {Zache}}, \bibinfo {author} {\bibfnamefont {F}~\bibnamefont {Hebenstreit}},
  \bibinfo {author} {\bibfnamefont {F}~\bibnamefont {Jendrzejewski}}, \bibinfo
  {author} {\bibfnamefont {M~K}\ \bibnamefont {Oberthaler}}, \bibinfo {author}
  {\bibfnamefont {J}~\bibnamefont {Berges}}, \ and\ \bibinfo {author}
  {\bibfnamefont {P}~\bibnamefont {Hauke}},\ }\bibfield  {title} {\enquote
  {\bibinfo {title} {Quantum simulation of lattice gauge theories using wilson
  fermions},}\ }\href {\doibase 10.1088/2058-9565/aac33b} {\bibfield  {journal}
  {\bibinfo  {journal} {Quantum Science and Technology}\ }\textbf {\bibinfo
  {volume} {3}},\ \bibinfo {pages} {034010} (\bibinfo {year}
  {2018})}\BibitemShut {NoStop}%
\bibitem [{\citenamefont {Halimeh}\ and\ \citenamefont
  {Hauke}(2020{\natexlab{b}})}]{Halimeh2020b}%
  \BibitemOpen
  \bibfield  {author} {\bibinfo {author} {\bibfnamefont {Jad~C.}\ \bibnamefont
  {Halimeh}}\ and\ \bibinfo {author} {\bibfnamefont {Philipp}\ \bibnamefont
  {Hauke}},\ }\bibfield  {title} {\enquote {\bibinfo {title} {Staircase
  prethermalization and constrained dynamics in lattice gauge theories},}\
  }\href@noop {} {\  (\bibinfo {year} {2020}{\natexlab{b}})},\ \Eprint
  {http://arxiv.org/abs/2004.07248} {arXiv:2004.07248 [cond-mat.quant-gas]}
  \BibitemShut {NoStop}%
\bibitem [{\citenamefont {Halimeh}\ and\ \citenamefont
  {Hauke}(2020{\natexlab{c}})}]{Halimeh2020c}%
  \BibitemOpen
  \bibfield  {author} {\bibinfo {author} {\bibfnamefont {Jad~C.}\ \bibnamefont
  {Halimeh}}\ and\ \bibinfo {author} {\bibfnamefont {Philipp}\ \bibnamefont
  {Hauke}},\ }\bibfield  {title} {\enquote {\bibinfo {title} {Origin of
  staircase prethermalization in lattice gauge theories},}\ }\href@noop {} {\
  (\bibinfo {year} {2020}{\natexlab{c}})},\ \Eprint
  {http://arxiv.org/abs/2004.07254} {arXiv:2004.07254 [cond-mat.str-el]}
  \BibitemShut {NoStop}%
\end{thebibliography}%
\end{document}